\documentclass[a4paper,12pt]{article}

\usepackage{graphicx}
\usepackage{amssymb,amsmath,subfloat}
\usepackage{amsthm}
\usepackage{setspace}
\usepackage{geometry}
\usepackage{longtable}
\usepackage{multirow}
\usepackage{caption}
\usepackage{pdflscape}

\DeclareCaptionLabelFormat{cont}{#1~#2\alph{ContinuedFloat}}
\begin{document}


\begin{center}
\textbf{Study of vibrational spectra of polycyclic aromatic hydrocarbons with phenyl side group}
~\\
~\\
Anju Maurya, Rashmi Singh and Shantanu Rastogi*
\\
Department of Physics, D.D.U. Gorakhpur University, Gorakhpur, India - 273009
\\
* e-mail: shantanu\_r@hotmail.com

\end{center}

\begin{abstract}

Computational study of polycyclic aromatic hydrocarbons (PAHs) with phenyl side group substituted at different positions is reported. The infrared spectral variations due to the position of phenyl substitution, ionization state and the size of the molecules are discussed and possible contribution of phenyl-PAHs to the mid-infrared emission features from astrophysical objects is analyzed. Structurally phenyl group substitution at 2$^{nd}$ position gives more stable species compared to substitution at other positions. Phenyl-PAHs exhibit new aromatic bands near 695 and 741~cm$^{-1}$ (14.4 and 13.5 $\mu$m), due to contribution from quintet C-H wag, that compare well with minor features at 14.2 and 13.5 $\mu$m observed in several astrophysical objects. Just as in plain PAHs, the C-C stretch vibrational modes ($\sim$1600~cm$^{-1}$) have negligible intensity in neutrals, but the cations of all phenyl-PAHs exhibit significantly strong phenyl group C-C stretch peak close to class B type 6.2 $\mu$m astrophysical band. In 2-phenylpyrene, it is the neutral molecule that exhibits this strong feature in the 6.2 $\mu$m range along with other features that match with sub-features at 6.66 and 6.9 $\mu$m, observed in astronomical spectra of some late type objects. The substitution of phenyl side group at solo position shifts the C-C stretch mode of parent PAH close to the region of 6.2 $\mu$m astrophysical band. The results indicate possibility of phenyl-PAHs in space and the bottom-up formation of medium sized compact PAHs with phenyl side group in carbon rich cool circumstellar shells. Phenyl-PAHs need to be considered in modelling mid-infrared emission spectra of various astrophysical objects.

\end{abstract}

{\bf keywords:}

PAH; Aromatic Infrared Bands; Interstellar molecules; DFT calculations; Astrochemistry


\section{Introduction}

Polycyclic Aromatic Hydrocarbons (PAHs) are considered to be ubiquitous and among abundant species in the interstellar medium (ISM) \cite{Leger84, Allamandola85, Tielens2008, Li2020}. This is due to the presence of mid-infrared emission bands, which are most prominent at 3.3, 6.2, 7.7, 8.6, 11.2 and 12.7~$\mu$m (3030, 1610, 1300, 1160, 890 and 790~cm$^{-1}$), in a wide variety of astrophysical sources \cite{Cohen89, ISO96, Lutz98, Sajina2007, Matsuura2018, Lacy2020}. These features correspond to the vibrational transitions in aromatic molecules and are, therefore, referred to as aromatic infrared bands (AIBs). PAHs in the ISM form around carbon-rich AGB stars and evolve through harsh astrophysical environments leaving only the most stable species in the ISM \cite{Tielens2008, Tielens2013}. The mechanism of infrared emission involves absorption of ultraviolet photons from background sources (viz. hot central remnant in evolved stars or hot young O/B associations in star forming regions) and redistribution of this energy into the vibrational modes that fluoresces in the infrared \cite{Allamandola89, Leger89}. The resulting AIBs carry signatures of the collection of PAHs present in the environment.

The variation in the profile of AIB features correlates with type of astrophysical object, indicating different PAH types and populations in different regions \cite{Peeters2002, Diedenhoven2004, Pathak2008, Peeters2017}. The modelled emission spectra using quantum chemical data for a large set of PAHs show good match for the 7.0 - 9.0~$\mu$m AIB range from different sources \cite{Pathak2008, Peeters2017, Andrews2015, Boersma2018}. Despite the general similarity of AIBs to PAH vibrational spectrum and some feature correlations, the simultaneous matching of all AIBs has not been possible. In particular the 6.2~$\mu$m feature and few features between 10 - 15~$\mu$m remain unmatched. When plain PAHs are considered the aromatic C-C stretching modes fall short by 30 - 40~cm$^{-1}$ of the 6.2~$\mu$m AIB \cite{Pathak2008}. To address these problems studies incorporating derivatives such as substituted PAHs, hydrogenated PAHs, dehydrogenated PAHs, protonated PAHs, PAHs with nitrogen heterocycles, vinyl--PAHs etc. \cite{Langhoff98, Beegle2001, Buragohain18, Tsuge2016, Hudgins2005, Maurya2012, Maurya2015, Pino2008, Galue2017} have been attempted with none or partial improvement.

The emissions between 10 - 15~$\mu$m reveal weak features \cite{Andrews2015, Boersma2018} that too remain unexplained by the C-H out-of-plane modes of peripheral hydrogen atoms in plain PAHs. The IR activity of C-H out-of-plane bend modes is governed by the number of dangling H atoms in an aromatic ring \cite{Hudgins1999}. The bands in 13.0 - 13.7~$\mu$m region indicate quartet (four adjacent Hydrogens in a ring) while quintet (five adjacent Hydrogens in a ring) have additional features in 14.0 - 14.5~$\mu$m range \cite{Hudgins1999, Hony2001, Bauschlicher2009}. Quintet modes are possible only in phenyl group. The phenyl radical has infrared spectrum similar to benzene and other neutral PAHs, with a very weak C-C stretch mode at 1624~cm$^{-1}$ \cite{Friderichsen2001}. But, the same phenyl as a side group, on aliphatic compounds and even in complex systems \cite{Aspartate96, Ngaojampa2015}, has a medium intense feature close to the 6.2~$\mu$m AIB.

The formation of aromatics in space is interpreted via combustion chemistry \cite{Cherchneff2011}, wherein phenyl radical is an important intermediate and may play a major role in chemistry of the ISM \cite{McMahon2003, Parker2012}. Modelling protoplanetary nebula conditions, of dense-warm gas irradiated by an strong UV field, has shown to produce benzene \cite{Woods2002}. The presence of CN could further lead to the formation of benzonitrile \cite{Woods2002}, which is reported to have been detected in Taurus Molecular Cloud \cite{McGuire2018}. Thus, the presence of phenyl group in the ISM is strongly indicated. Pyrolysis of simple aromatics could form polyphenyls \cite{Badger61, Shukla2008, Talbi2012}. Phenyl-PAHs are also obtained in laboratory study of shock against benzene \cite{Mimura2003}. Such shock conditions are possible in the expanding envelopes of protoplanetary nebulae, suggesting a finite possibility for PAHs acquiring phenyl side group in these environments.

Considering these aspects of phenyl substitution, the vibrational spectra of biphenyl, phenylnaphthalenes, phenylanthracenes, phenylpyrenes, phenylcoronene and phenylovalene are computed. Analysis of spectral feature variations is performed for phenyl group attached at different positions on the plain PAHs. Both neutral and cation state of the molecules are considered. The possible correlations of modes close to the astrophysical AIBs are identified and discussed in the present work.

\section{Computation and Results}

The density functional theory (DFT) calculations have proved propitious in evaluating infrared spectra of PAHs \cite{Langhoff96, Pathak2005, Pathak2006, Pathak2007, Bauschlicher2008, Bauschlicher2010, Ricca2012} and in using them to model AIBs from different sources \cite{Pathak2008, Peeters2017, Rosenberg2014}. For the study of phenyl-PAHs the DFT-B3LYP method in combination with the 6-31G(d) basis set is used to obtain optimized molecular structure and infrared vibrational frequencies. The GAMESS ab initio program \cite{Gamess1993} is used for the calculations. Study of ten phenyl-PAH molecules shown in Figure~\ref{1Fig} is performed, which include biphenyl, 1-phenylnaphthalene, 2-phenylnaphthalene, 2-phenylanthracene, 9-phenylanthracene, 1-phenylpyrene, 2-phenylpyrene, 4-phenylpyrene, 1-phenylcoronene and 14-phenylovalene. As molecules in the ISM are mostly ionized, study is performed for both neutrals and cations.

In all the molecules, the phenyl moiety is out-of-plane with respect to the parent PAH. The torsion angle for different position of phenyl substitution is in general different due to different steric hindrances. The molecular energy is computed for different out-of-plane angles of the phenyl moiety to study the behaviour near optimized minima. In biphenyl and for 2$^{nd}$ position substituted phenyl-PAHs, there is a sharp minima at torsion angle $\sim$40$^\circ$ (Figure~\ref{2FigA}a). While in 1 and 4 position substituted phenyl-PAHs, the optimized phenyl torsion angle is $\sim$55$^\circ$ out-of-plane of the parent PAH (Figure~\ref{2FigB}b), but the minima are not sharp and there is very little increment in energy upon increasing the torsion angle beyond the minima. The environment of the phenyl group is similar in biphenyl and 2- substituted phenyl PAHs, with symmetric placement of two adjacent hydrogen atoms. Whereas, in phenyl-PAHs substituted at 1 and 4 position there is asymmetric hydrogen placement i.e., there is only one hydrogen on the same ring as the phenyl group (Figure~\ref{1Fig}). In solo position substituted phenyl-PAHs (9-phenylanthracene and 14-phenylovalene), where there is no hydrogen on the ring with the phenyl group, the steric interaction is negligible and the torsion angle has little effect on the molecular energy. Hence, a shallow minima is obtained (Figure~\ref{2FigB}b) when the plane of phenyl group is almost perpendicular to the plane of the parent PAH. The molecular energy variation with torsion angle in cations follow a trend similar to their corresponding neutrals.

The values of final optimized phenyl torsion angles are given in Table~\ref{tab1} for both neutrals and cations. The result for neutral biphenyl agrees closely with the torsion angle reported earlier using various methods \cite{Bastiansen85, furuya98} and also with the internal rotation ($\sim$44$^\circ$) between phenyl groups in poly(p-phenylene) \cite{Soto92}. Upon ionization the torsion angle is reduced in all the molecules, with maximum reduction, by a factor of 2, in biphenyl. As the size increases, the change in torsion angle gets smaller upon ionization. This structural change upon ionization becomes negligible in 14-phenylovalene ($\sim$0.1$^\circ$). The variation in the torsion angle upon ionization depends on the change in steric interaction due to change in charge distribution on the hydrogen atoms of the parent moiety near the phenyl group. As the cation size increases the additional charge is distributed over a larger number of atoms, so large cations have smaller charge variation from neutrals \cite{Pathak2006} and hence lesser change in torsion angle. Table~\ref{tab1} also shows optimized energies and the change in energy upon ionization, i.e. removal of one electron. The energy change on ionization is nearly equal for all molecules, but as the molecule size increases the energy change gets slightly smaller.

To select suitable level of DFT theory and the size of the basis set it is important to compare the computed vibrational frequencies with experimental values. The use of B3LYP/4-31G for calculations on PAHs is well established and the vibrational frequencies show good comparison with experimental data \cite{Langhoff96, Pathak2005}. But use of higher basis sets will give better orbital representation leading to accurate and better results. Report on 2-vinyl-anthracene and 9-vinyl-anthracene \cite{Maurya2012} indicated the suitability of 6-31G(d) basis set for side group substituted PAHs. The computed frequencies need to be scaled for comparison with experimental spectrum and the scale factor depends on the level of theory used \cite{Langhoff96, Bauschlicher97, Maurya2012}. The current report uses 6-31G(d) basis and the two level frequency scaling procedure \cite{Maurya2012, Bauschlicher2018}. For the C-H stretching vibrations computed around 3000~cm$^{-1}$ the scale factor considered is 0.9603 and for the 1800~-~500~cm$^{-1}$ range it is 0.9697. 

\begin{figure}
\centerline{\includegraphics[width=0.8\textwidth]{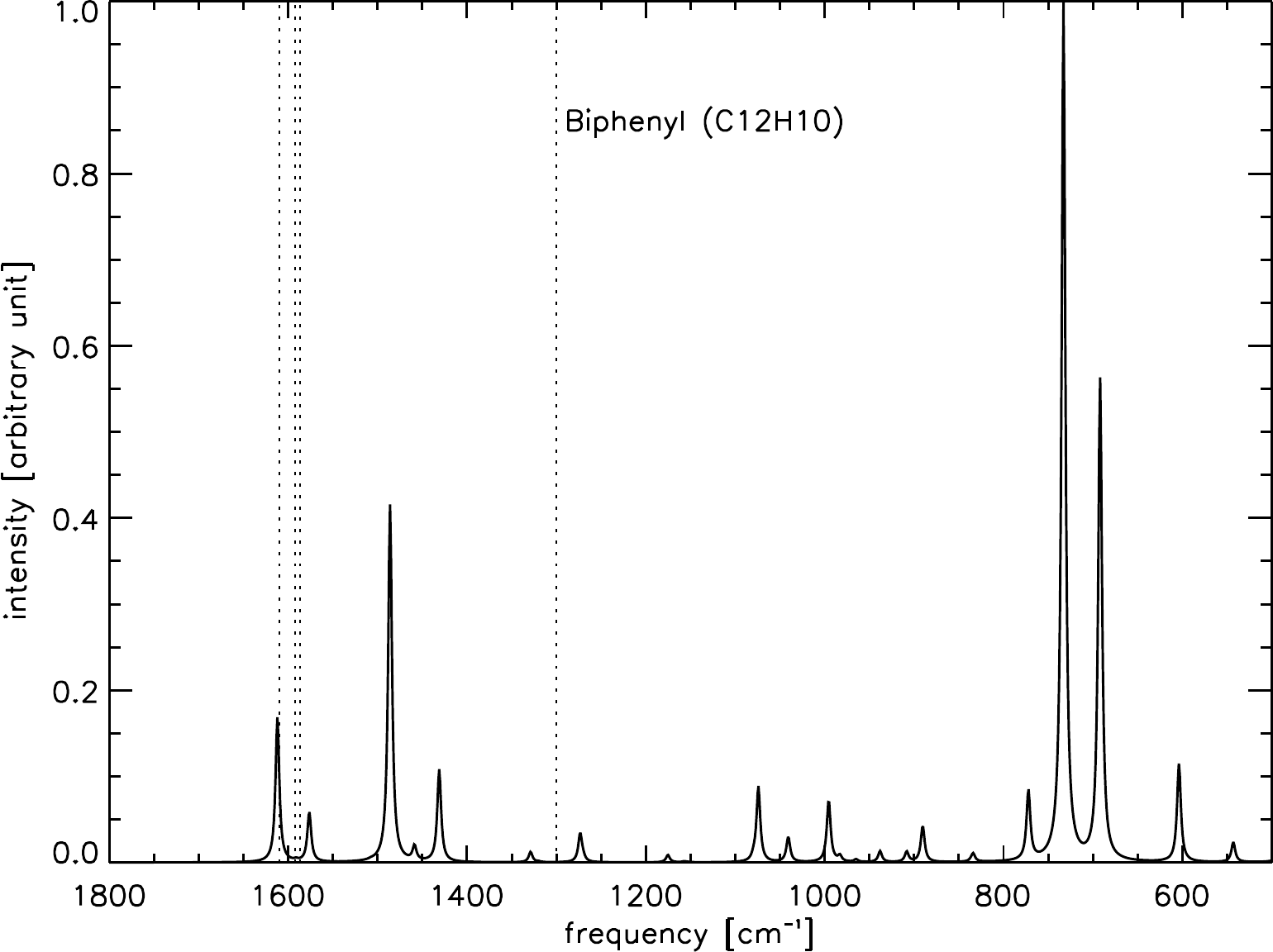}}
\centerline{\includegraphics[width=0.8\textwidth]{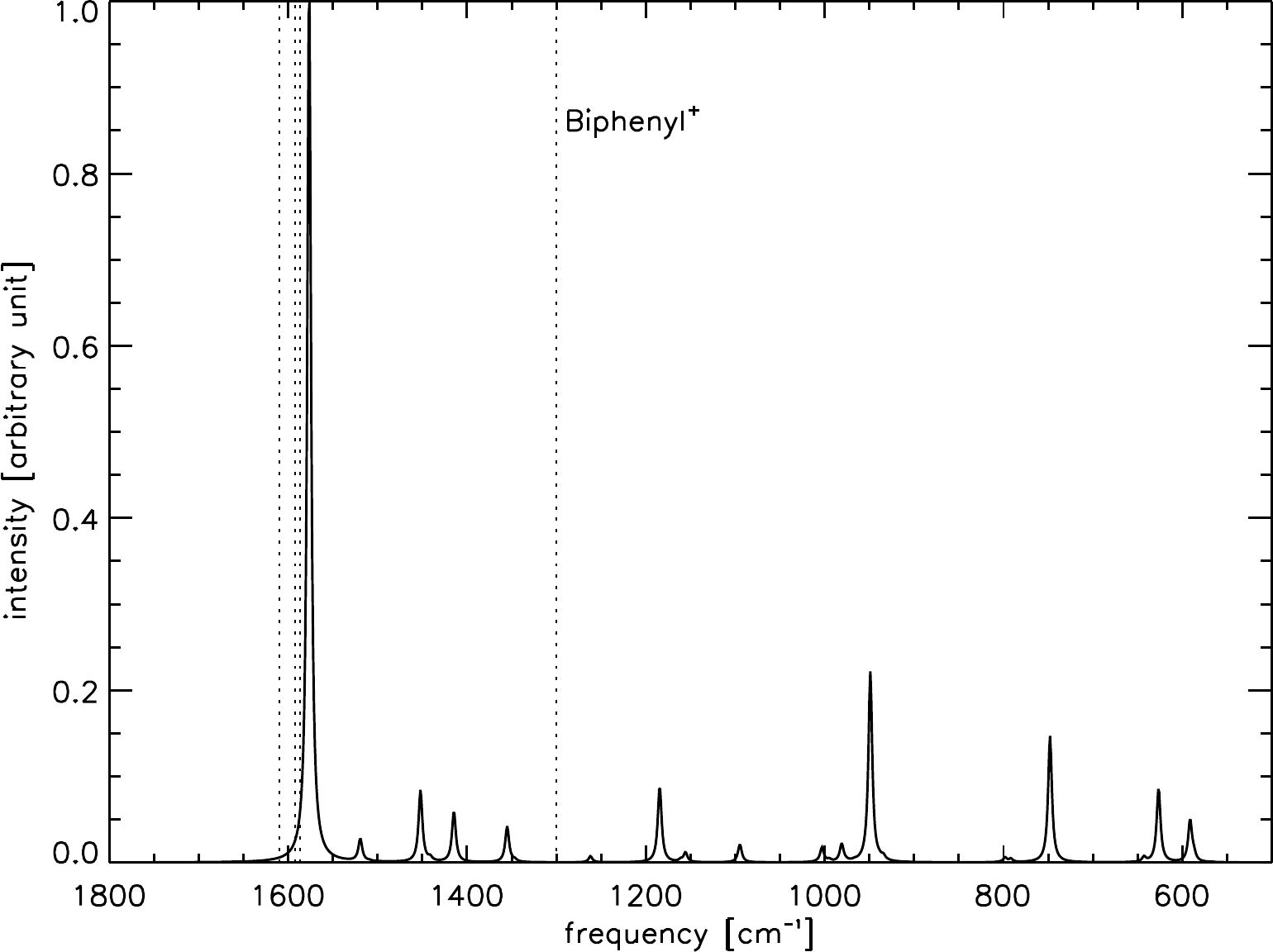}}
\caption{Computed infrared spectra of Biphenyl neutral and cation.}
\label{3Fig}
\end{figure}

\begin{figure}
\centerline{\includegraphics[width=0.8\textwidth]{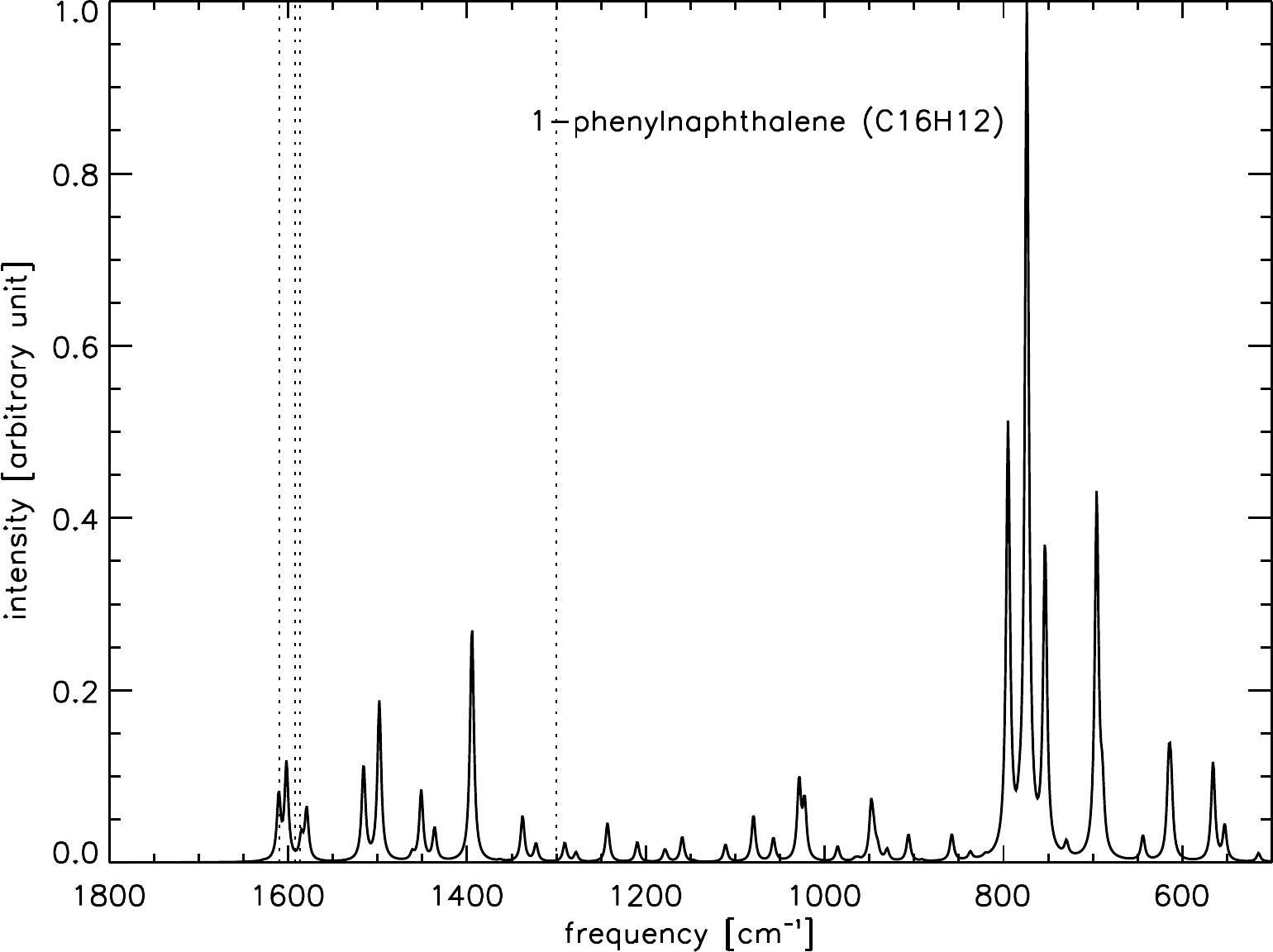}}
\centerline{\includegraphics[width=0.8\textwidth]{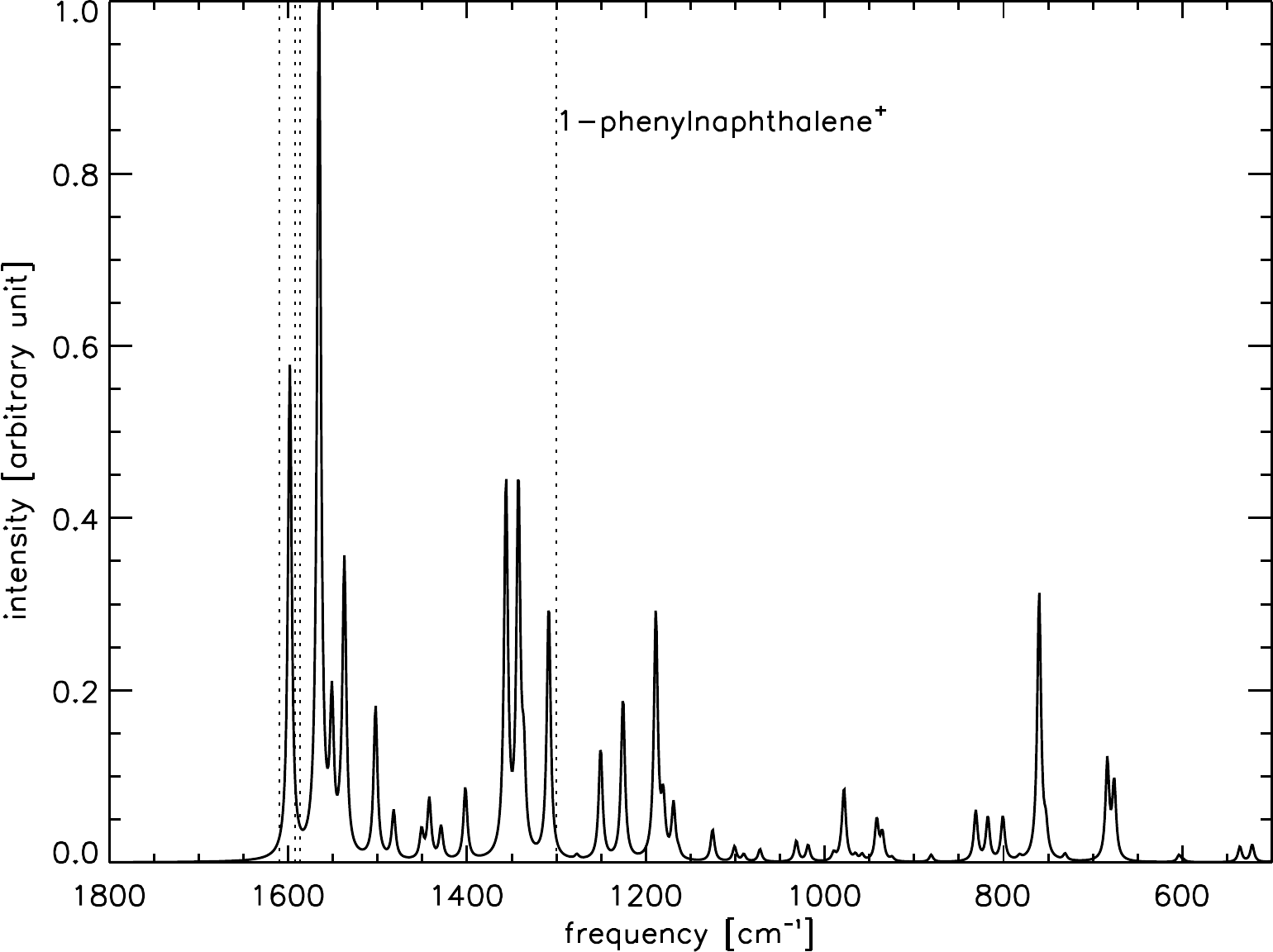}}
\caption{Computed infrared spectra of 1-phenylnaphthalene neutral and cation.}
\label{4Fig}
\end{figure}

\begin{figure}
\centerline{\includegraphics[width=0.8\textwidth]{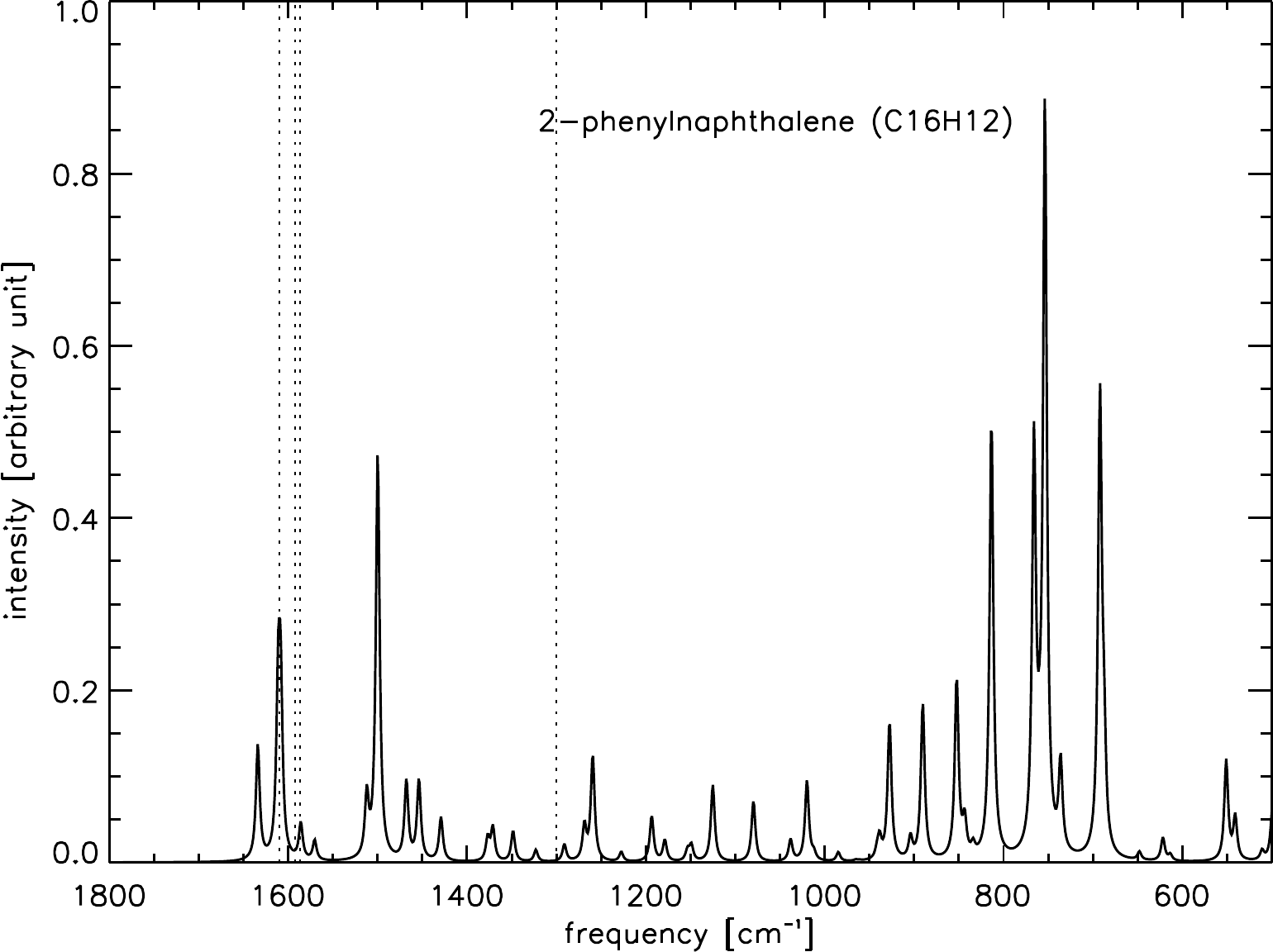}}
\centerline{\includegraphics[width=0.8\textwidth]{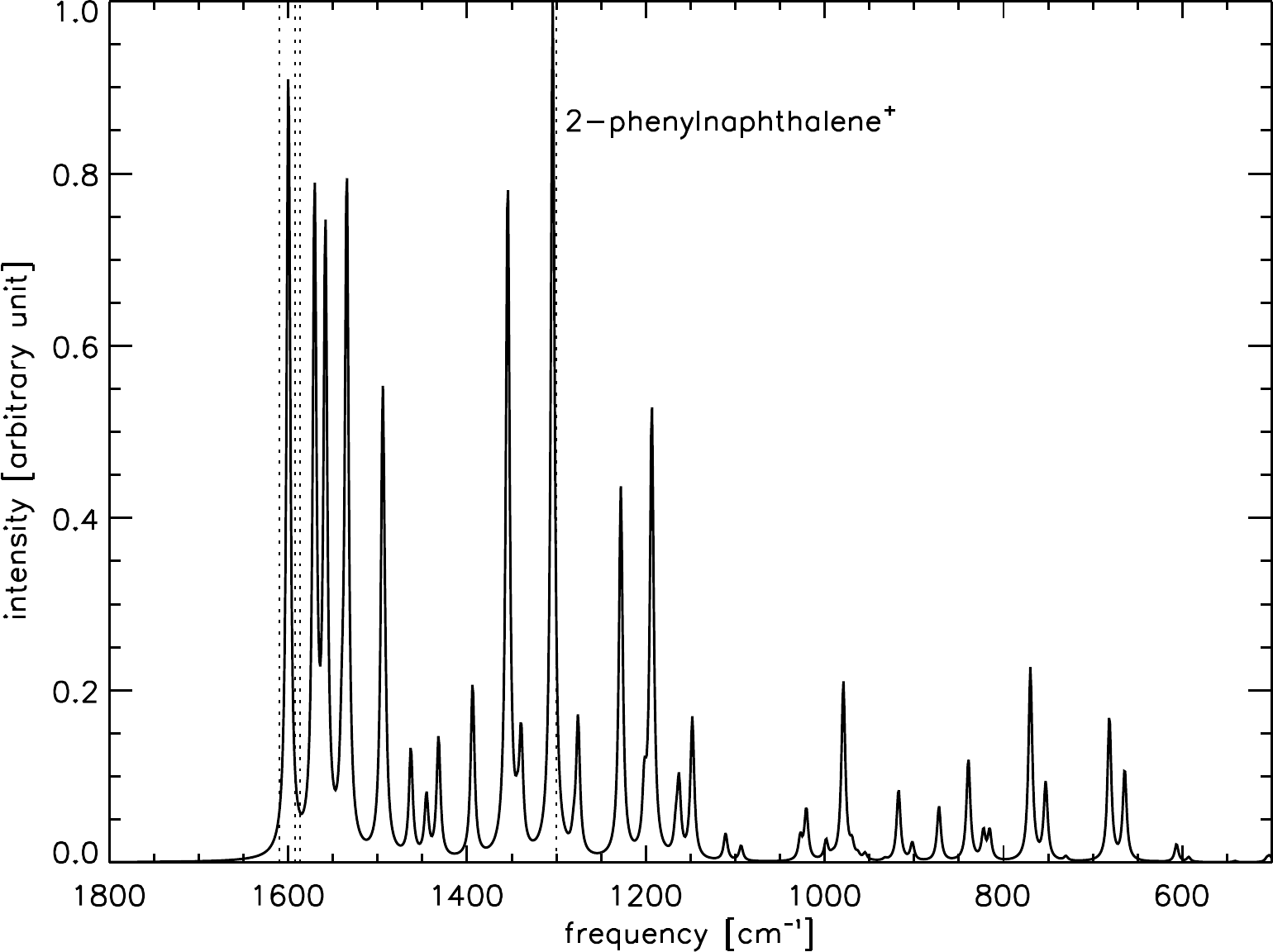}}
\caption{Computed infrared spectra of 2-phenylnaphthalene neutral and cation.}
\label{5Fig}
\end{figure}

\begin{figure}
\centerline{\includegraphics[width=0.8\textwidth]{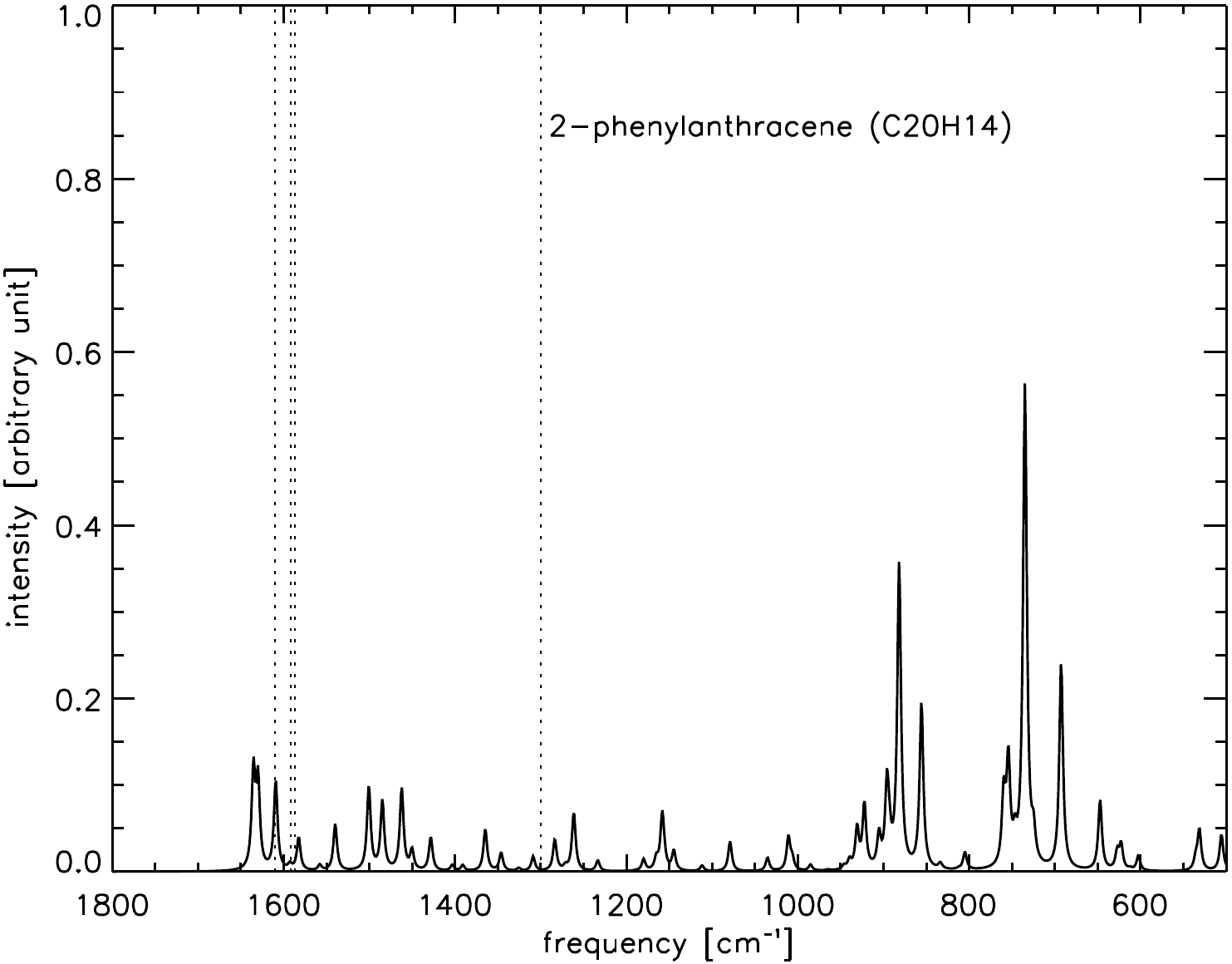}}
\centerline{\includegraphics[width=0.8\textwidth]{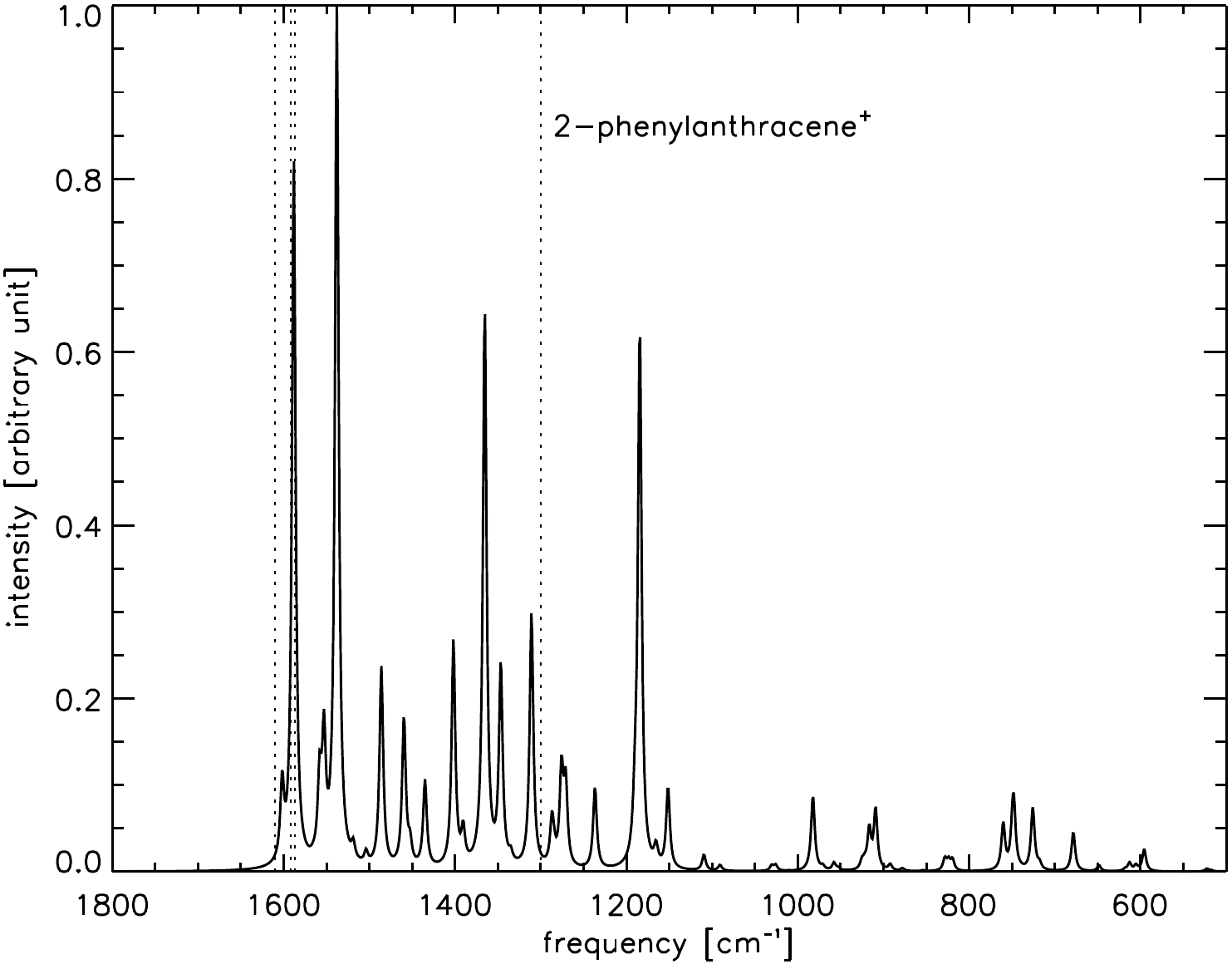}}
\caption{Computed infrared spectra of 2-phenylanthracene neutral and cation.}
\label{5Figa}
\end{figure}

\begin{figure}
\centerline{\includegraphics[width=0.8\textwidth]{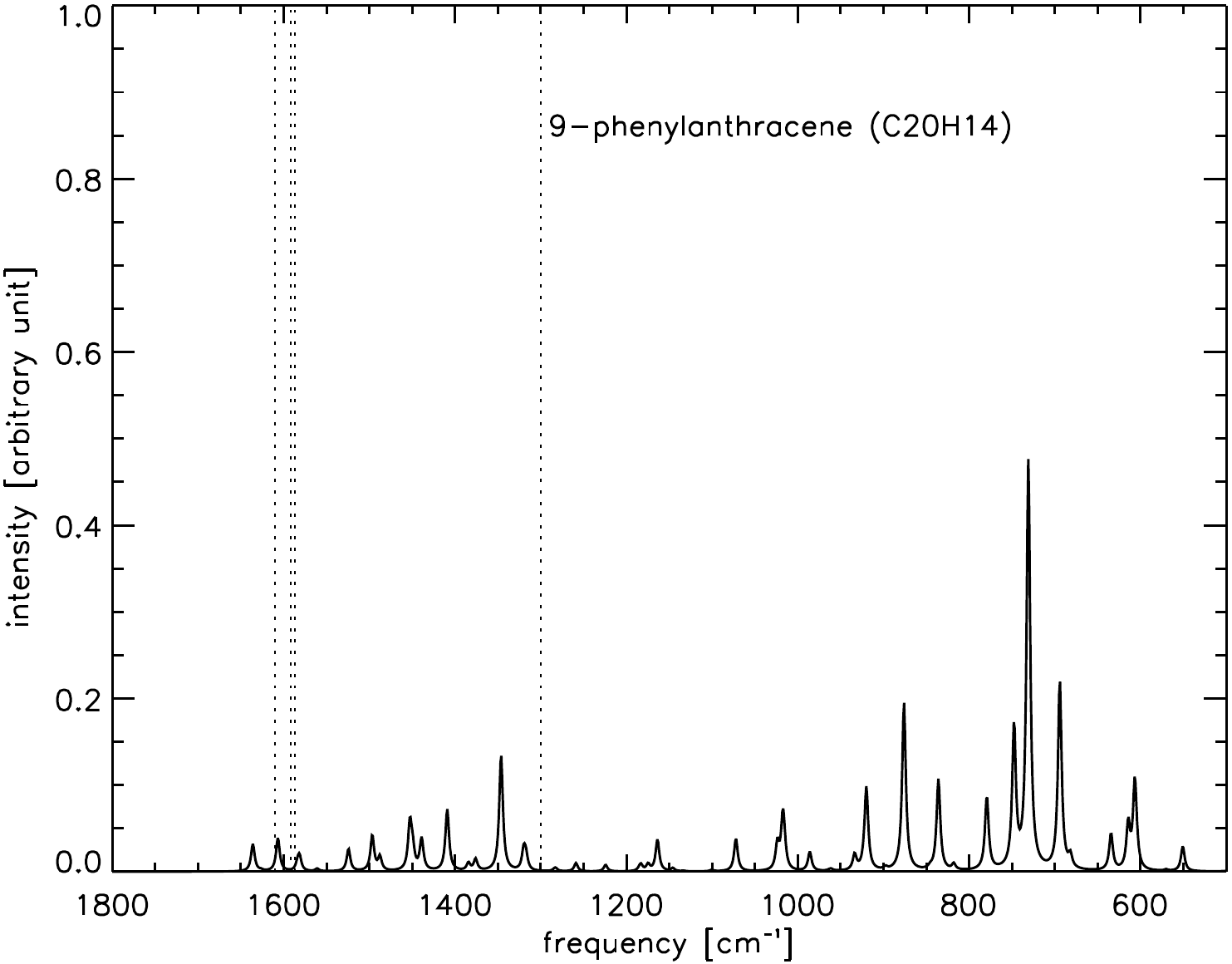}}
\centerline{\includegraphics[width=0.8\textwidth]{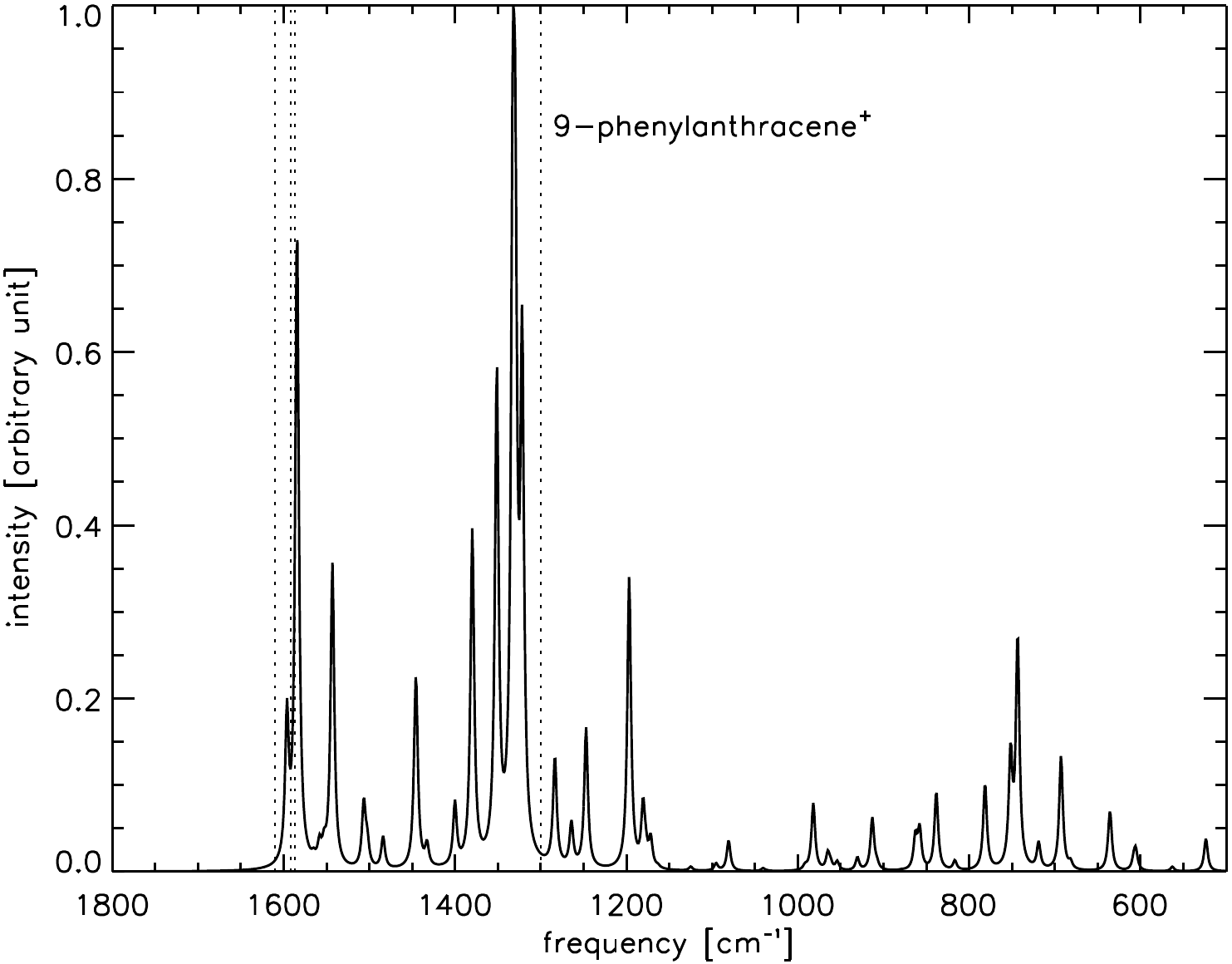}}
\caption{Computed infrared spectra of 9-phenylanthracene neutral and cation.}
\label{5Figb}
\end{figure}

\begin{figure}
\centerline{\includegraphics[width=0.8\textwidth]{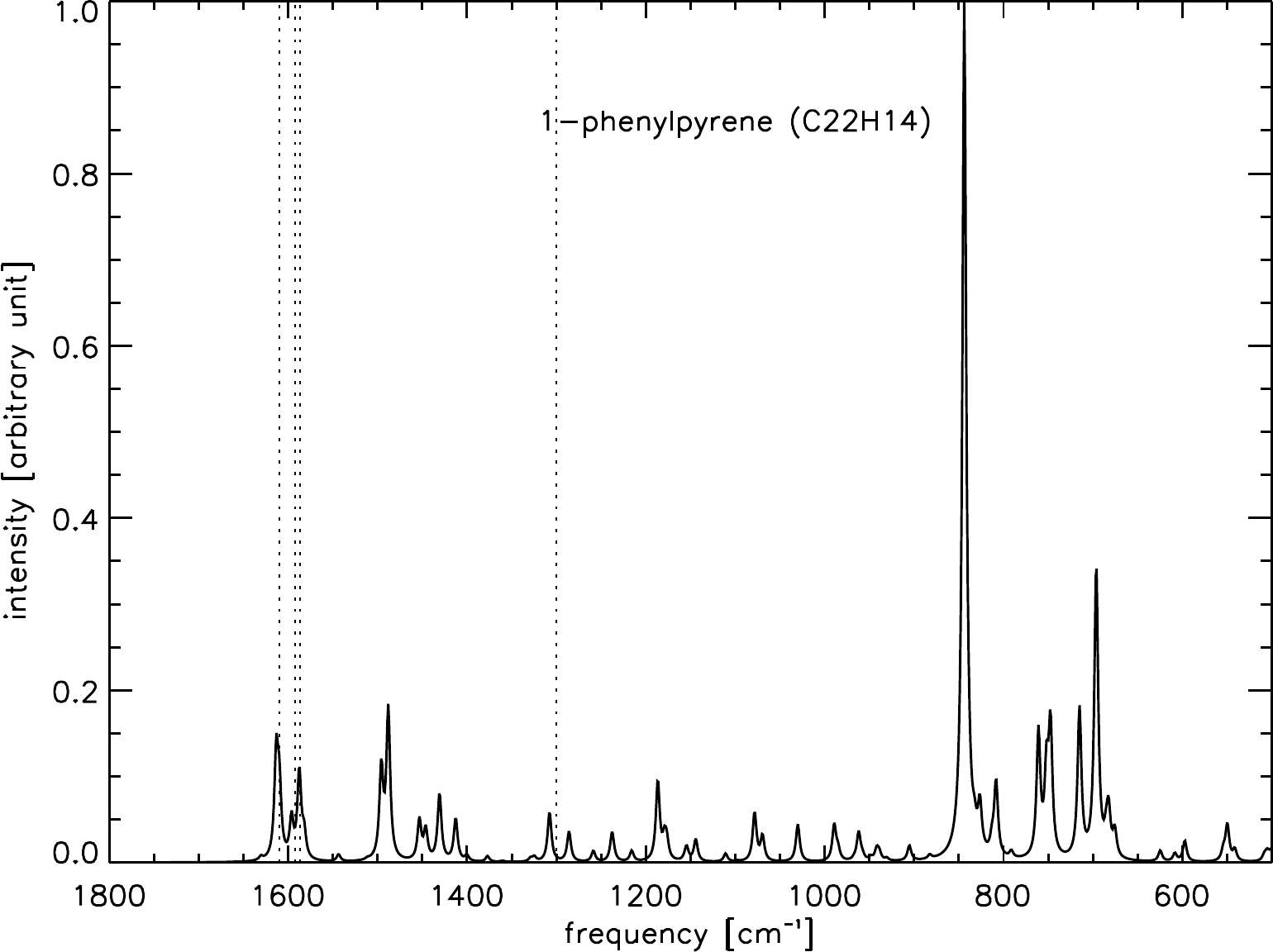}}
\centerline{\includegraphics[width=0.8\textwidth]{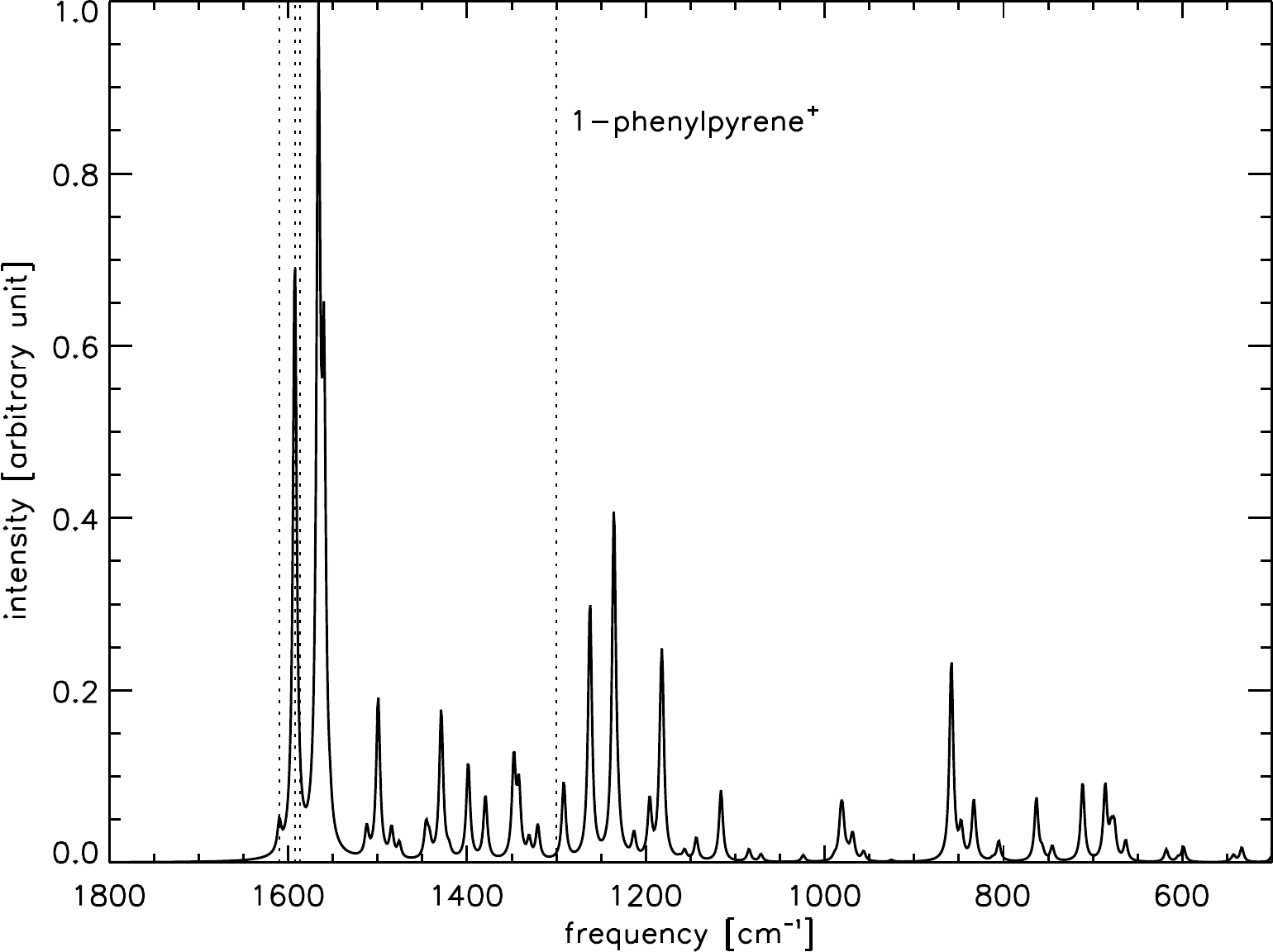}}
\caption{Computed infrared spectra of 1-phenylpyrene neutral and cation.}
\label{6Fig}
\end{figure}

\begin{figure}
\centerline{\includegraphics[width=0.8\textwidth]{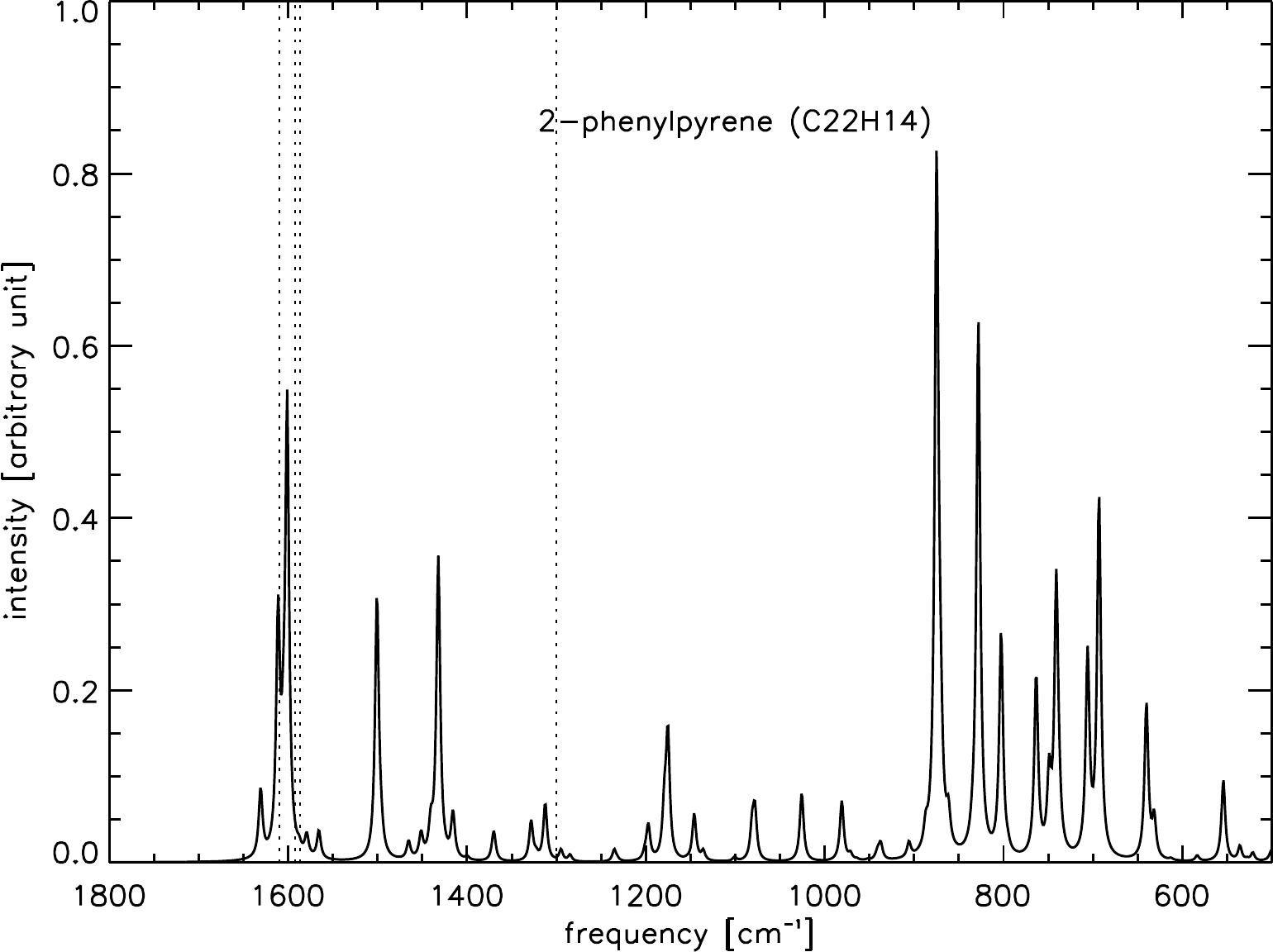}}
\centerline{\includegraphics[width=0.8\textwidth]{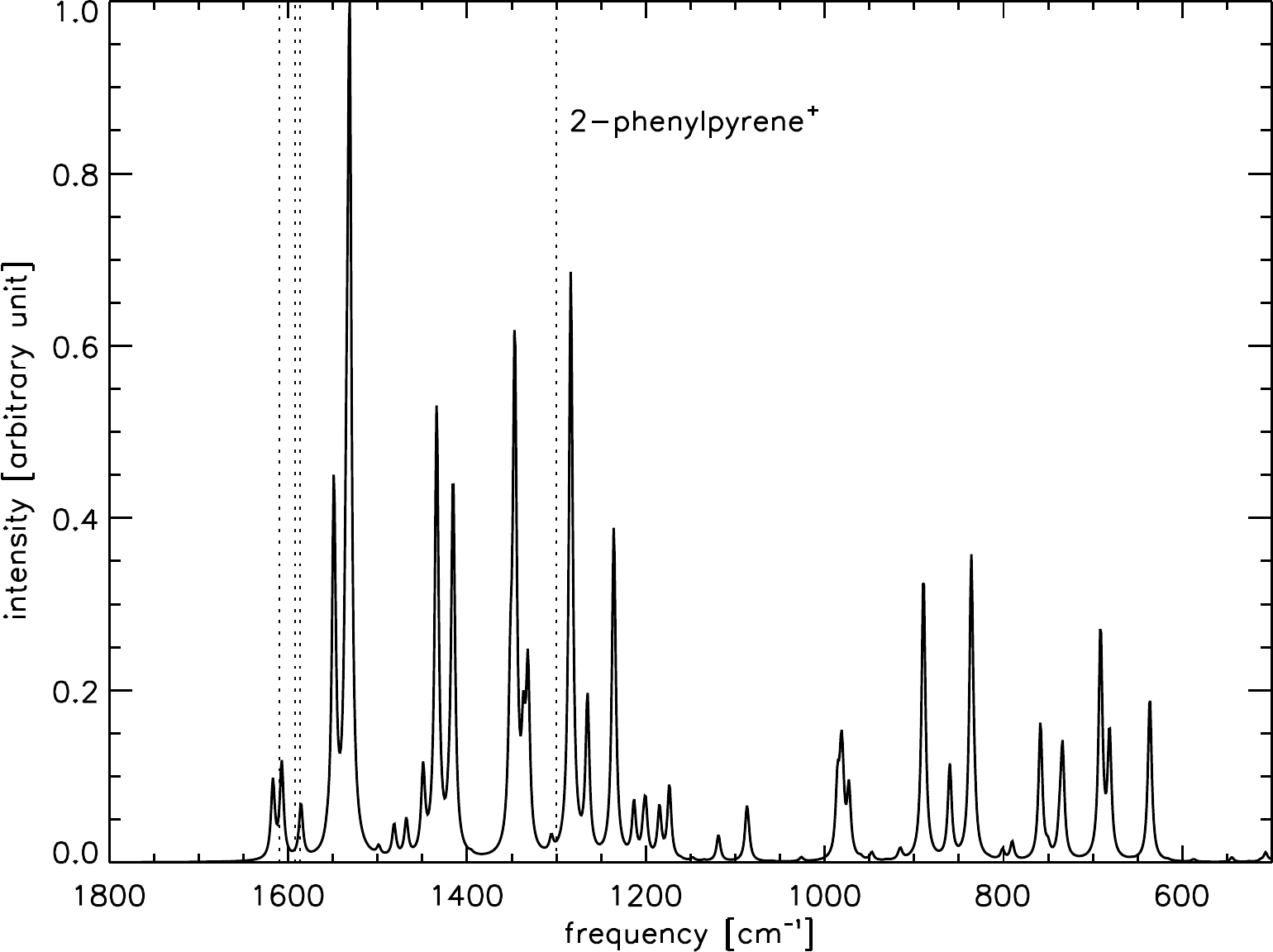}}
\caption{Computed infrared spectra of 2-phenylpyrene neutral and cation.}
\label{7Fig}
\end{figure}

\begin{figure}
\centerline{\includegraphics[width=0.8\textwidth]{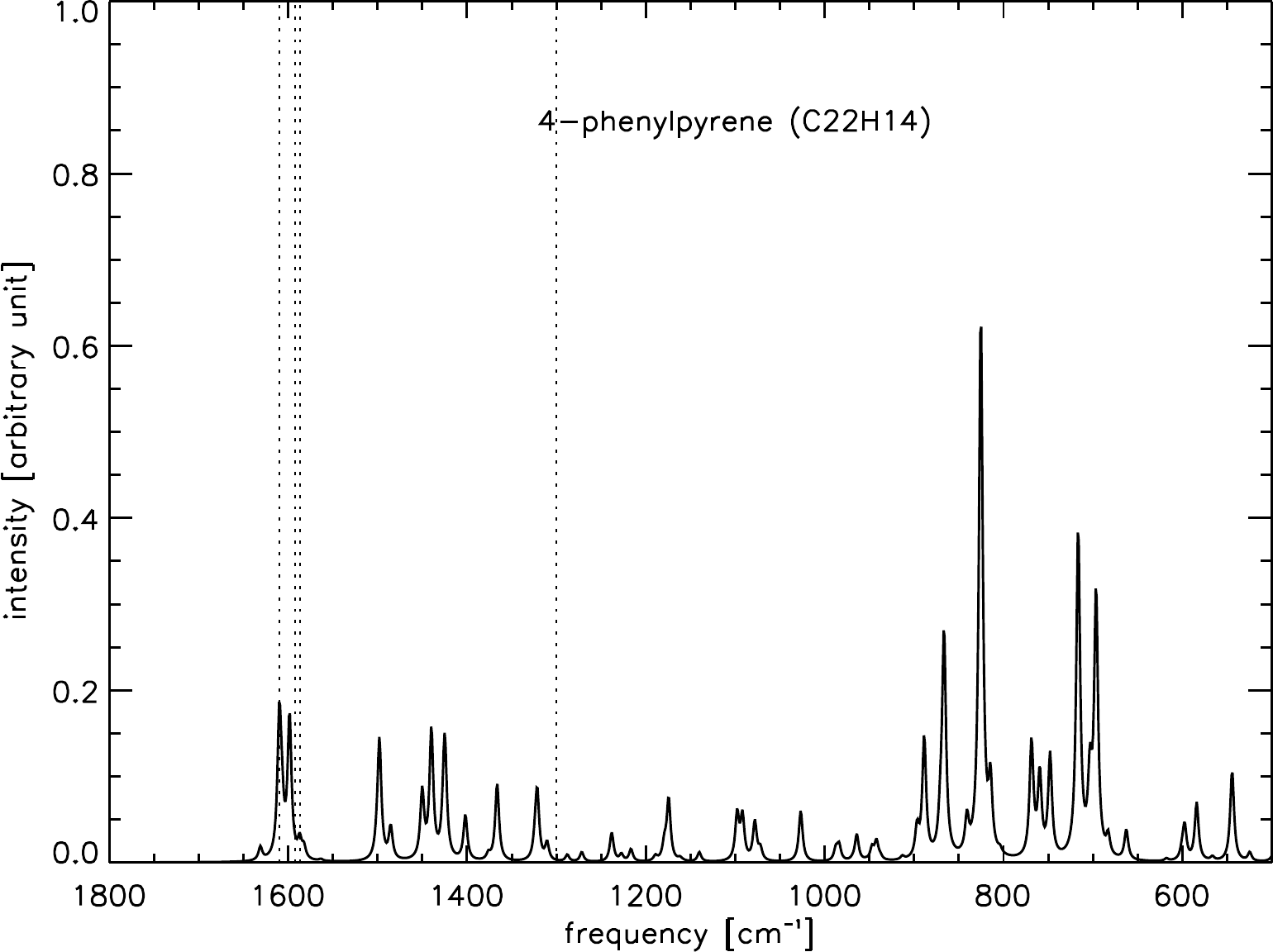}}
\centerline{\includegraphics[width=0.8\textwidth]{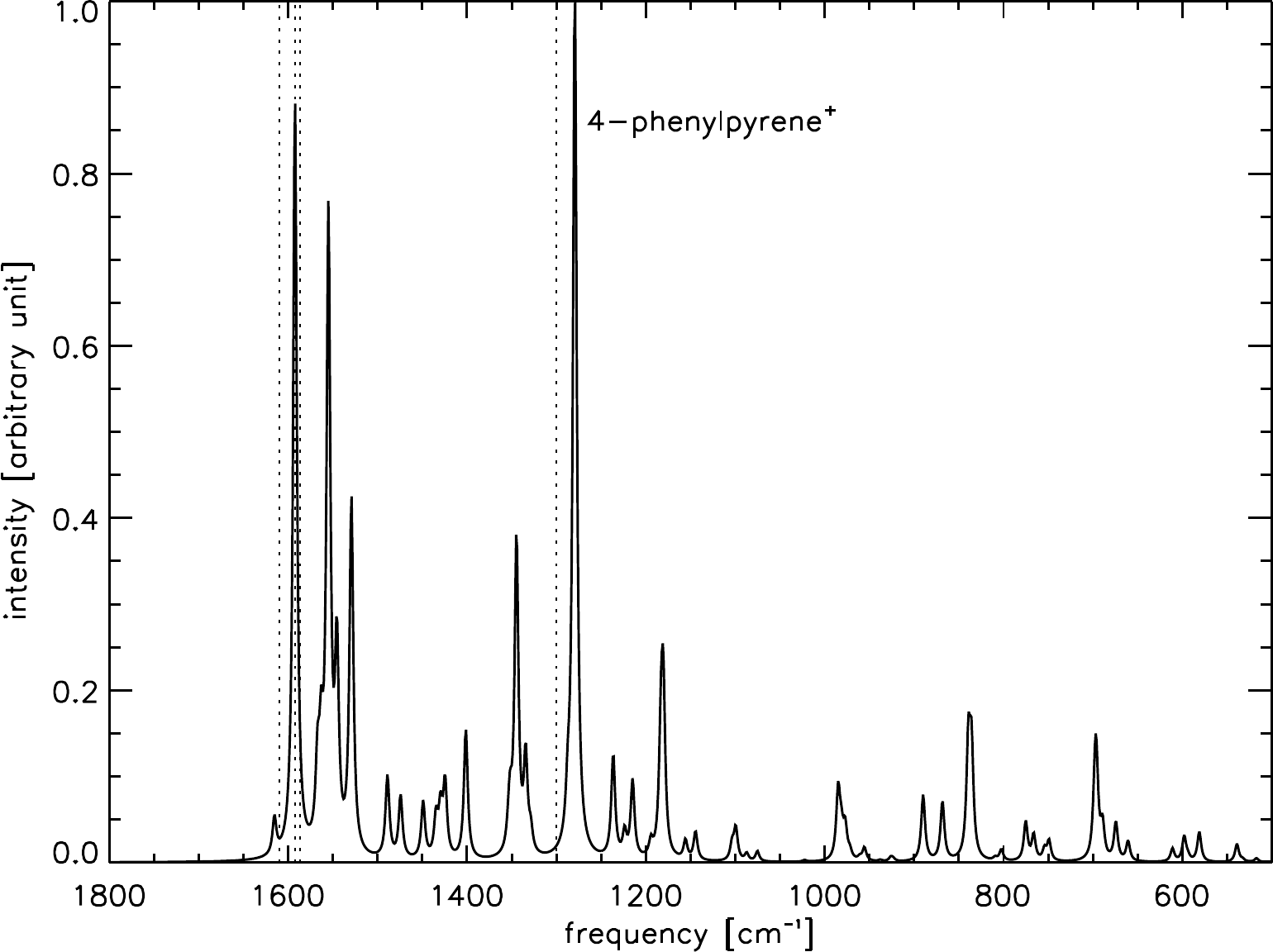}}
\caption{Computed infrared spectra of 4-phenylpyrene neutral and cation.}
\label{8Fig}
\end{figure}

\begin{figure}
\centerline{\includegraphics[width=0.8\textwidth]{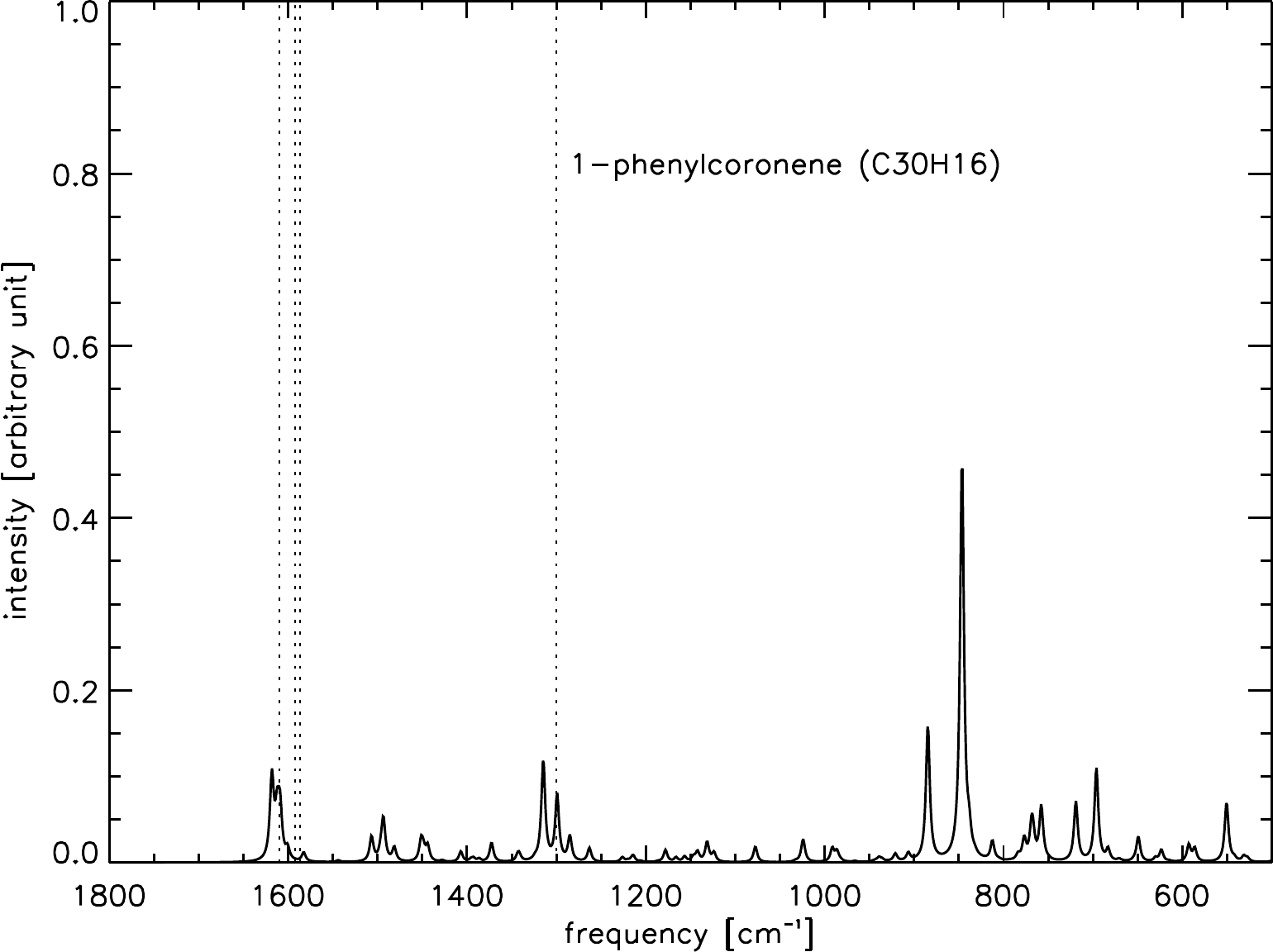}}
\centerline{\includegraphics[width=0.8\textwidth]{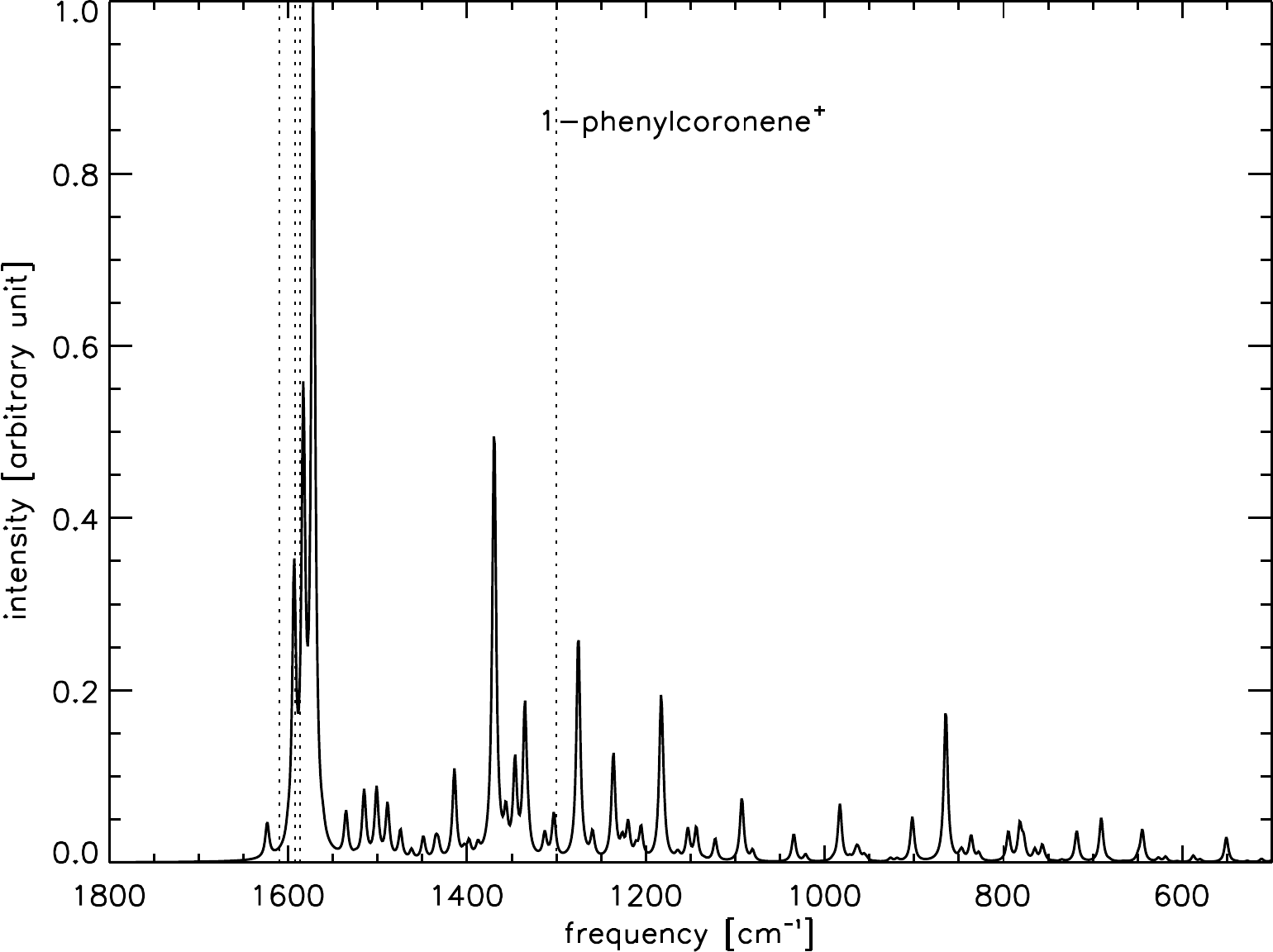}}
\caption{Computed infrared spectra of 1-phenylcoronene neutral and cation.}
\label{9Fig}
\end{figure}

The computed and scaled vibrational frequencies and the corresponding intensities are used to simulate infrared spectrum of each molecule. In each spectrum Lorentzian profile peaks are obtained considering FWHM of 5~cm$^{-1}$ and the intensities are taken relative to the most intense mode. These spectra for both neutrals and cations are shown in Figures~\ref{3Fig}~--~\ref{9Figa} for the range 1800~-~500~cm$^{-1}$. The vertical dotted lines show 6.2 and 7.7~$\mu$m AIB positions. The C-H stretch modes in the 3000~cm$^{-1}$ range fall very close to each other and collocate into a complex feature in the simulated spectra. These are shown in Figures~\ref{10Fig}(a - d). The frequencies and corresponding infrared intensities for each molecule and its cation can be obtained upon request.

In general, the spectral features in phenyl-PAHs are not very different from plain PAHs, even the spectral change upon ionization is similar. For example, in the spectra of neutral phenyl-PAHs the most intense bands correspond to C-H stretch (3030~cm$^{-1}$) and C-H out-of-plane bend (890~cm$^{-1}$) modes, but upon ionization, i.e. in cations, the intensity of C-H stretch modes is drastically reduced and the intensity of C-C stretch and C-H in-plane bend modes (1600~--~1100~cm$^{-1}$) gets enhanced. This variation due to ionization is related to the change in charge distribution around the peripheral hydrogens in plain PAHs \cite{Pathak2005, Pathak2006} and even in PAHs with side groups \cite{Langhoff98, Maurya2015}. The prominent spectral features in 1800~-~500~cm$^{-1}$ range for each molecule are discussed below individually.

\textbf{Biphenyl} - The computed spectrum (Figure~\ref{3Fig}) compares very well with the reported gas phase infrared spectrum \cite{NIST} and also with that reported earlier at the same theoretical level \cite{Talbi2012, furuya98}. The most intense peak in the neutral molecule is at 733.0~cm$^{-1}$, which is a combined peak due to neighbouring C-H out-of-plane modes at 732.9 and 733.2~cm$^{-1}$. Another strong C-H out-of-plane wag appears at 692.0~cm$^{-1}$. Medium intense peaks at 1612.0 and 1486.0~cm$^{-1}$ are due to the C-C stretch and C-H in-plane bend modes. In the spectrum of the cation the most prominent band is the C-C stretch mode at 1576.4~cm$^{-1}$.

\textbf{1-phenylnaphthalene} - In the spectrum of neutral 1-phenylnaphthalene (Figure~\ref{4Fig}) the most prominent band lies at 774.0~cm$^{-1}$, due to C-H wag motion of three hydrogens in the ring that also contains the phenyl group. This and the band at 1498.0~cm$^{-1}$ correlate well with the corresponding bands at 788.8 and 1502.5~cm$^{-1}$ in naphthalene \cite{Hudgins1998a, Pathak2005}. In the neutral there are other C-H out-of-plane wag modes appearing at 695.9, 753.6 and 794.9~cm$^{-1}$, while in the cation only one intense band is present related to this mode at 760.2~cm$^{-1}$. The cation band is also similar to that observed in plain naphthalene cation at 767.9~cm$^{-1}$ \cite{Pathak2005}. The highest intensity mode in cation spectrum is at 1565.4~cm$^{-1}$, along with a strong peak at 1598.1~cm$^{-1}$ and a medium intensity peak near 1300~cm$^{-1}$.

\textbf{2-phenylnaphthalene} - Four distinct peaks for the C-H out-of-plane bend modes are present at 692.3, 753.7, 765.9 and 813.5~cm$^{-1}$, with the most intense one at 753.7~cm$^{-1}$ (Figure~\ref{5Fig}). The 692.3~cm$^{-1}$ mode is symmetric C-H wag motion in the phenyl side group and the 813.5~cm$^{-1}$ mode is purely of the naphthalene moiety. For neutral, the C-C stretch mode gives two bands of medium intensity at 1611.0 and 1633.9~cm$^{-1}$ due to the phenyl and naphthalene moiety respectively. Ionization results in a more complex spectrum with several intense features appearing in the 1100--1600~cm$^{-1}$ range. The most intense feature lies at 1304.1~cm$^{-1}$ due to C-C stretch vibration at the junction of phenyl and the main naphthalene unit. Another significant peak is at 1599.7~cm$^{-1}$ due to C-C stretch within the phenyl side group. In both phenylnaphthalenes several modes of the parent PAH remain only slightly disturbed, but the effect of phenyl group is more prominent on substitution at 2$^{nd}$ position.

\textbf{2-phenylanthracene} - The computed spectrum of neutral (Figure~\ref{5Figa}) is in good agreement with the reported experimental solid-state spectrum in KBr pellet \cite{NIST2PA}. Number of intense C-H out-of-plane bend modes appear with contributions from solo, quarto and quintet hydrogens. The prominent peaks are at 692.7, 735.1, 855.7 and 881.8 cm$^{-1}$, with the most intense one due to quarto C-H wag motion at 735.1 cm$^{-1}$. The mode at 881.8 cm$^{-1}$ is due to solo C-H wag motion. These quarto and solo modes correlate well with the corresponding bands at 734.5 and 891.4 cm$^{-1}$ in anthracene \cite{Pathak2005}. The 692.7 cm$^{-1}$ peak corresponds to C-H wag in phenyl group and the 855.7 cm$^{-1}$ peak is due to mixing of phenyl C-H wag motion and solo C-H wag motion. There are also two moderate intensity peaks at 1609.3 and 1630.0~cm$^{-1}$ due to C-C stretch motion in phenyl unit and in anthracene unit respectively.

Upon ionization the quarto and solo modes are blue shifted by 12.7 and 27.4 cm$^{-1}$, respectively, just as in plain anthracene molecule \cite{Hudgins1999}. In the spectrum of cation the strongest band corresponds to the C-C stretch motion in anthracene unit at 1538.0 cm$^{-1}$. The strong band at 1588.2 cm$^{-1}$ is due to C-C stretch motion in both anthracene and phenyl moieties. Other prominent peaks present at 1184.4 and 1365.4 cm$^{-1}$ arise due to C-C stretch and C-H in-plane bend motion in anthracene unit.

\textbf{9-phenylanthracene} - The computed spectrum of neutral 9-phenylanthracene (Figure~\ref{5Figb}) matches well with the reported gas phase infrared spectrum \cite{NIST9PA}. The most intense peak at 731.0 cm$^{-1}$ is due to C-H out-of-plane bend motion of quarto hydrogens and correlates well with the corresponding band at 734.5 cm$^{-1}$ in anthracene \cite{Pathak2005}. This mode is unaffected even with side group position as it is at the same position in 2-phenylanthracene. Other intense features due to C-H wag motions appear at 694.3, 747.7 and 876.1 cm$^{-1}$. The 694.3 and 747.7 cm$^{-1}$ modes correspond to the C-H out-of-plane motion in the phenyl unit while the 876.1 cm$^{-1}$ mode is due to solo C-H out-of-plane motion. 

Upon ionization the 694.3 and 747.7 cm$^{-1}$ modes are nearly unaltered and the quarto and solo C-H wag modes are blue shifted by 12.5 cm$^{-1}$ and 37.0 cm$^{-1}$, respectively. In the cation, the peak at 1331.2 cm$^{-1}$ is strongest. This peak results from two close modes, one at 1329.8 cm$^{-1}$, attributed to the C-C stretch and C-H in-plane bends motion in both anthracene and phenyl units, and the other at 1332.6 cm$^{-1}$ due to C-C stretch vibration at the junction of phenyl and the main anthracene unit. The band at 1196.8 cm$^{-1}$ correlates with the band at 1188.6 cm$^{-1}$ in plain anthracene \cite{Hudgins1998a}. An interesting strong feature at 1584.4 cm$^{-1}$ corresponds to C-C stretch motion in both anthracene and phenyl units.

\textbf{1-phenylpyrene} - The neutral 1-phenylpyrene spectrum (Figure~\ref{6Fig}) has a prominent feature at 843.9~cm$^{-1}$ due to C-H wag in pyrene. This mode is relic of the strong feature at 850~cm$^{-1}$ in plain pyrene \cite{Pathak2006}. The other strong peak at 696.4~cm$^{-1}$ is due to the C-H out-of-plane wag motion of phenyl group hydrogens. This mode corresponds to the strongest peak in the spectrum of phenyl radical observed at 706~cm$^{-1}$ \cite{Friderichsen2001}. The cation spectrum has strong C-C stretch features, at 1566.1~cm$^{-1}$ due to the pyrene unit and a band at 1592.4~cm$^{-1}$ due to the phenyl group. The vibration of C-C bond at the junction of phenyl and pyrene moiety contributes to the peak at 1262.4~cm$^{-1}$.

\textbf{2-phenylpyrene} - There are a number of intense features in both neutral and cation (Figure~\ref{7Fig}). In the neutral, besides several C-H out-of-plane modes between 900--650~cm$^{-1}$, there are two strong peaks at 1611.2 and 1601.1~cm$^{-1}$. These features belong to C-C stretch within the phenyl side group and pyrene moiety respectively. The relative intensity of the band at 1611.2~cm$^{-1}$ is reduced in cation. The change in the relative intensity of this mode in the cation is not related to ionization, it is due to strong enhancement of intrinsic intensity of other modes in this frequency range upon ionization. The value of absolute intensity of this feature is 0.36 Debye$^{2}$/AMU \AA$^{2}$ in neutral and is 0.33 Debye$^{2}$/AMU \AA$^{2}$ in the cation. The strong peaks at 874.8 and 828.2~cm$^{-1}$ correspond to the C-H wag motion in the pyrene unit. While the intense peak at 693.3~cm$^{-1}$ is due to the C-H out-of-plane wag motion of phenyl group hydrogens and the peaks at 741.0 and 763.5~cm$^{-1}$ arise from the C-H out-of-plane vibrations in both units. These three modes remain unaffected in the cation at 691.6, 734.1 and 758.7~cm$^{-1}$. The most intense band in the cation spectrum, however, is at 1531.2~cm$^{-1}$ with a companion at 1548.8~cm$^{-1}$, both due to the C-C stretch vibrations within the pyrene unit.

\textbf{4-phenylpyrene} - The spectral features in 4-phenylpyrene (Figure~\ref{8Fig}) are close to those in 1-phenylpyrene. The neutral exhibits a strong peak at 825.5~cm$^{-1}$ attributed to C-H out-of-plane wag in the pyrene unit. Other C-H wag modes appear at 696.6, 716.6 and 866.6~cm$^{-1}$. In 1600~cm$^{-1}$ range medium intensity C-C stretch modes peak at 1598.4 and 1609.8~cm$^{-1}$. In the cation the most intense band is at 1279.4~cm$^{-1}$. There is a strong C-C stretch mode at 1592.3~cm$^{-1}$. Another strong feature at 1554.9~cm$^{-1}$ is due to pyrene moiety C-H in-plane bend mode. Comparing the spectra of three different substituent positions of phenyl group in pyrene, it is observed that the substitution at 2$^{nd}$ position shows greater effect and enhances the infrared activity of several modes.

\textbf{1-phenylcoronene} - The spectrum of neutral and cation (Figure~\ref{9Fig}) are very similar to those of plain coronene \cite{Pathak2006}. In neutral a prominent feature is at 846.3~cm$^{-1}$, due to the C-H out-of-plane wag motion of the coronene duo hydrogens. The effect of phenyl substitution results in additional medium intensity mode at 884.6~cm$^{-1}$. The cation spectrum consist of some intense features in the range 1100~-~1600~cm$^{-1}$, with the most prominent band at 1571.9~cm$^{-1}$ corresponding to C-C stretch and C-H in-plane bending vibrations in coronene. The presence of phenyl group does induce additional features at 1582.9 and 1593.3~cm$^{-1}$.

\textbf{14-phenylovalene} - The spectrum of neutral 14-phenylovalene, like neutral ovalene \cite{Pathak2006}, has two intense peaks for C-H out-of-plane bend vibrations at 835.8 and 870.3 cm$^{-1}$, corresponding respectively to the duo and solo hydrogens (Figure~\ref{9Figa}). Additional peaks of moderate intensity are present at 695.2 and 705.7 cm$^{-1}$ due to phenyl C-H wag motions. For neutral, the 1200 - 1600 cm$^{-1}$ region has four medium intensity features at 1235.2, 1327.5, 1609.9 and 1629.7 cm$^{-1}$. The 1235.2 cm$^{-1}$ mode is mainly C-H in-plane bending, while the 1327.5 cm$^{-1}$ feature is due to aromatic C-C stretch vibrations in ovalene unit and stretching of the single C-C bond that joins the phenyl and ovalene units. In cation, intense peaks appear at 1227.8, 1322.0, 1341.0, 1544.7 and 1590.3 cm$^{-1}$ along with several moderate intensity features. The strongest peak at 1590.3 cm$^{-1}$ corresponds to C-C stretch vibrations in ovalene moiety.

\textbf{\textit{The C-H stretch mode}} features are of significance only in the neutral molecules as in the cations these modes are more than an order of magnitude less intense. In Figure~\ref{10Fig}, the intensity scale for cations is expanded so that the C-H stretch features may be visible. The frequency position in cations is slightly blue shifted with respect to the neutrals. As the phenyl group is also aromatic all hydrogens have similar bonding and give rise to a collective feature, with multiple peaks. Astronomical observations reveal that the 3.3~$\mu$m (3030~cm$^{-1}$) band is accompanied by a plateau and sub-features \cite{Maltseva2016}. The profile and position of the peaks observed along different astronomical objects is different. This difference is attributed to the different size and peripheral structure of PAHs \cite{Candian2012, Maltseva2016}.

In neutral molecules with phenyl group substituted at 1$^{st}$ position, an interesting peak near 3110~cm$^{-1}$ (3.2~$\mu$m) is observed. This mode corresponds to the motion of hydrogen that is asymmetrically placed with respect to the phenyl group. This mode has enhanced frequency possibly due to the hindered hydrogen motion. Although 4-phenylpyrene has a similar asymmetry, as reflected in its similar energy~--~angle variation (Figure~\ref{2FigB}b) it does not exhibit this enhanced feature. This may be because in 4-phenylpyrene the position and inclination of phenyl group makes the surrounding hydrogens spaced more at ease as compared to that in 1- substitution. The solo position phenyl substituted molecules (14-phenylovalene and 9-phenylanthracene) have energy~--~angle variation similar to that in 1$^{st}$ position substituted phenyl PAHs (Figure~\ref{2FigB}b). But, only 14-phenylovalene exhibits a peak near 3110 cm$^{-1}$ which corresponds to the vibrational motions in symmetrically placed hydrogen atoms with respect to the phenyl group.

\section{Discussion}

The possibility of phenyl substituted PAHs in the ISM is indicated via combustion pyrolysis of simple aromatic molecules \cite{Badger61, Woods2002, Mimura2003, Shukla2008, Cherchneff2011, Talbi2012, Zhao2019}. Like unsubstituted plain PAHs, phenyl substituted PAHs also exhibit features near the AIB regions. But these species introduce additional interesting features in their infrared spectra, that need to be examined for suitable incorporation in composite PAH-AIB models. In this section, variation in peak position and intensity of C-H out-of-plane and C-C stretch vibrational bands are analysed. Attempt is made to identify any correlation of the variations with different shapes, sizes, and ionization of studied phenyl-PAHs. 

\subsection{C-H out-of-plane vibrations}

The peak positions of prominent C-H out-of-plane features for neutrals and corresponding cations of all molecules are listed in the Table~\ref{tab1a}. It is observed that generally the C-H out-of-plane vibrations in the phenyl moiety, marked with superscript '$a$', and those in the parent plain PAHs can be distinctly identified. Therefore, the two types of features are discussed separately.

\subsubsection{Phenyl C-H wag}
The quintet C-H present in the phenyl moiety produces out-of-plane vibrational bands in 692 -- 706 cm$^{-1}$ (14.5 -- 14.1 $\mu$m) and 730 -- 770 cm$^{-1}$ (13.7 -- 13.0 $\mu$m) regions \cite{Hudgins1999}. A common feature in all studied neutral systems is the peak near 695 cm$^{-1}$ (Table~\ref{tab1a}). Similar feature has been reported in neutral biphenyl and polyphenyls \cite{Talbi2012}. This mode is attributed to the in phase out-of-plane vibration of all five phenyl group hydrogens. Only in neutral 14-phenylovalene, instead of a single peak there is a doublet with peaks at 695.2 and 705.7 cm$^{-1}$.

\begin{figure}
\centerline{\includegraphics[width=0.9\textwidth]{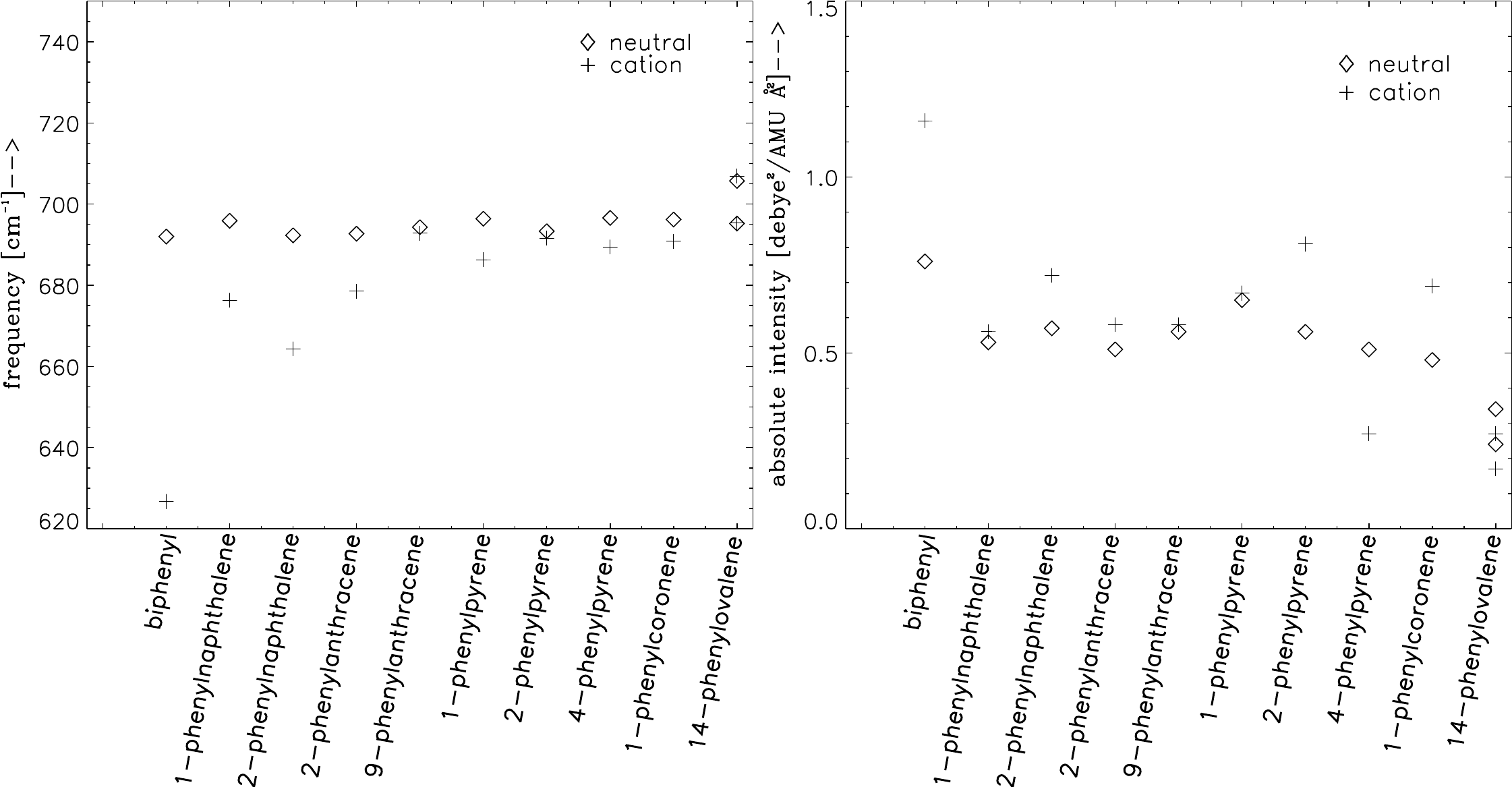}}
\caption{The frequency and the absolute intensity of the phenyl C-H wag mode in 710 - 690 cm$^{-1}$ region is plotted for both neutral and cation.}
\label{11aFig}
\end{figure}

The band position and intensity of this feature is plotted against the size of the phenyl-PAH for neutrals and their corresponding cations in Figure~\ref{11aFig}. As the PAH size increases the frequency of this mode in neutrals is nearly unaltered i.e., the mode is independent of the size of the molecule. Upon ionization the mode is red shifted, with largest shift for biphenyl (65 cm$^{-1}$). The red shift decreases with the size of the PAH and it is negligible in 14-phenylovalene. The shift in the band position in cations can be attributed to the change in fractional charge on the phenyl hydrogen atoms, as the PAH size increases the change in charge reduces \cite{Pathak2005, Pathak2006}.

The absolute intensity of this quintet phenyl C-H wag mode is maximum for biphenyl, it decreases gradually with PAH size and is minimum for 14-phenylovalene (Figure~\ref{11aFig}). Ionization results in the increase of absolute intensity of this mode, except in 4-phenylpyrene and 14-phenylovalene. The fractional change in absolute intensity upon ionization is less than 0.50 for this mode (Table~\ref{tab1a}).

The higher frequency phenyl C-H wag mode in the 730 -- 770 cm$^{-1}$ region corresponds to the C-H out-of-plane vibrations of the front three hydrogens in the phenyl moiety. This mode is clearly observed in all except 2$^{nd}$ substituted phenyl-PAHs (Table~\ref{tab1a}). The parent PAH's quarto C-H wag and the tail of the trio C-H wag modes, being in the same frequency range, get mixed with the phenyl quintet C-H modes. Upon ionization, this feature shows very small shift ($\Delta$$\nu$ $<$ 8 cm$^{-1}$) except in biphenyl, and the absolute intensity of the mode is maximum in biphenyl cation (Table~\ref{tab1a}).

\subsubsection{C-H wag in parent unit}
The substitution of phenyl group has little effect on the main C-H out-of-plane modes of the parent PAH. A comparison of main peaks of the phenyl-PAH in C-H out-of-plane bend region to the features in corresponding parent PAH is presented in Table~\ref{tab1b}. The modes related to C-H wag in the parent unit of phenyl-PAHs are blue shifted upon ionization, just as in plain PAHs \cite{Hudgins1999,Pathak2007}. In contrast the quintet phenyl C-H wag modes are normally red shifted upon ionization. The quarto and trio modes of parent unit that have mixed contributions from the phenyl quintet C-H out-of-plane motion show both blue and red shifting, depending on the fraction of mixing.

The quarto C-H wag exists in single layered plain PAHs. Besides the features similar to those in parent PAHs, additional features due to interaction and mixing of phenyl C-H wag appear in phenylnaphthalenes at 774.0, 753.7 and 765.9 cm$^{-1}$. Upon ionization, the 774.0 cm$^{-1}$ peak of 1-phenylnaphthalene and 765.9 cm$^{-1}$ peak of 2-phenylnaphthalene shift redwards, while the 753.7 cm$^{-1}$ peak of 2-phenylnaphthalene is blue shifted.

The trio C-H out-of-plane vibrations present in phenylpyrenes also have mixed contributions of phenyl C-H out-of-plane vibrations. These modes fall in the range 741 -- 764 cm$^{-1}$ (Table~\ref{tab1b}). The mixing of modes does not much affect the peak positions and only slightly alter the intensity patterns.

The duo C-H wag modes show split in peaks but fall in the range similar to corresponding plain PAHs \cite{Hudgins1999,Pathak2006}. These modes are intense features and peak between 828 -- 875 cm$^{-1}$. The solo C-H wag modes present at 881.8, 876.1 and 870.3 cm$^{-1}$ in 2-phenylanthracene, 9-phenylanthracene and 14-phenylovalene, respectively, are akin to the corresponding parent PAH features and show similar blue shift in the cations \cite{Hudgins1999,Pathak2005,Pathak2006}.

\subsection{The C-C stretch modes}

The strong and intense features associated with C-C stretch vibrations appear normally in PAH cations. The addition of phenyl group introduces a significant peak of the {\it phenyl group C-C stretch}, in the astrophysically relevant 1600~cm$^{-1}$ (6.2~$\mu$m) spectral region, except in cations of 14-phenylovalene and 2-phenylpyrene. The frequency position of the significant peaks in each molecule are shown in Figure~\ref{11Fig}. The phenyl group vibrations in cations are marked with `$\times$'. Except in biphenyl, these modes lie well inside the 6.2~$\mu$m AIB region, demarcated by horizontal lines in the figure. In 2-phenylpyrene it is the neutral that shows a significant peak in this spectral region, marked by `{\scriptsize $\triangle$}' symbol. These may be considered favourable results, suitable for incorporation in astrophysical models. But in all cases, there is a strong feature mostly due to C-C stretch in the parent aromatic moiety, marked as `o' in the figure. In all molecules this is the more intense mode compared to the higher frequency phenyl group C-C stretch mode, except in 2-phenylnaphthalene, where it's intensity is slightly less. This parent PAH C-C stretch is very much affected by the position of phenyl group substitution. The phenyl substitution at 1 position enhances this plain PAH mode to around 1570~cm$^{-1}$ (6.3~$\mu$m), while phenyl substitution at other positions vary randomly from molecule to molecule. The two solo position phenyl substitutions, as in 9-phenylanthracene and 14-phenylovalene, show large enhancement in the mode bringing it within the 6.2~$\mu$m AIB region.

\subsection{Astrophysical significance}

The infrared emission bands observed from diverse astrophysical objects depend on the excitation and internal conversion mechanisms in possible PAHs and result from transitions in higher vibrational levels. This may cause the emission features to be broader and slightly ($\sim$10 cm$^{-1}$) red shifted with respect to gas phase absorption bands. The exact magnitude of this effect vary between molecules and between vibrational modes and both are less than the natural line width of the interstellar emitters, i.e. $\sim$30 cm$^{-1}$ \cite{Brenner1992, Joblin1995, Cook1998}. There are few laboratory emission studies available which show no drastic changes between emission and absorption bands \cite{Kim2001, Kim2002} and are also in good agreement with theoretical calculations. Therefore, the spectral data presented here do not include correction for the redshift expected between the calculated absorption peaks and the emission frequencies of vibrationally excited PAHs. Comparison of interstellar IR spectra with the calculated absorption spectra could give an idea about the PAH type which may contribute in infrared emission models \cite{Mulas2006} of different astrophysical sites.

The astronomical emission bands in 10 - 15 $\mu$m range arise from the C-H out-of-plane bending vibrations \cite{Hony2001}. The intensities and positions of various peaks for these features depend on the number of adjacent hydrogen atoms on the periphery of the PAH ring \cite{Hudgins1999}. The frequency position of solo C-H wag is highest and is respectively lowered for duo, trio, quarto, and quintet C-H wags. While the solo and duo C-H out-of-plane vibrations of parent PAHs remain unmodified in phenyl-PAHs, there are additional features due to quintet C-H. Thus, astrophysical observation of any emission feature at lower frequencies may indicate phenyl substitution. The quintet C-H out-of-plane vibrations in phenyl group contribute to peaks in two regions, 692 - 706 cm$^{-1}$ (14.4 - 14.1 $\mu$m) and 770 - 730 cm$^{-1}$ (13.0 - 13.7 $\mu$m). The quintet features for neutrals lie near 695 cm$^{-1}$ (14.4 $\mu$m) and except in biphenyl the cations show small red shift (Section 3.1.1, Figure~\ref{11aFig}). The position of this mode is independent of the size and shape of phenyl-PAHs and is also unaffected by the position of phenyl group attachment. A feature between 14.2 - 14.3 $\mu$m is observed in a variety of sources \cite{Hony2001, Andrews2015, Boersma2018}, which may be the quintet C-H wag contribution from phenyl-PAHs. Therefore, the environments of these objects may contain phenyl-PAHs and must be included in emission models.

Another weak infrared emission band at 13.5 $\mu$m (741 cm$^{-1}$), observed in different astrophysical sites \cite{Hony2001}, falls in the region of the other quintet C-H out-of-plane feature, i.e., 13.0 - 13.7 $\mu$m range. This quintet phenyl C-H wag mode is mixed with quartet or trio C-H wags in phenyl-PAHs (section 3.1.1). Among the phenyl-PAHs in the present work, the phenylpyrenes and neutral 2-phenylpyrene, in particular, show good match with the feature at 741.0 cm$^{-1}$ (13.5$\mu$m) (Table~\ref{tab1b}). This is a trio C-H wag that has mixed contribution from phenyl C-H out-of-plane bending. There is negligible red shift of this feature upon ionization. The mixing of quintet and trio/quarto C-H out-of-plane modes in phenyl-PAHs may provide prominent features in this spectral range, so more pyrene like compact PAHs with phenyl side groups need to be studied.

\begin{figure}
\centerline{\includegraphics[width=0.9\textwidth]{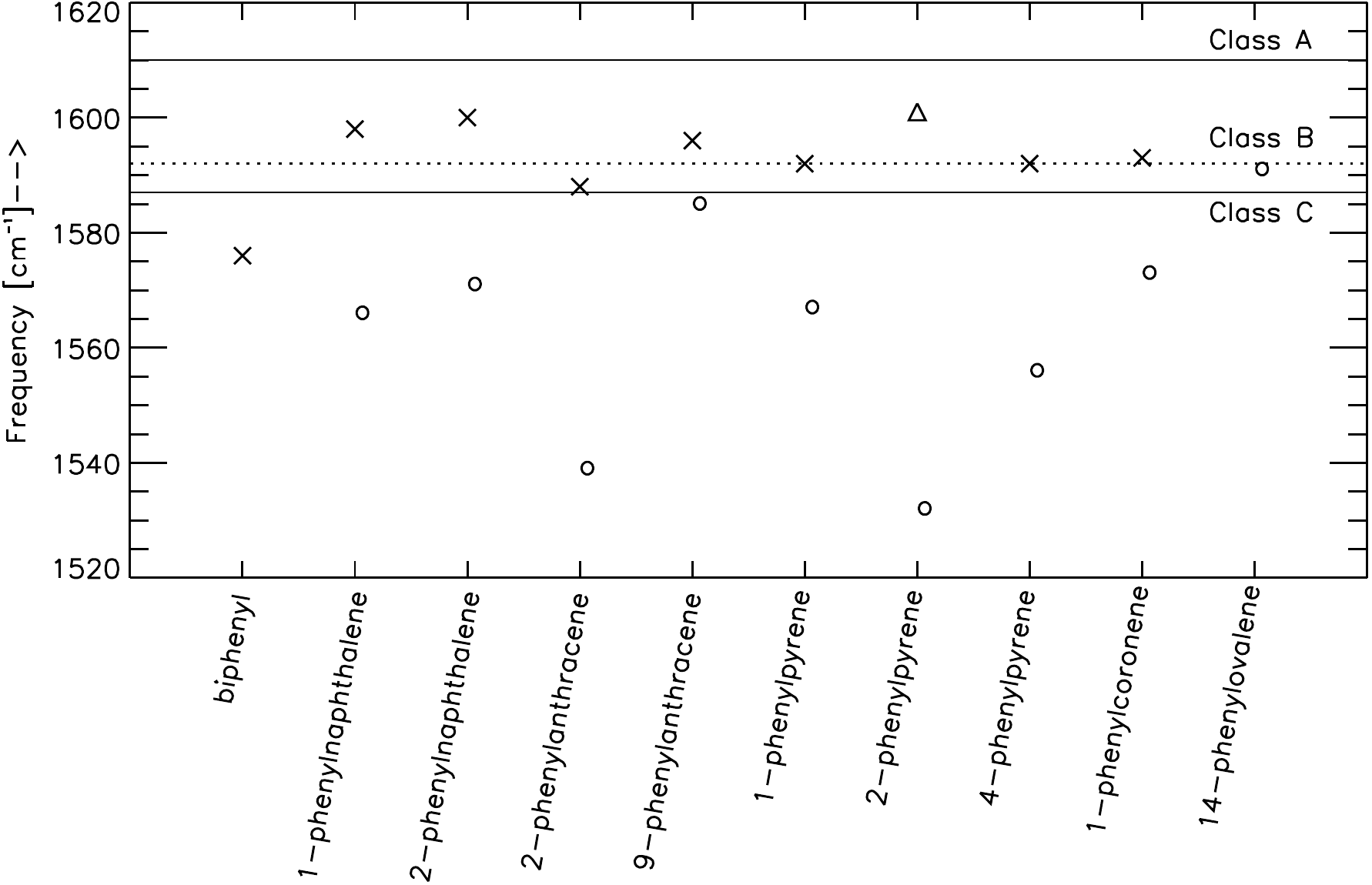}}
\caption{The position of modes near 6.2~$\mu$m AIB in phenyl-PAH cations; `$\times$' - C-C stretch in phenyl unit and `o' - C-C stretch in PAH moiety; `{\tiny $\triangle$}' - phenyl C-C stretch in neutral 2-phenylpyrene.}
\label{11Fig}
\end{figure}

Astrophysical objects exhibiting the 14.2 $\mu$m feature generally show class A type 6.2 $\mu$m AIB (peak in 6.20 - 6.22 $\mu$m) \cite{Peeters2002}. But some late type objects, viz. HD 44179 (Post- AGB star), HE 2-113 (Planetary nebulae), IRAS 17047 (Planetary nebulae), with 14.2 $\mu$m band show class B type 6.2 $\mu$m AIB (peak in 6.23 - 6.26 $\mu$m) \cite{Peeters2002}. In the present work, all the phenyl-PAH cations except 2-phenylpyrene exhibit a significantly strong peak (Figure~\ref{11Fig}) giving closer match to class B 6.2 $\mu$m AIB. In the case of 2-phenylpyrene it is the neutral which gives a strong phenyl C-C stretch peak that is also close to class B 6.2 $\mu$m AIB. This points towards the possibility of the presence of phenyl-PAHs in the environments of such late type stars. Neutral 2-phenylpyrene is also interesting due to additional strong modes in this range. While the 1432~cm$^{-1}$ (6.98~$\mu$m) peak falls close to a feature $\sim$ 6.9~$\mu$m, observed in protoplanetary nebulae and cool objects \cite{Hrivnak2000, Buss90}, the 1501~cm$^{-1}$ (6.66~$\mu$m) peak can correspond to observed sub-feature along IRAS 18434-0242 \cite{Peeters99} and in H~II region-like and seyfert~2 galaxies \cite{Laureijs2000}. The 1500~cm$^{-1}$ peak is also present in neutral 2-phenylnaphthalene.

The substitution of phenyl side group on PAHs does shift the C-C stretch mode of parent PAH towards higher frequencies that depend on the site of phenyl substitution. Astrophysical observations do not show any strong feature on the red side of the 6.2~$\mu$m AIB and specific object spectrum needs to be examined more closely for possible fitting of a broadened Class~C AIB feature. In 9-phenylanthracene and 14-phenylovalene, where phenyl substitution is at solo position, there is large enhancement in the mode bringing it within the 6.2~$\mu$m AIB region (Figure~\ref{11Fig}). 

The current NASA PAH data base \cite{Bauschlicher2018} contains a large variety of PAHs but PAHs with phenyl side group are not considered. The comparison with observations does indicate towards the possibility of phenyl-PAHs in the ISM. In particular species with substitution at 2$^{nd}$ position are relatively stable and compact 2-phenylpyrene type or solo substituted species appear to be more promising.

\section{Conclusions}

Structural analysis of a number of neutral and positively charged phenyl-PAHs is performed and their corresponding infra-red spectra are examined. Generally, the infra-red spectral features of phenyl-PAH neutrals and cations are similar to those in their corresponding plain PAHs. But there are a number of bands that are peculiar to phenyl-PAHs. 

The main conclusions of the study are summarized below:
\begin{itemize}
\item The variation of the energy of optimized molecular structure at different torsion angles, for the phenyl group, shows that the substitution of phenyl at 2$^{nd}$ position gives more stable specie than the substitution of phenyl at other positions in different PAH molecules. 
\item The quintet C-H out-of-plane vibrations, possible only in phenyl group, lie near 695 cm$^{-1}$ (14.4 $\mu$m) in all the studied phenyl-PAHs and is independent of the molecule's shape, size and position of phenyl substitution. This feature compares well with the observed emission feature between 14.2 - 14.3 $\mu$m in diverse astrophysical objects \cite{Hony2001, Andrews2015, Boersma2018}.
\item The quintet mixed with quartet or trio C-H out-of-plane bending modes produce minor band in 13.0 - 13.7 $\mu$m range. Among the studied phenyl-PAHs, phenylpyrenes and neutral 2-phenylpyrene in particular, show good match with the observed astrophysical feature at 741.0 cm$^{-1}$ (13.5$\mu$m) \cite{Hony2001}.
\item A significantly strong peak due to phenyl C-C stretch (Figure~\ref{11Fig}), giving close match to class B 6.2 $\mu$m AIB, is exhibited in all phenyl-PAH cations. In the case of 2-phenylpyrene it is the neutral which gives a strong phenyl C-C stretch peak that is also close to class B 6.2 $\mu$m AIB. This points towards the presence of phenyl-PAHs in the environments of late type stars, which exhibit class B 6.2 $\mu$m AIB.
\item Neutral 2-phenylpyrene is also interesting due to additional strong modes in the 1400 - 1500~cm$^{-1}$ range. While the 1432~cm$^{-1}$ (6.98~$\mu$m) peak falls close to a feature $\sim$ 6.9~$\mu$m, observed in protoplanetary nebulae and cool objects \cite{Hrivnak2000, Buss90}, the 1501~cm$^{-1}$ (6.66~$\mu$m) peak can correspond to observed sub-feature along IRAS 18434-0242 \cite{Peeters99} and in HII region like and seyfert~2 galaxies \cite{Laureijs2000}. This mode is also present in neutral 2-phenylnaphthalene.
\item The substitution of phenyl side group on PAHs also shift the C-C stretch mode of parent PAH towards higher frequencies that depend on the position of phenyl substitution. The shift in solo position substituted phenyl-PAHs is large enough to reach the 6.2 $\mu$m AIB region. 
\end{itemize}

On the basis of structural and spectral analysis, the possibility of phenyl-PAHs in carbon rich cool circumstellar shells is indicated. Medium sized compact phenyl-PAHs can form in high-density regions and shocks \cite{Woods2002, Mimura2003} that are possible in protoplanetary circumstellar shells. A bottom-up scenario for the formation of PAHs is suggested \cite{Cox2016} in dense photodissociation regions (PDRs) of planetary and protoplanetary nebulae. Hence, the PAH database, used to model AIB emission features for specific astrophysical objects, must include phenyl-PAHs.

\section{Acknowledgment}
Authors acknowledge the use of computing and library facilities of the Inter-University Centre for Astronomy and Astrophysics, Pune, India. A. Maurya acknowledges the CSIR-HRDG New Delhi for its RA fellowship.

\bibliography{pah-bib}
\bibliographystyle{elsarticle-num}


\clearpage
\begin{figure}
\centerline{\includegraphics[width=1.0\textwidth]{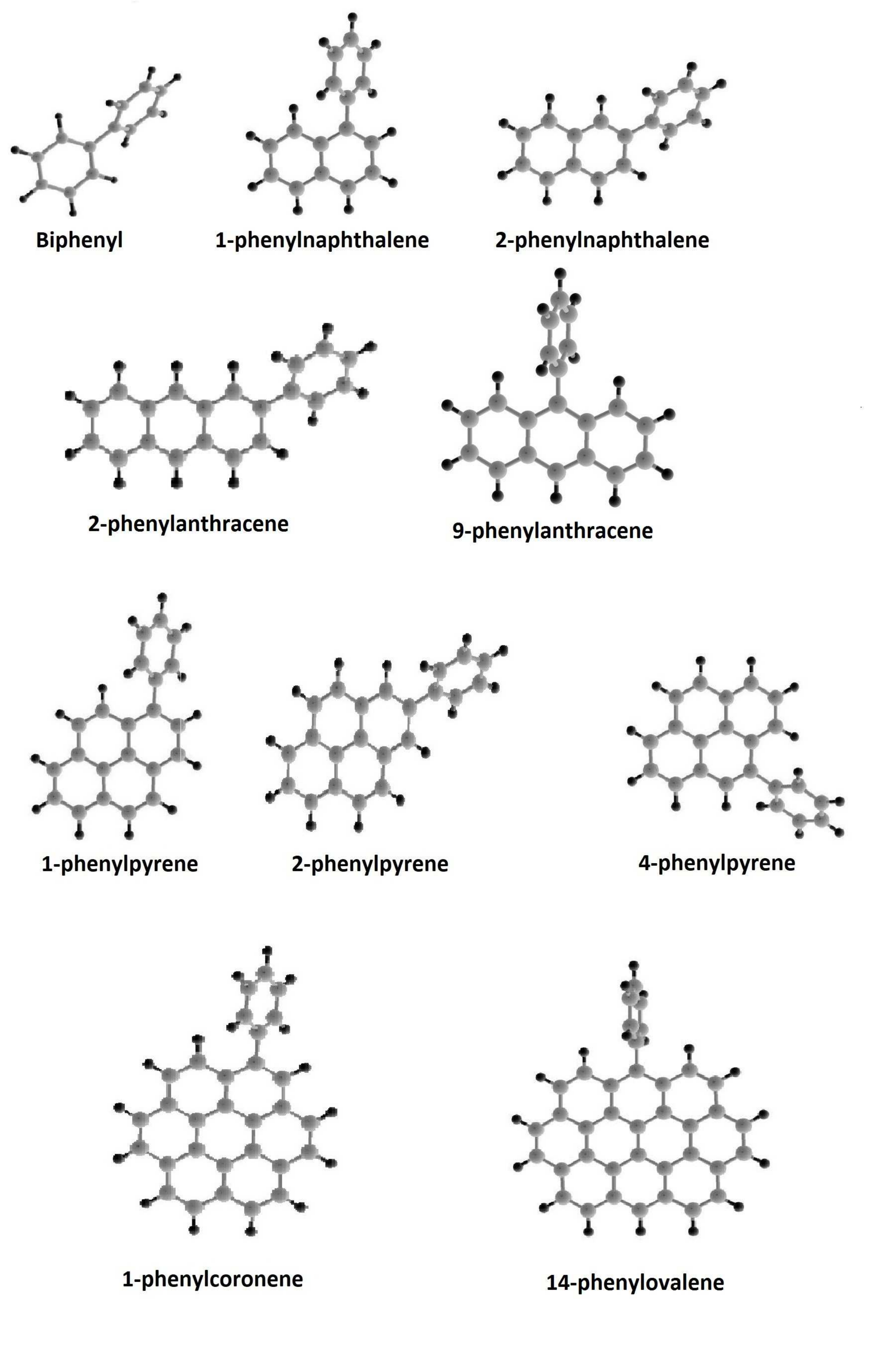}}
\caption{Molecular structures of the studied phenyl-PAHs.}
\label{1Fig}
\end{figure}

\clearpage
\begin{figure}
     \ContinuedFloat*
    \centerline{\includegraphics[width=0.5\textwidth]{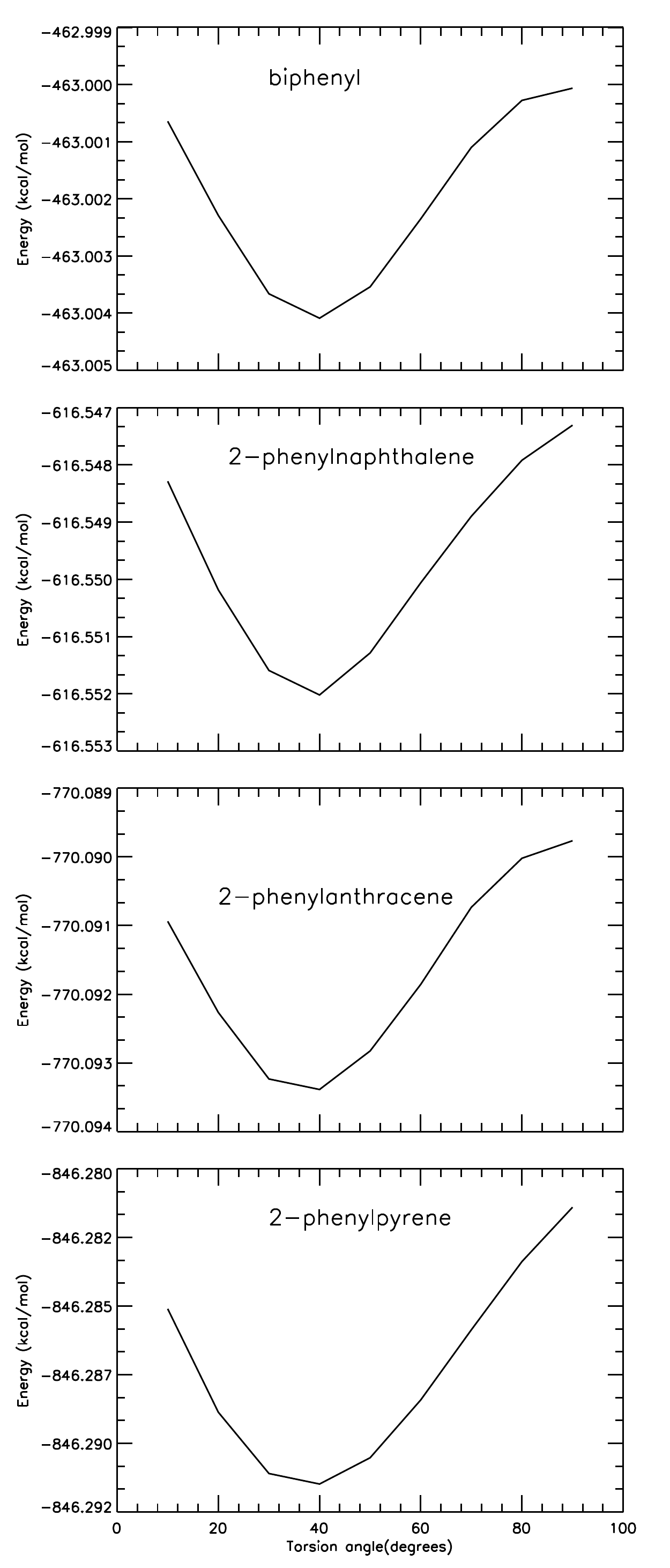}}
    \caption{Variation of optimization energy with torsion angle of the phenyl group substituted in symmetric H placement.}
    \label{2FigA}
\end{figure}

\clearpage
\begin{figure}
     \ContinuedFloat
    \centerline{\includegraphics[width=0.8\textwidth]{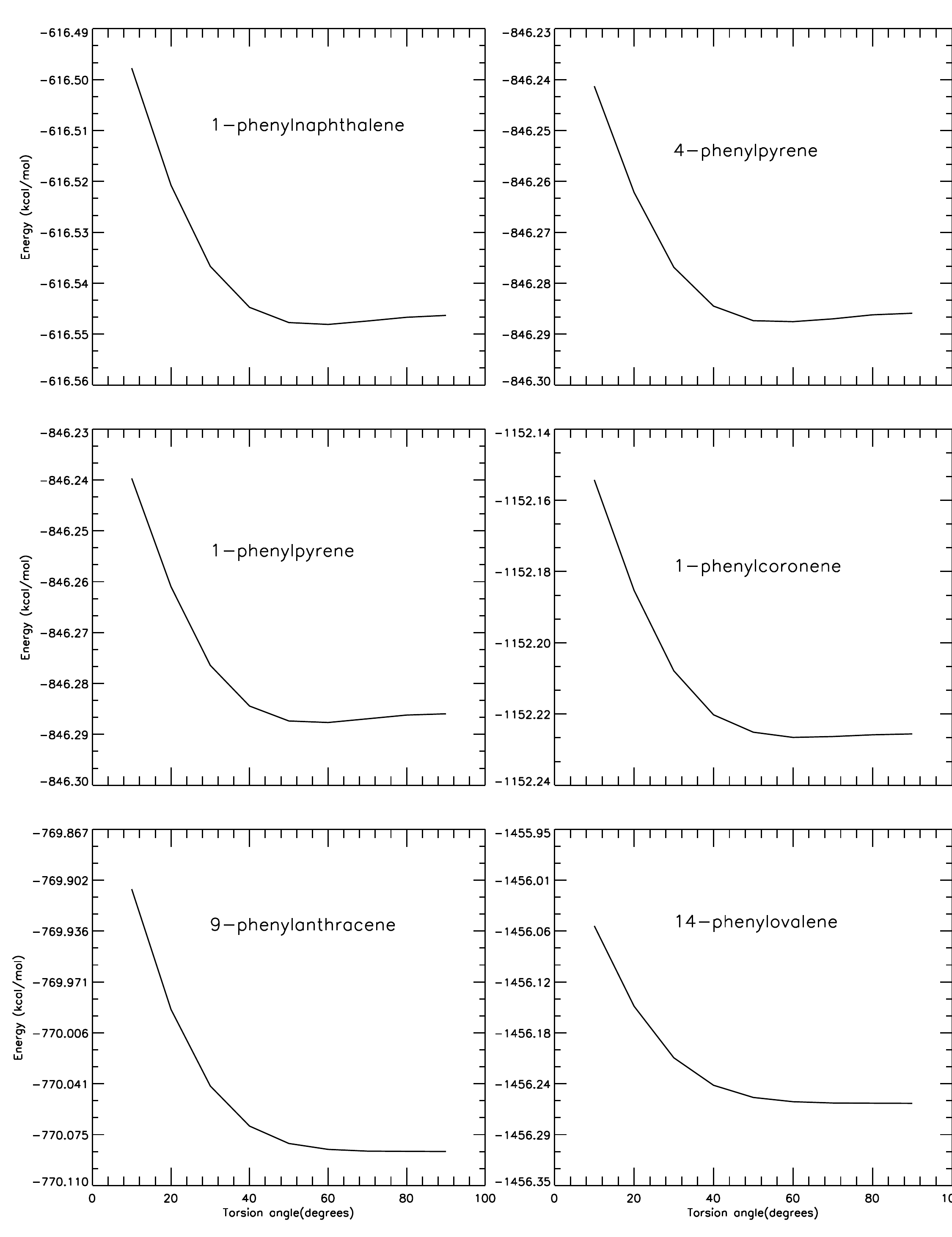}}
    \caption{Variation of optimization energy with torsion angle of the phenyl group substituted in asymmetric H placement and at solo position.}
    \label{2FigB}
\end{figure}

\clearpage

\clearpage

\clearpage

\clearpage

\clearpage

\clearpage

\clearpage

\clearpage

\clearpage

\clearpage
\begin{figure}
\centerline{\includegraphics[width=0.8\textwidth]{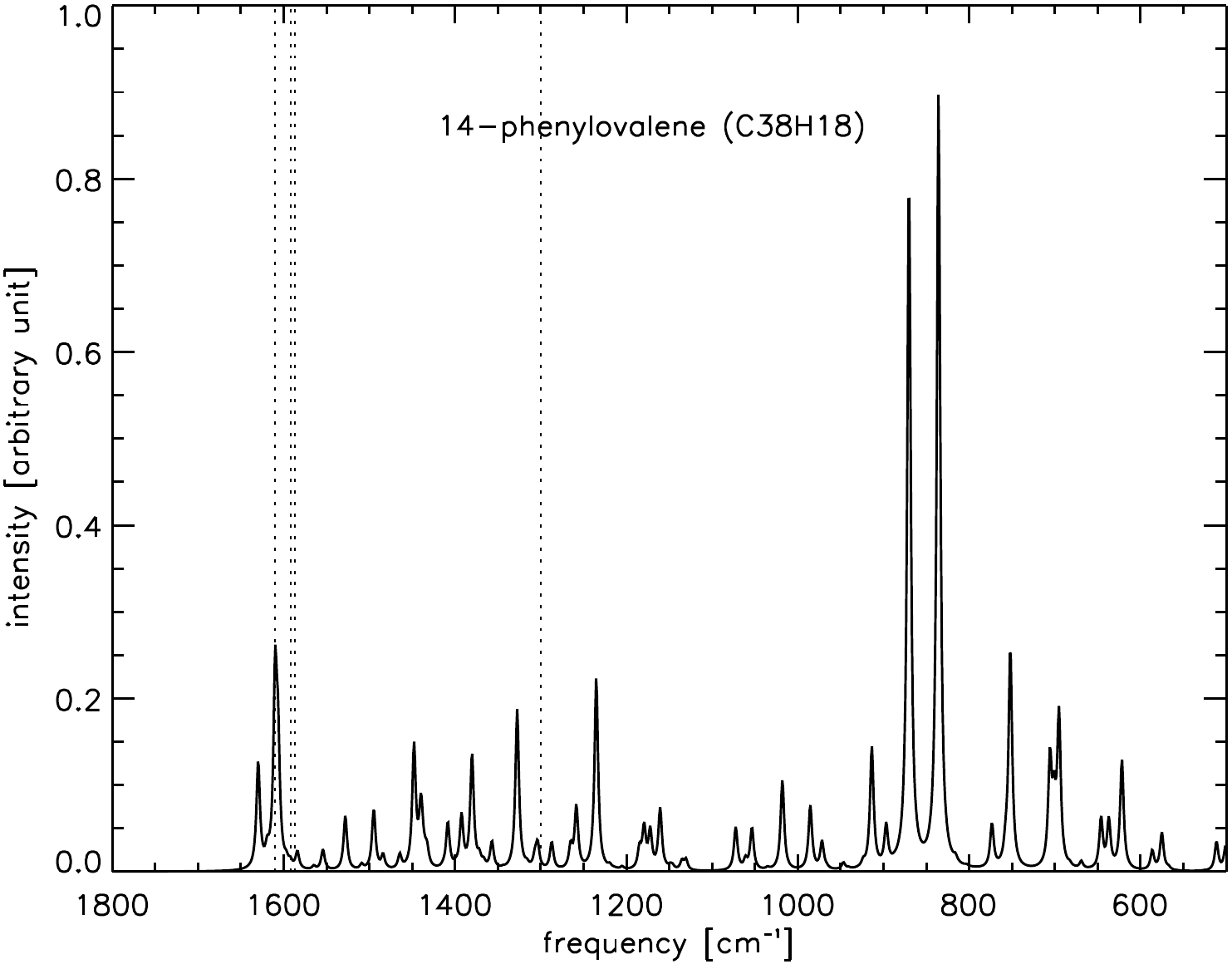}}
\centerline{\includegraphics[width=0.8\textwidth]{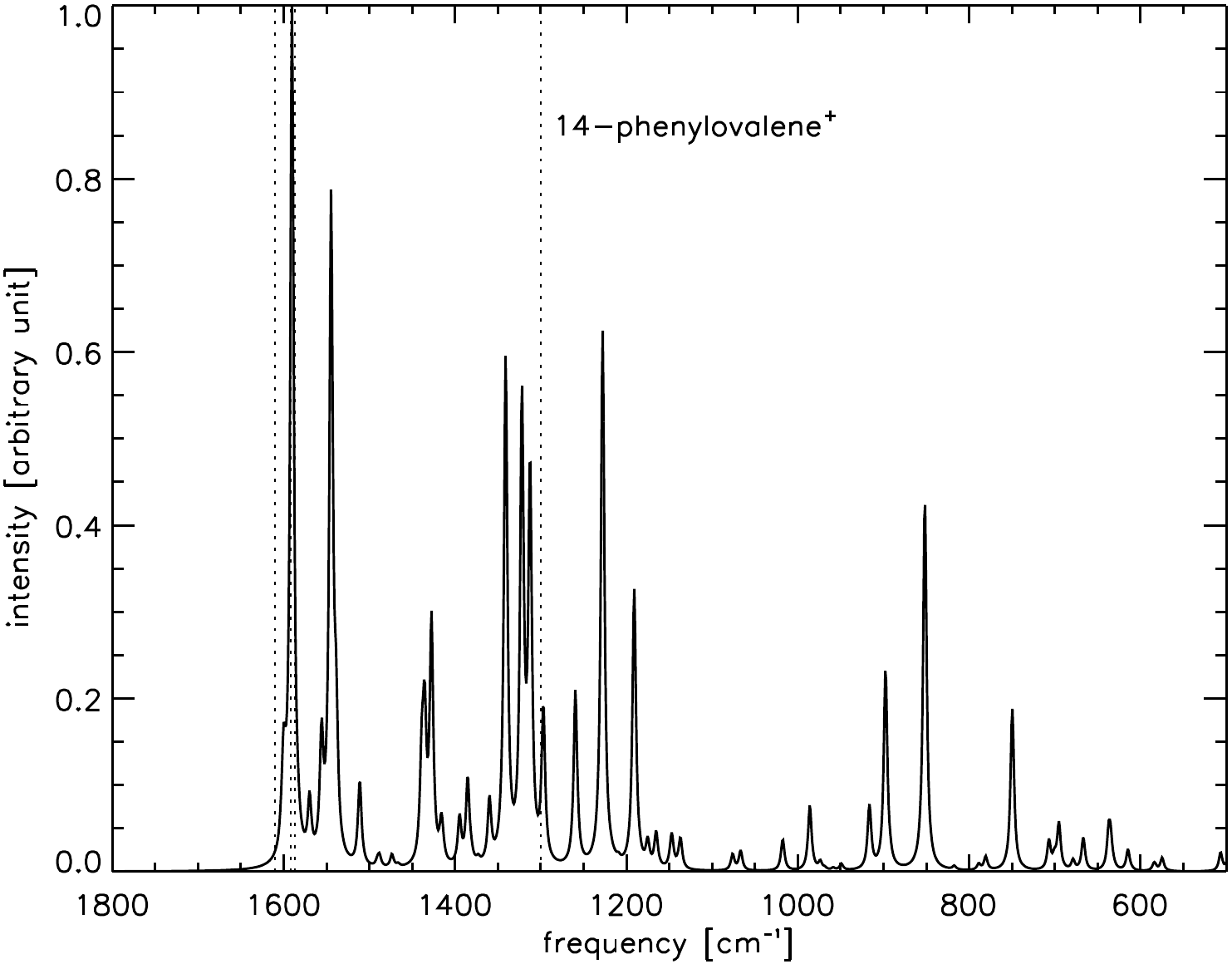}}
\caption{Computed infrared spectra of 14-phenylovalene neutral and cation.}
\label{9Figa}
\end{figure}

\clearpage
\begin{figure}
     \ContinuedFloat*
        \centering
        \begin{tabular}{cc}
                      \includegraphics[width=0.4\textwidth]{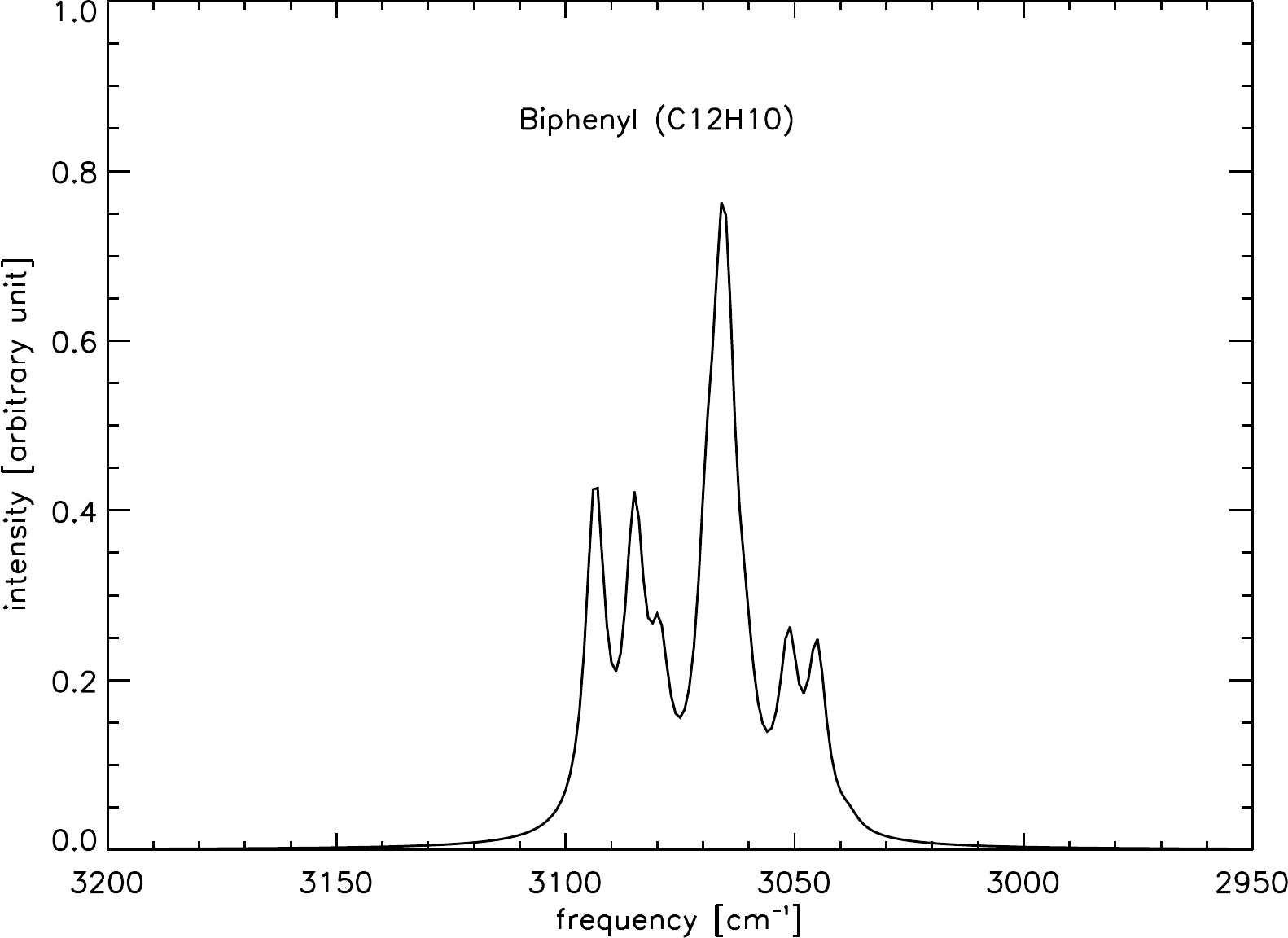}&
                      \includegraphics[width=0.4\textwidth]{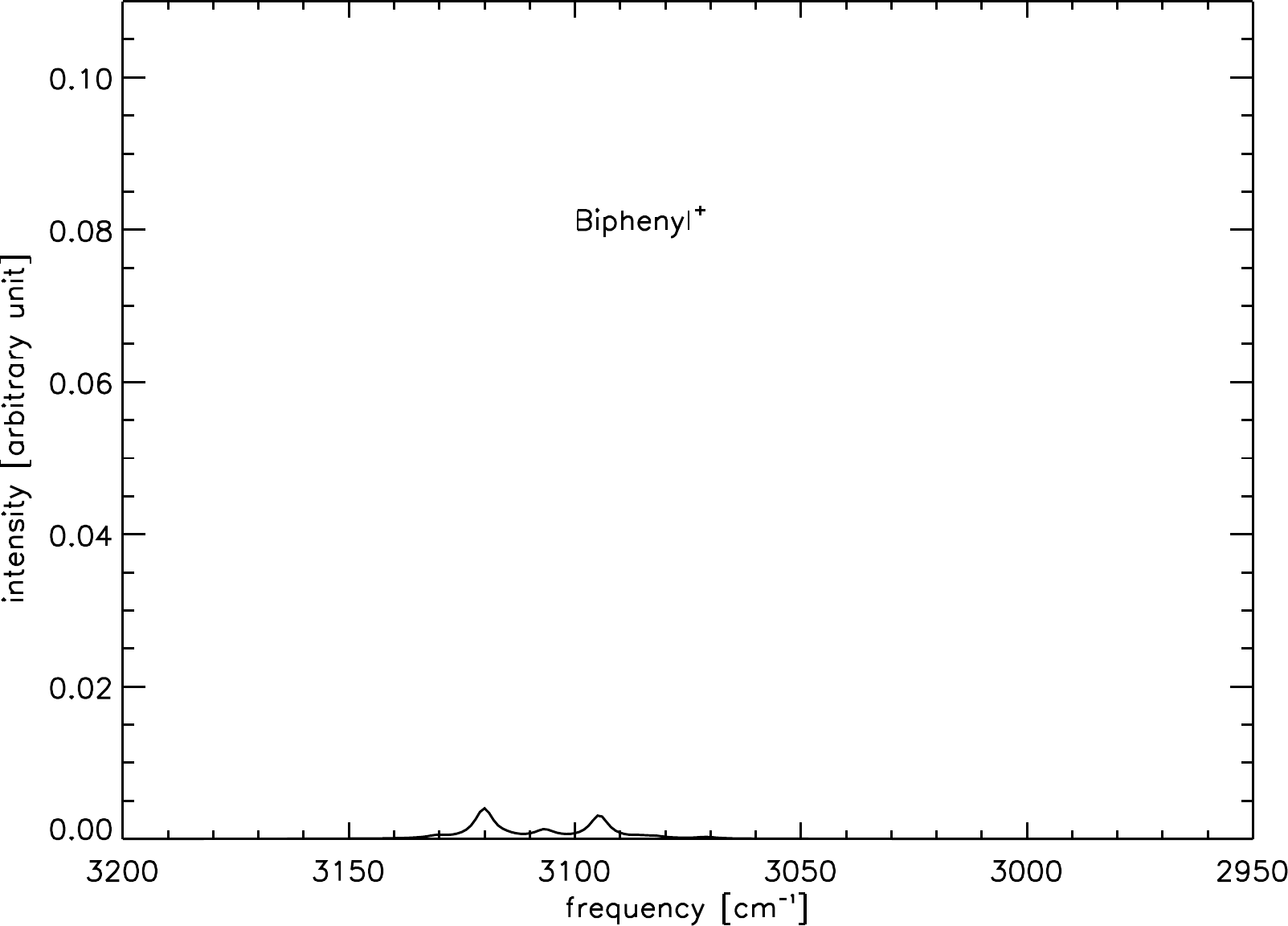}\\
                      \includegraphics[width=0.4\textwidth]{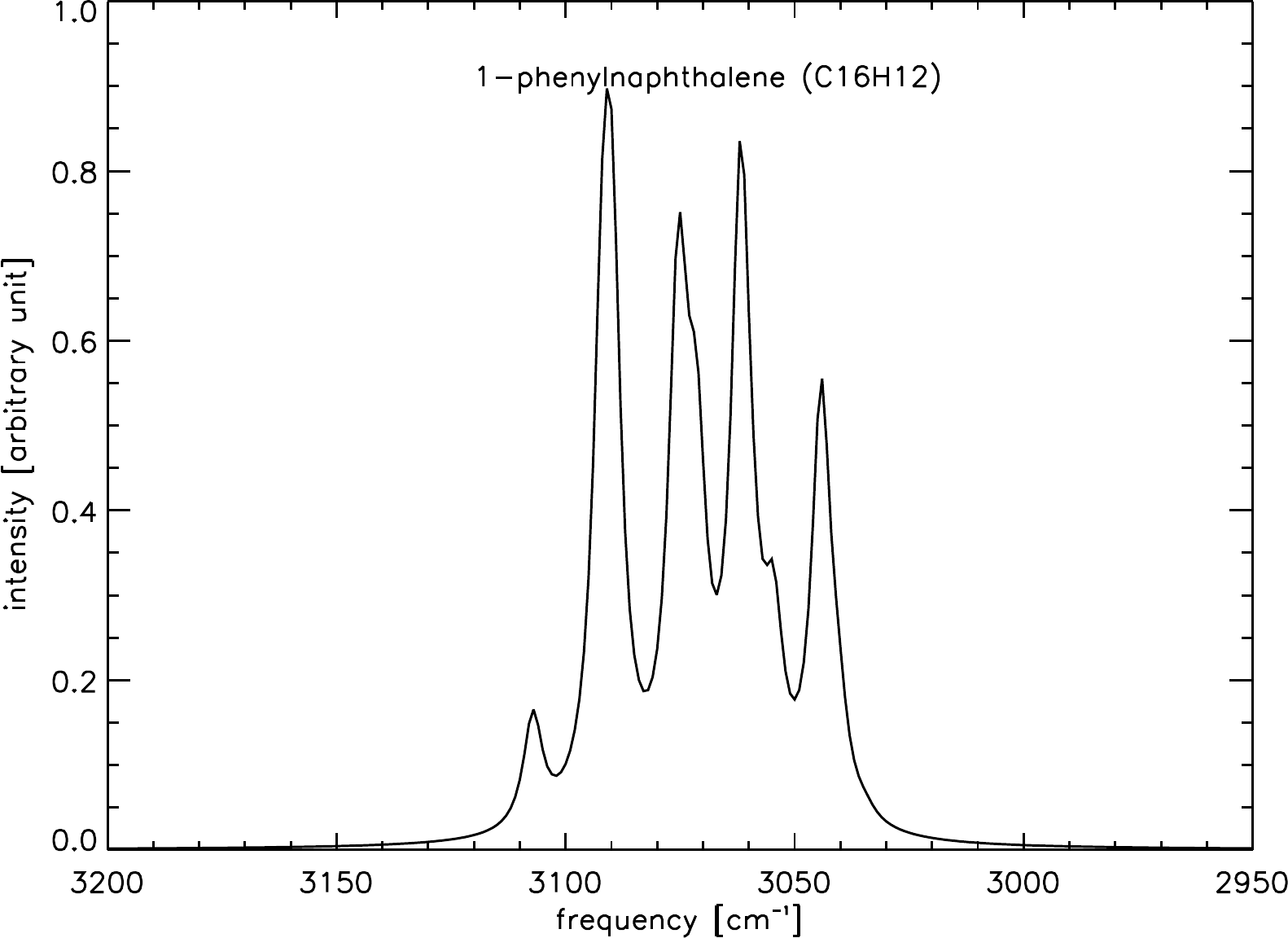}&
                      \includegraphics[width=0.4\textwidth]{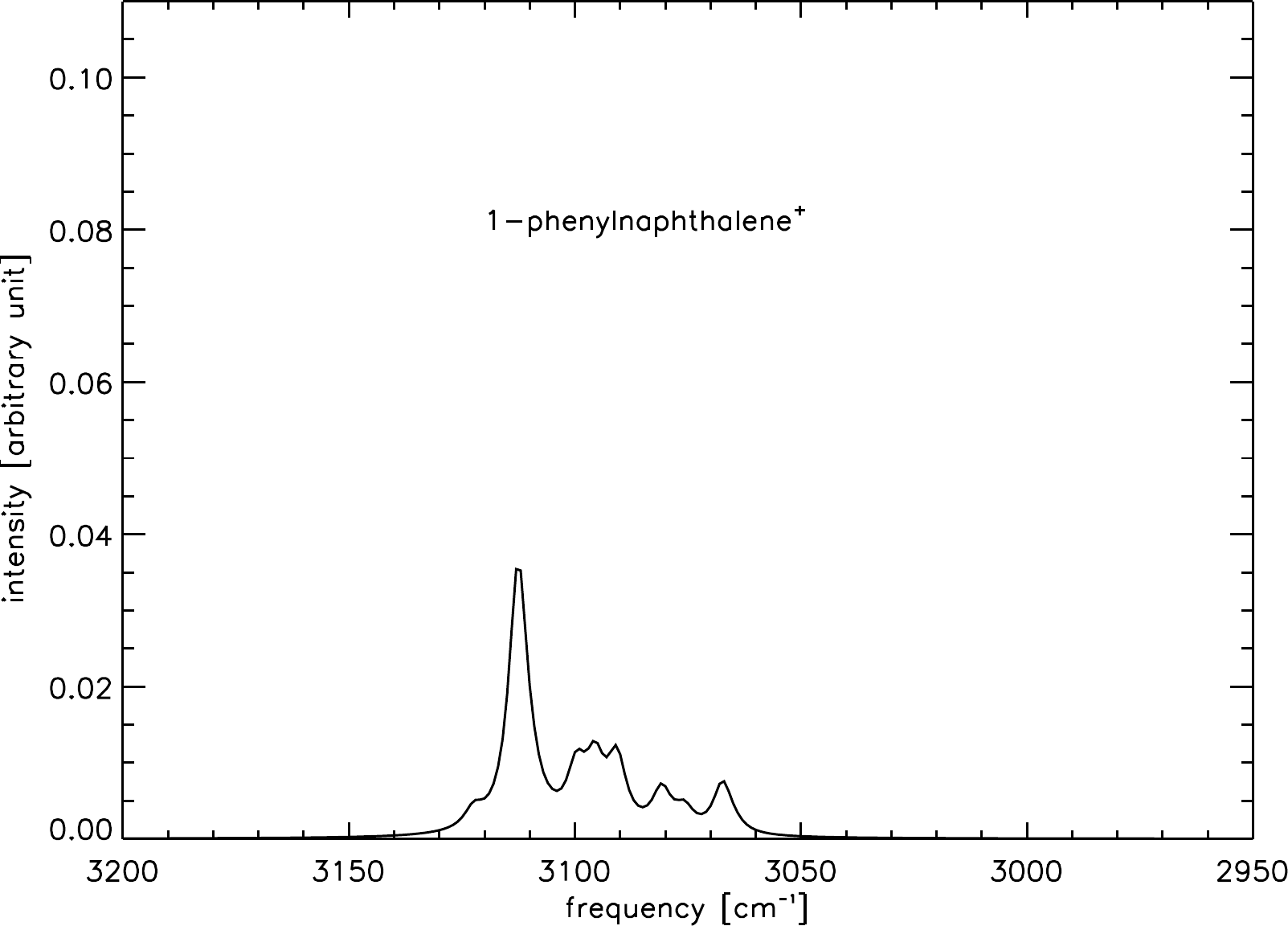}\\
                      \includegraphics[width=0.4\textwidth]{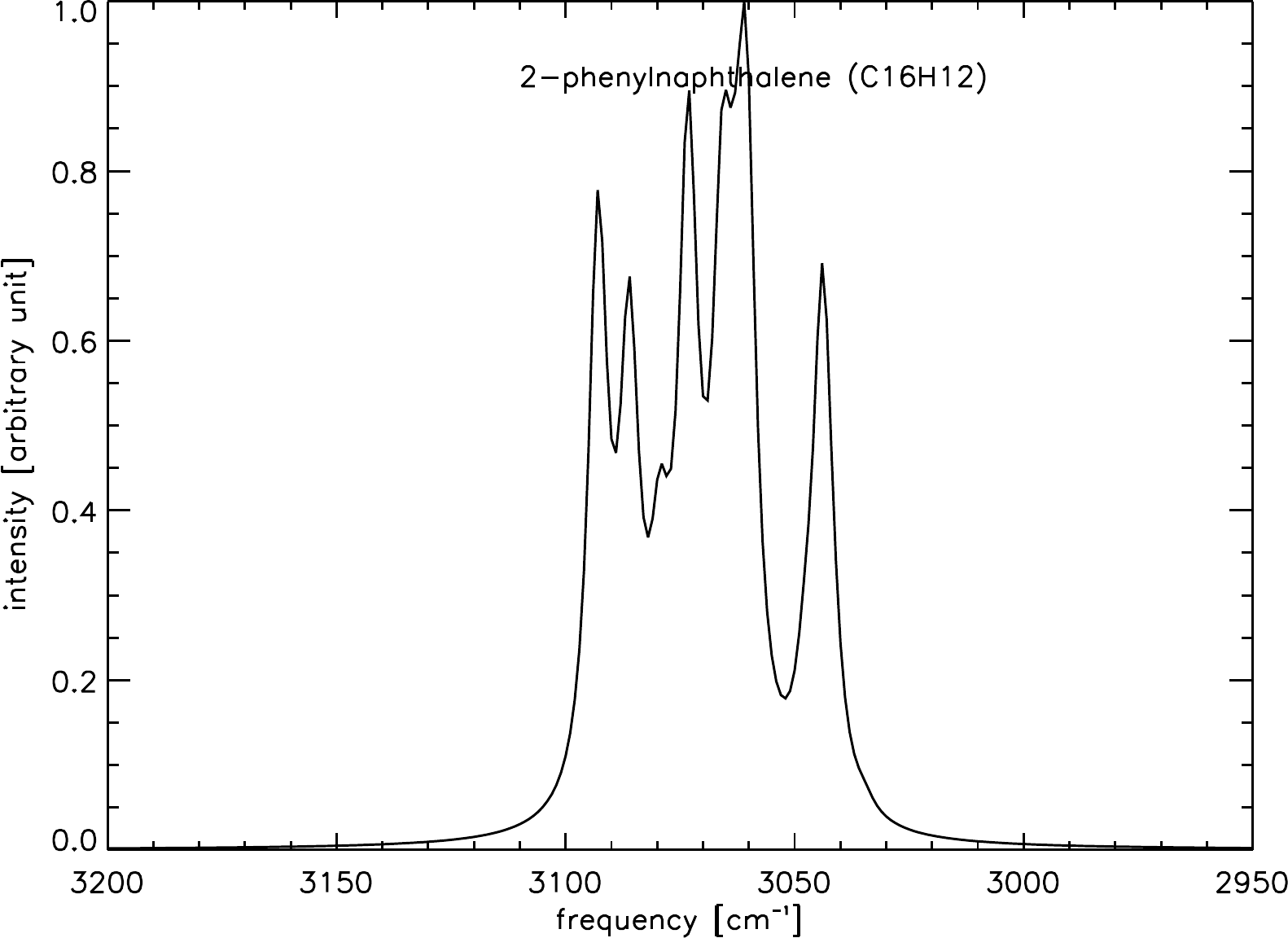}&
                      \includegraphics[width=0.4\textwidth]{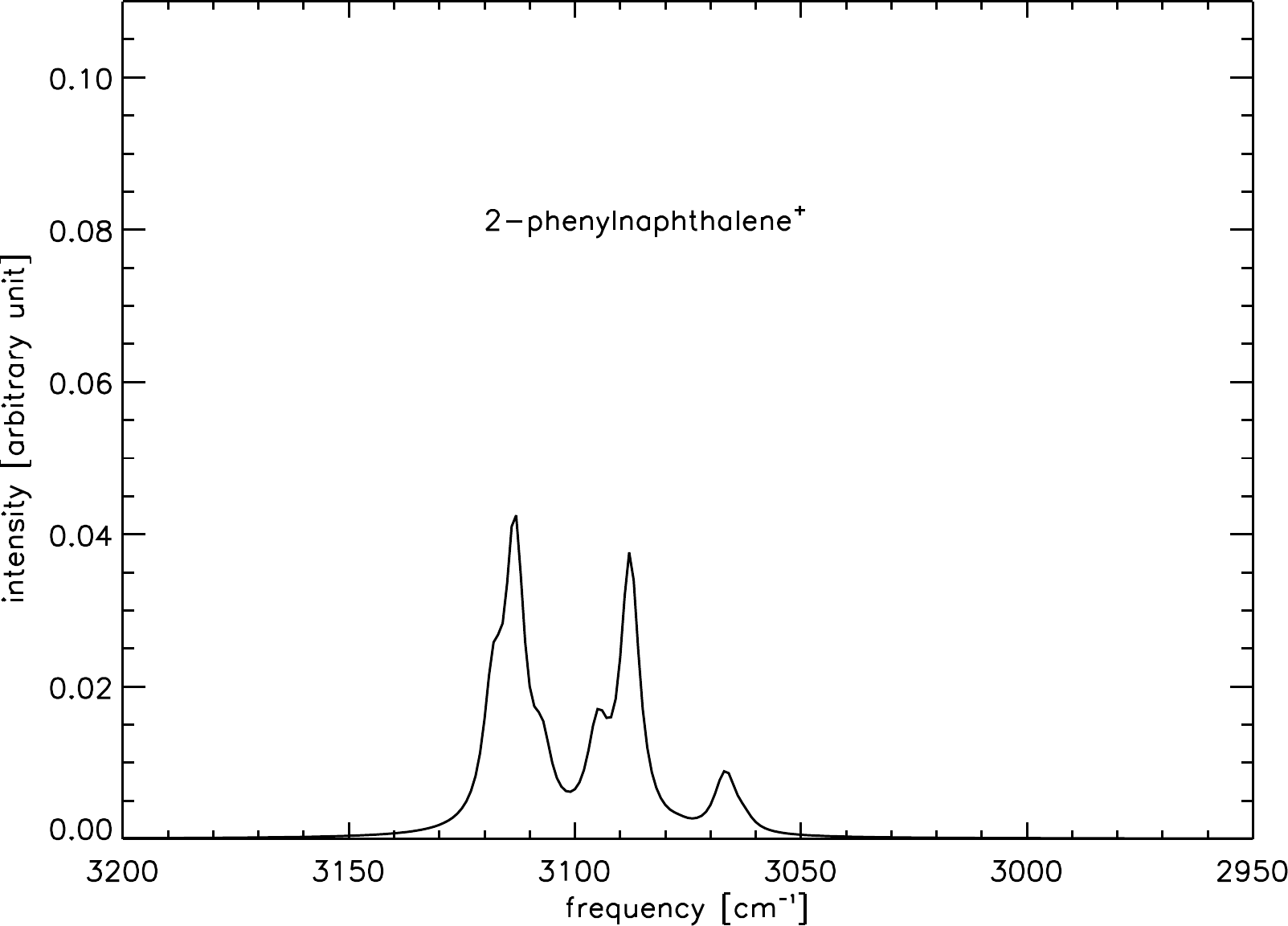}
                     
        \end{tabular}
\caption{The C-H stretch spectral region of infrared spectra in neutrals and cations of biphenyl and phenylnaphthalenes}
\label{10Fig}
\end{figure}

\clearpage
\begin{figure}
     \ContinuedFloat
        \centering
        \begin{tabular}{cc}
                      \includegraphics[width=0.4\textwidth]{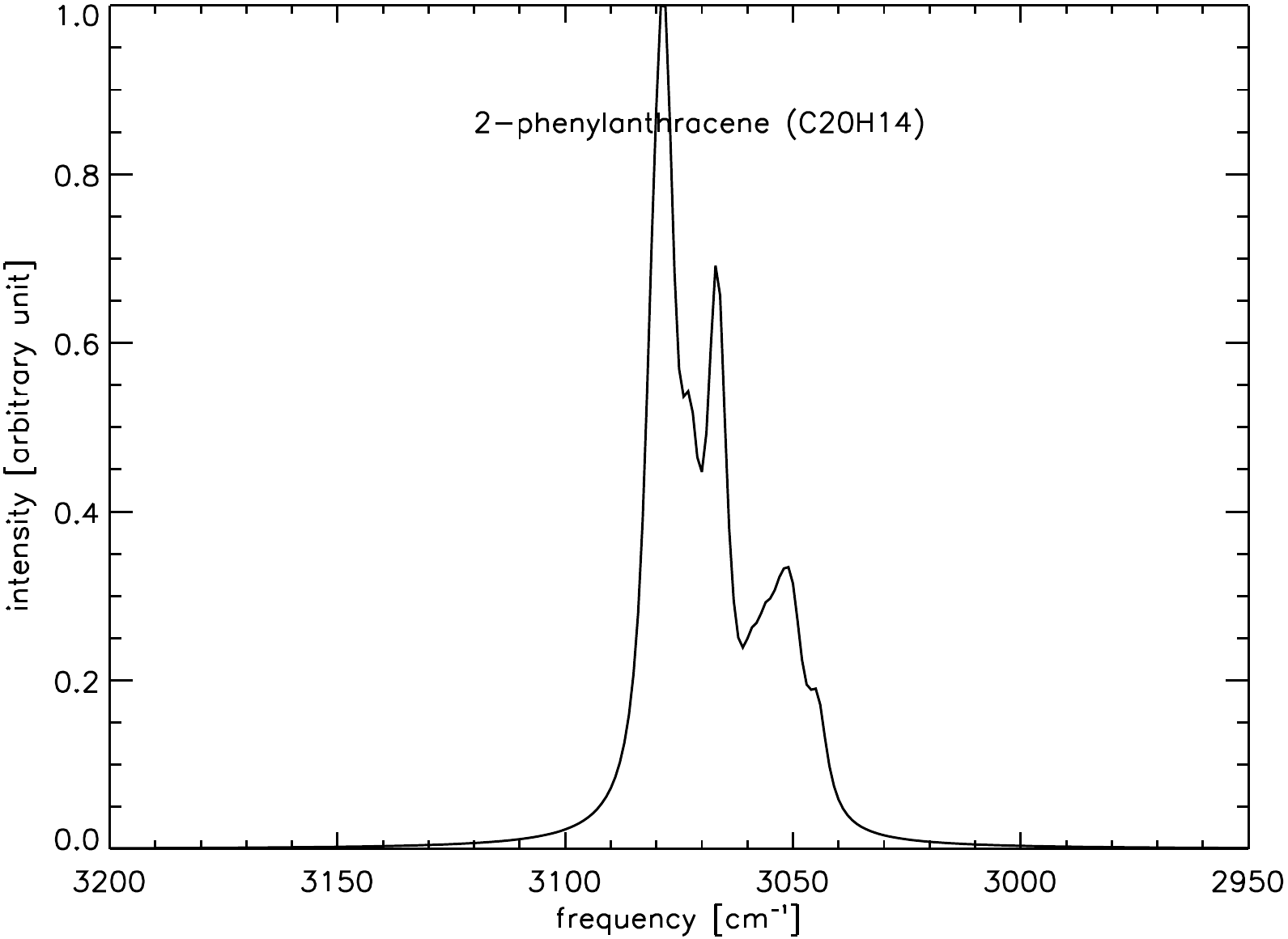}&
                      \includegraphics[width=0.4\textwidth]{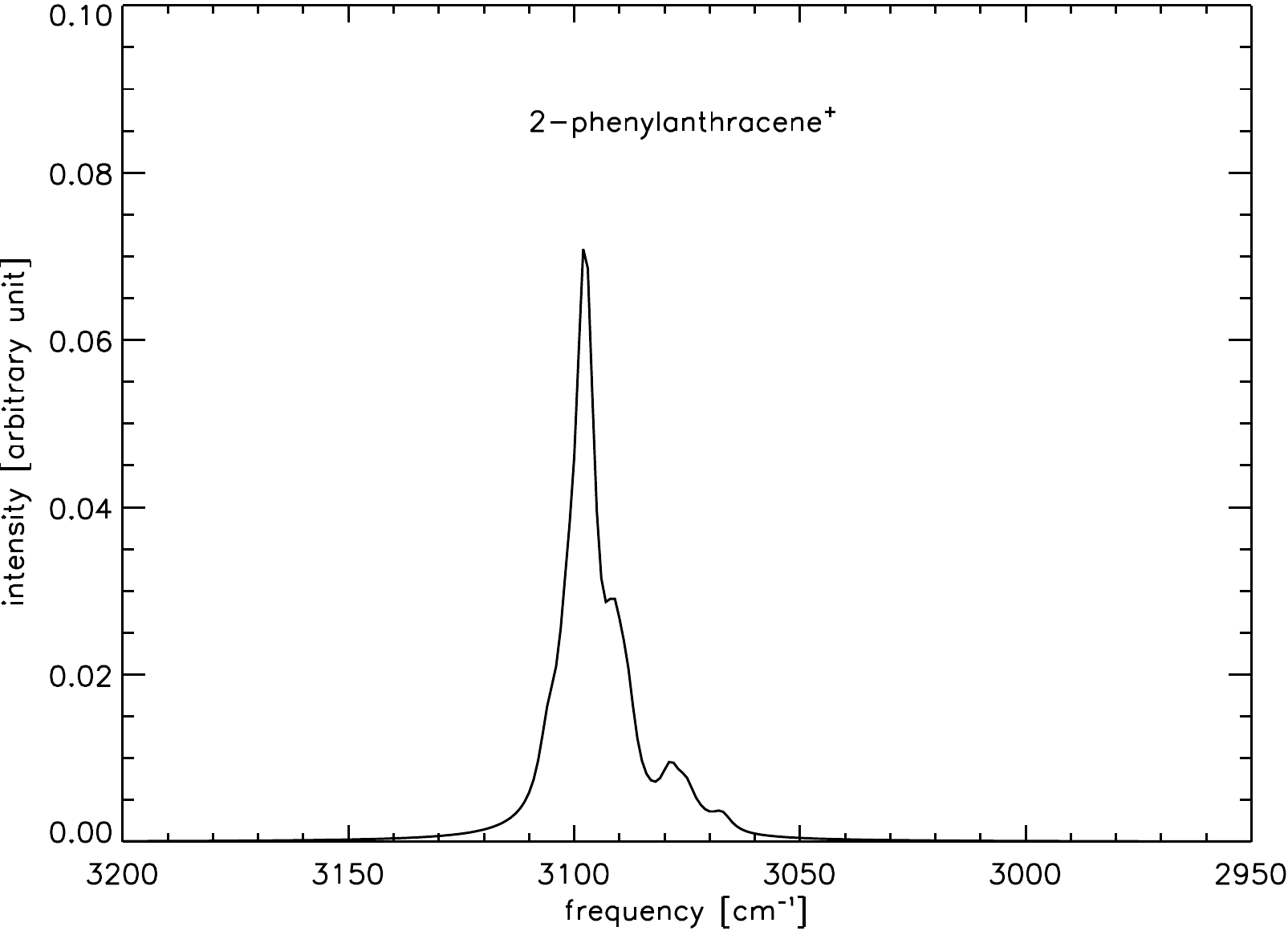}\\                     
                      \includegraphics[width=0.4\textwidth]{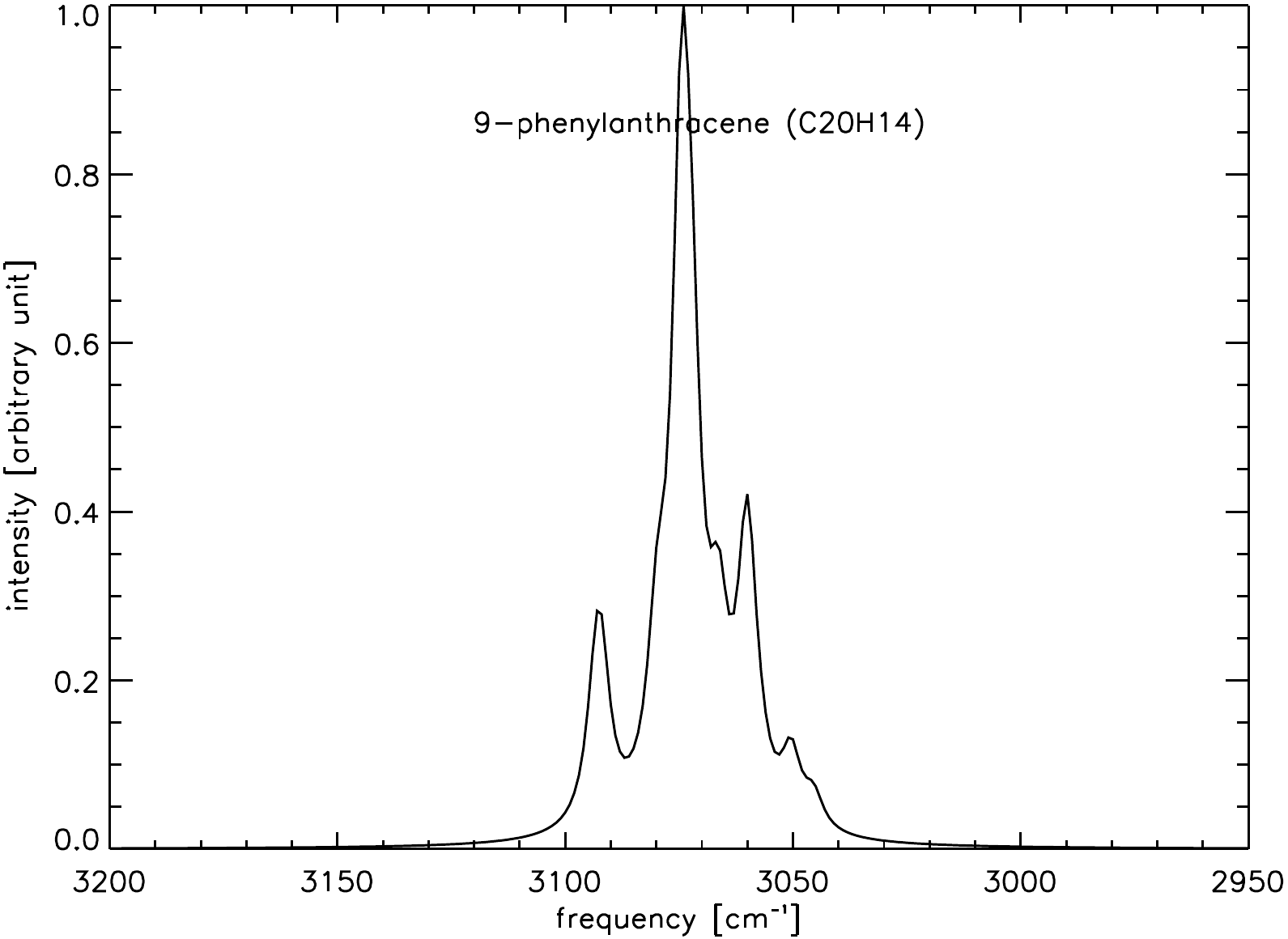}&
                      \includegraphics[width=0.4\textwidth]{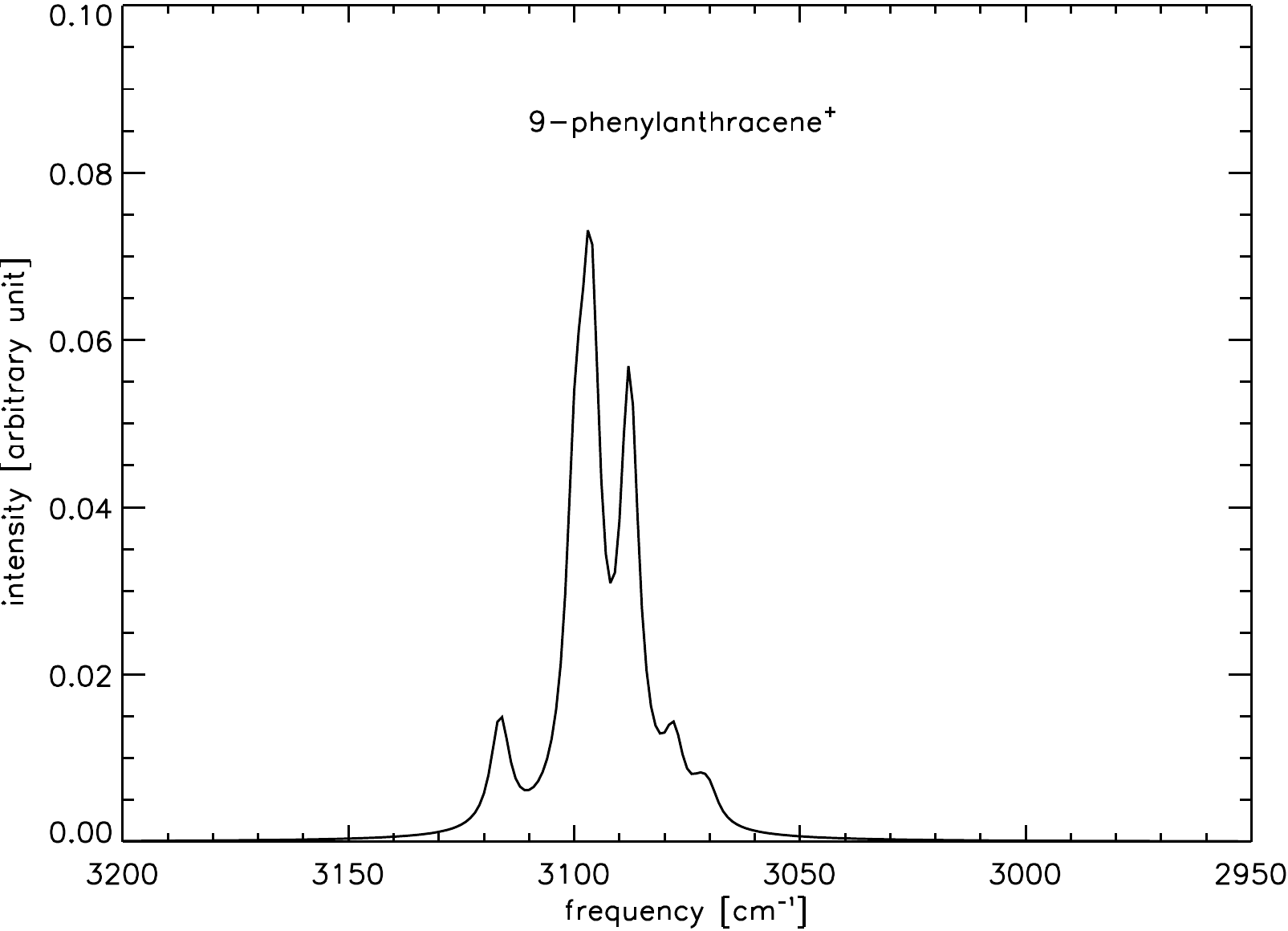}

 \end{tabular}
\caption{The C-H stretch spectral region of infrared spectra in neutrals and cations of phenylanthracenes.}
\end{figure}

\clearpage
\begin{figure}
     \ContinuedFloat

        \centering
        \begin{tabular}{cc}
                      \includegraphics[width=0.4\textwidth]{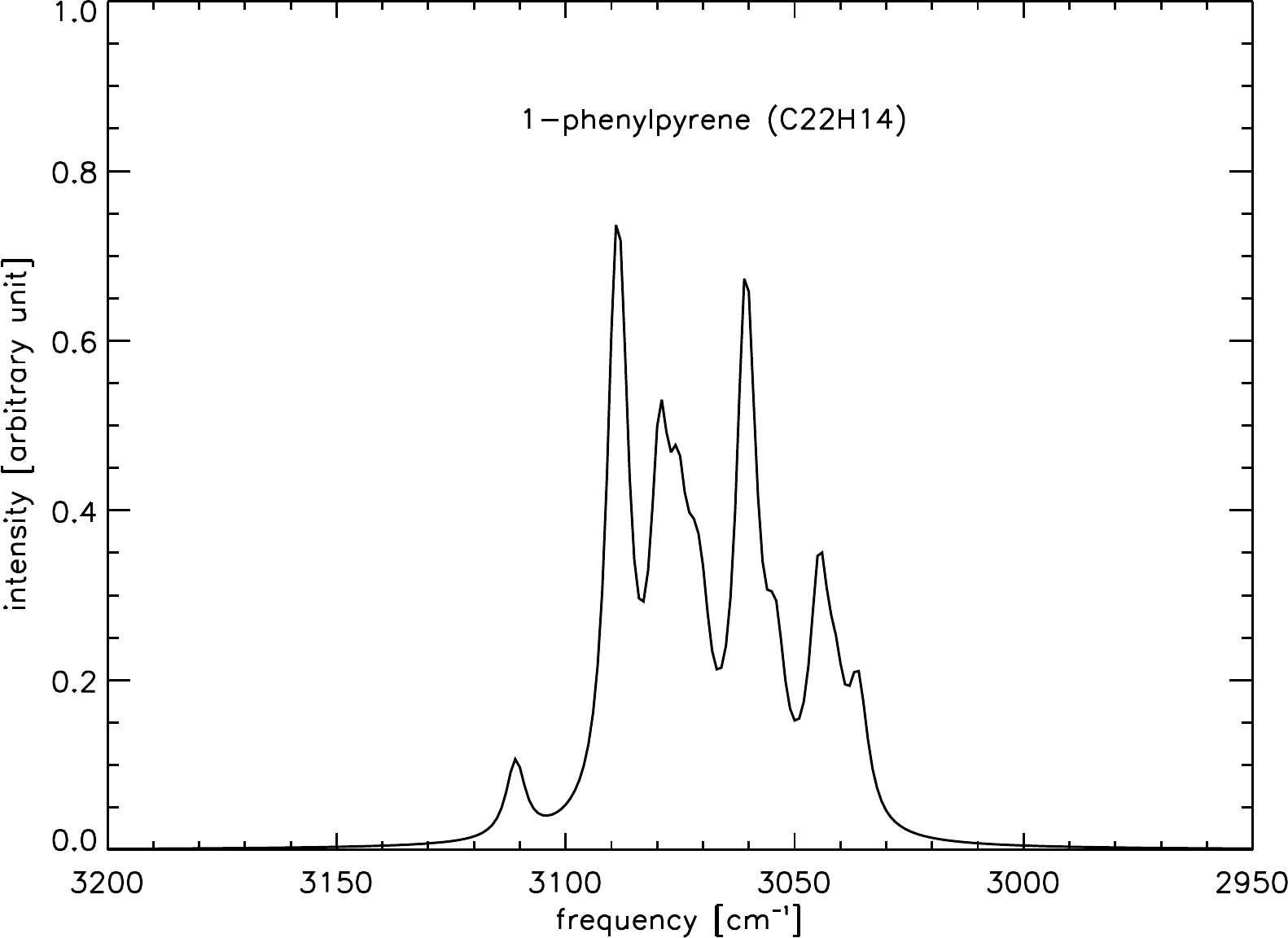}&
                      \includegraphics[width=0.4\textwidth]{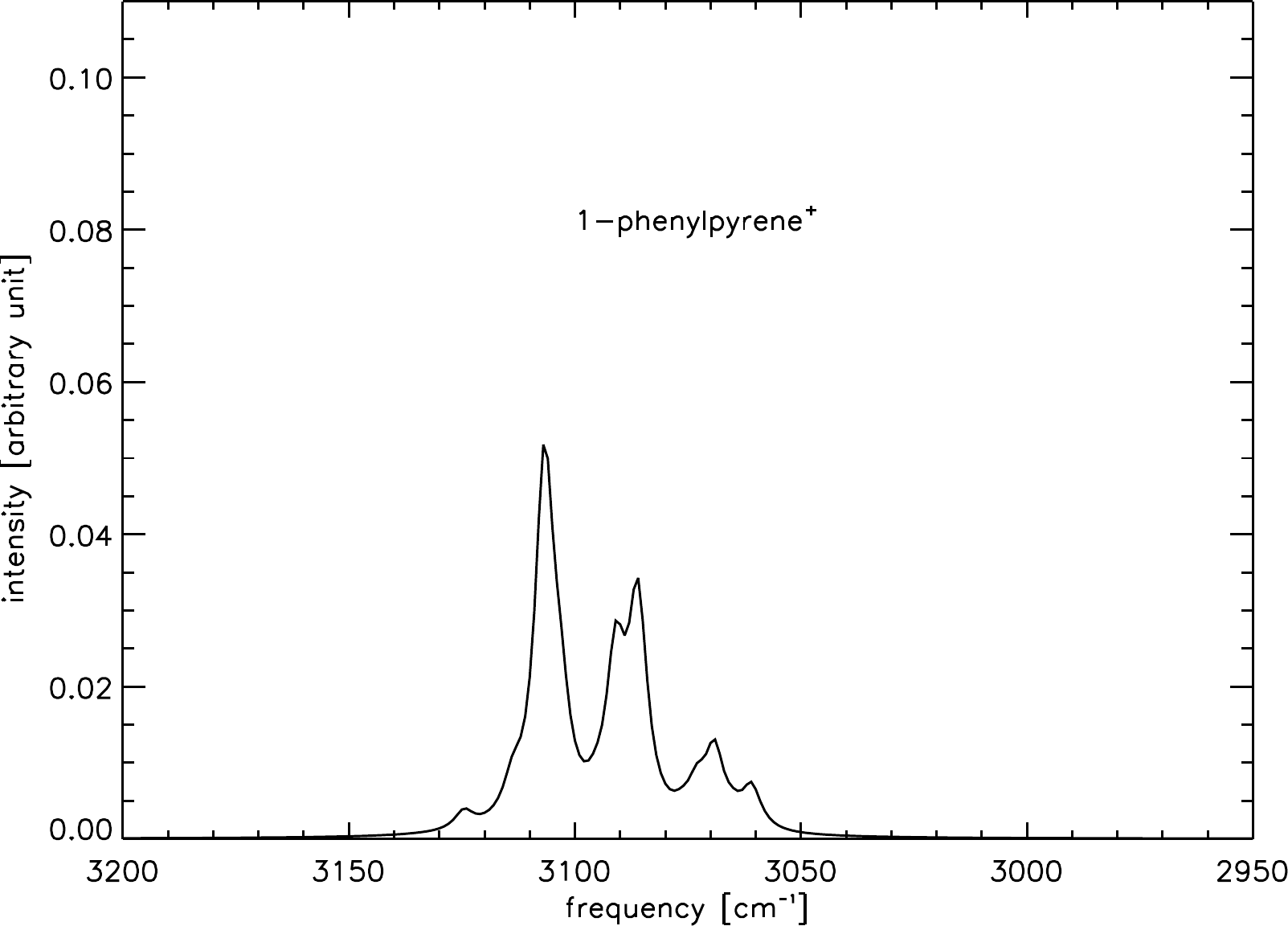}\\
                      \includegraphics[width=0.4\textwidth]{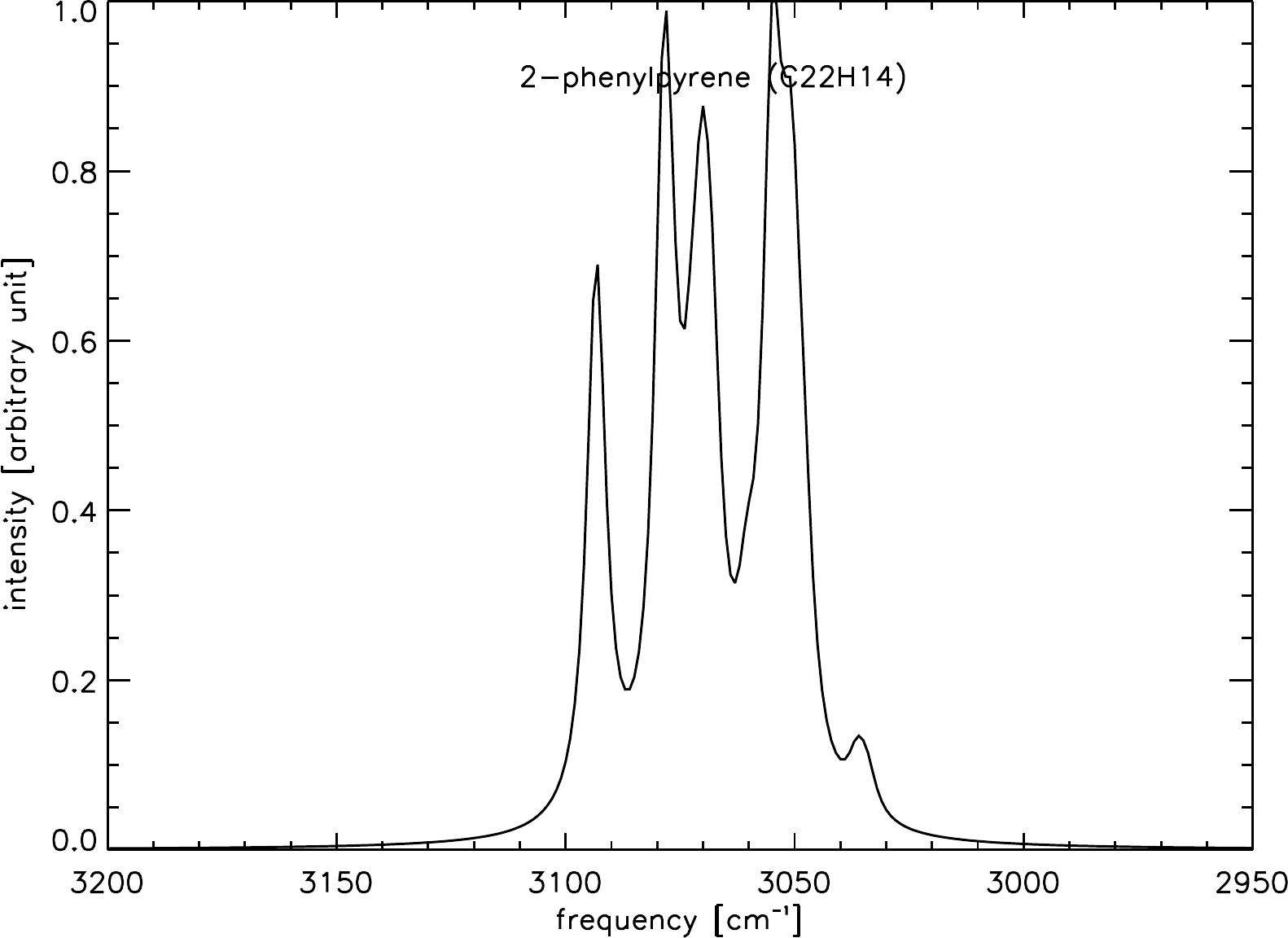}&
                      \includegraphics[width=0.4\textwidth]{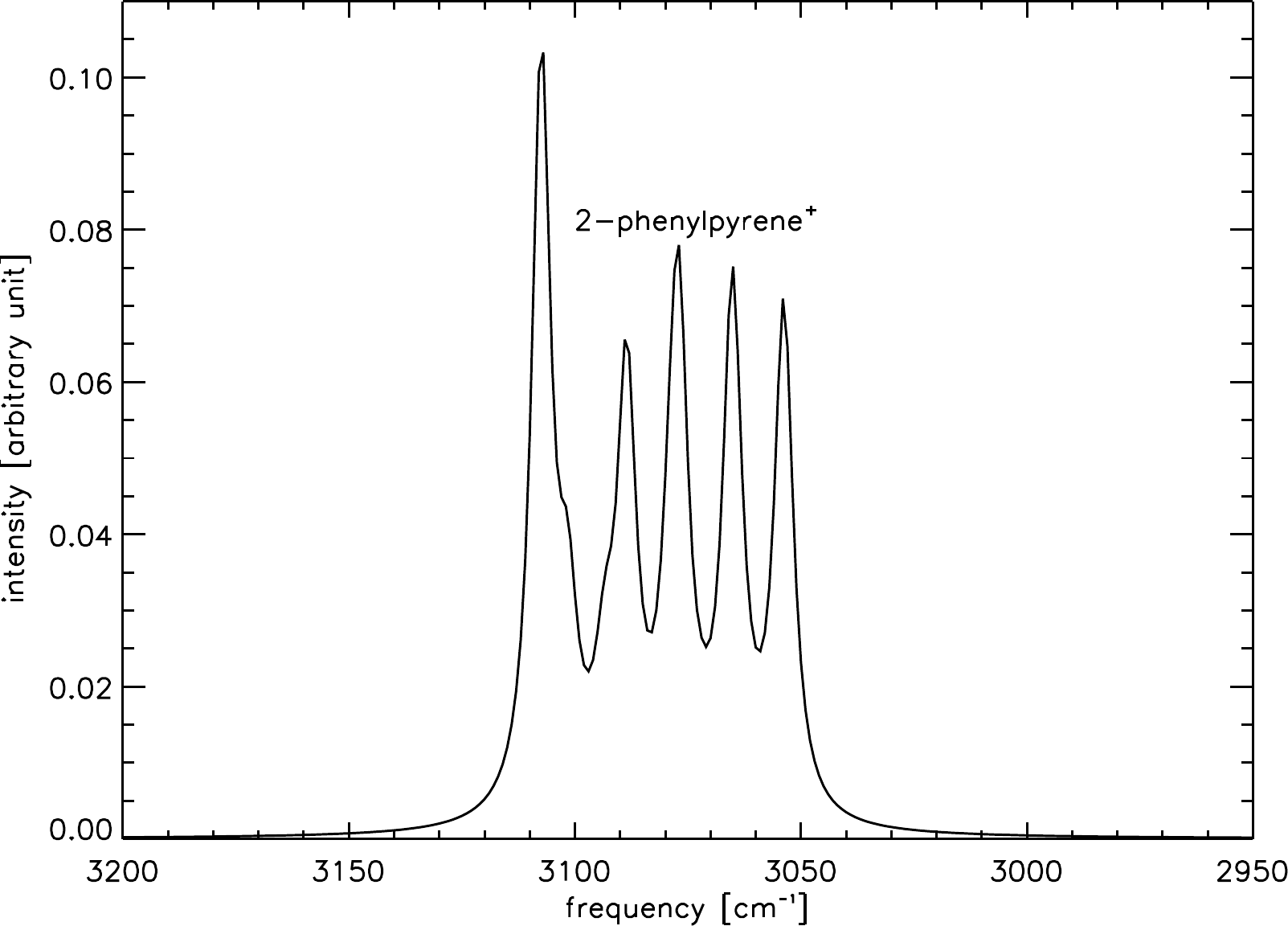}\\
                      \includegraphics[width=0.4\textwidth]{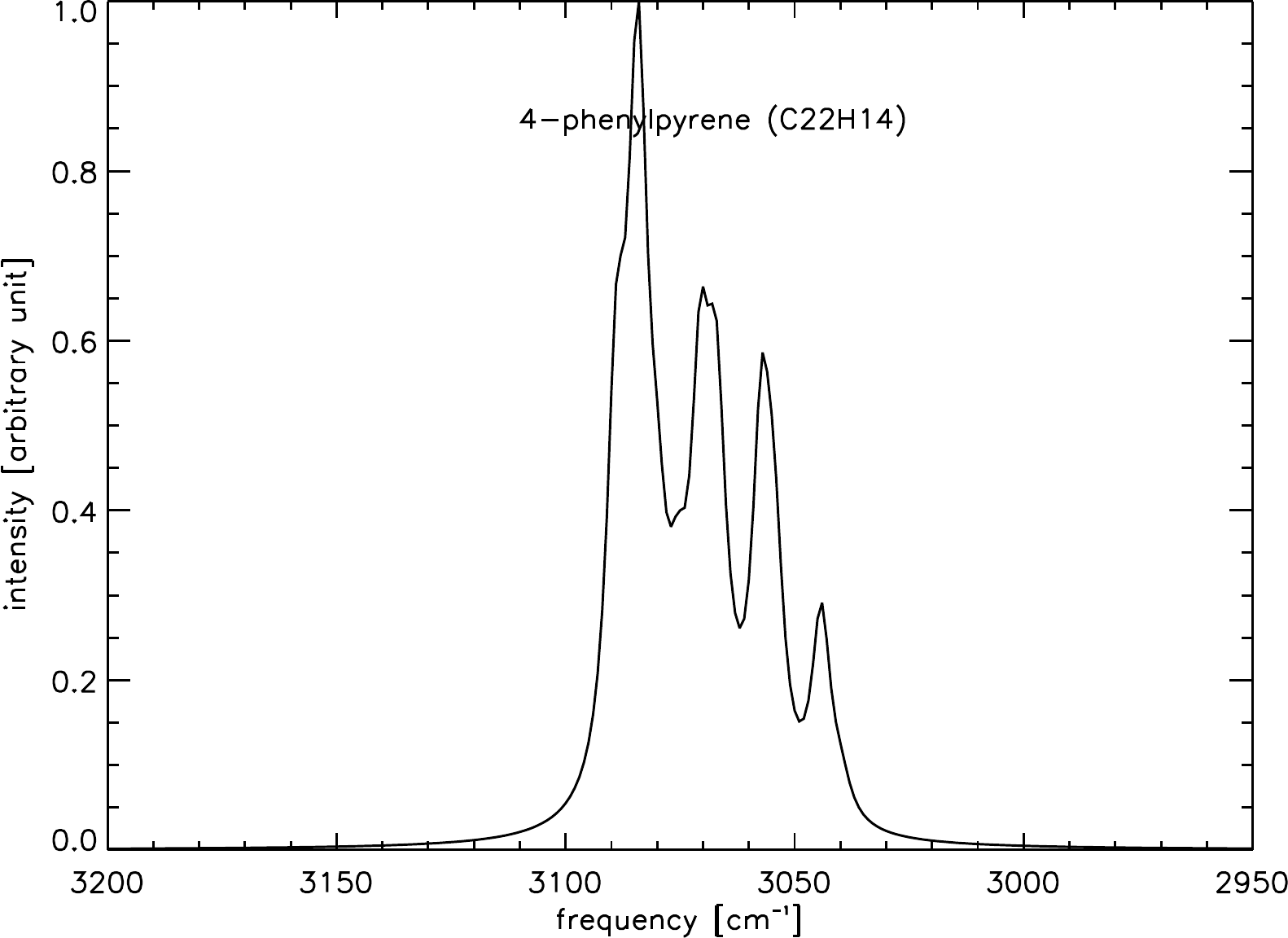}&
                      \includegraphics[width=0.4\textwidth]{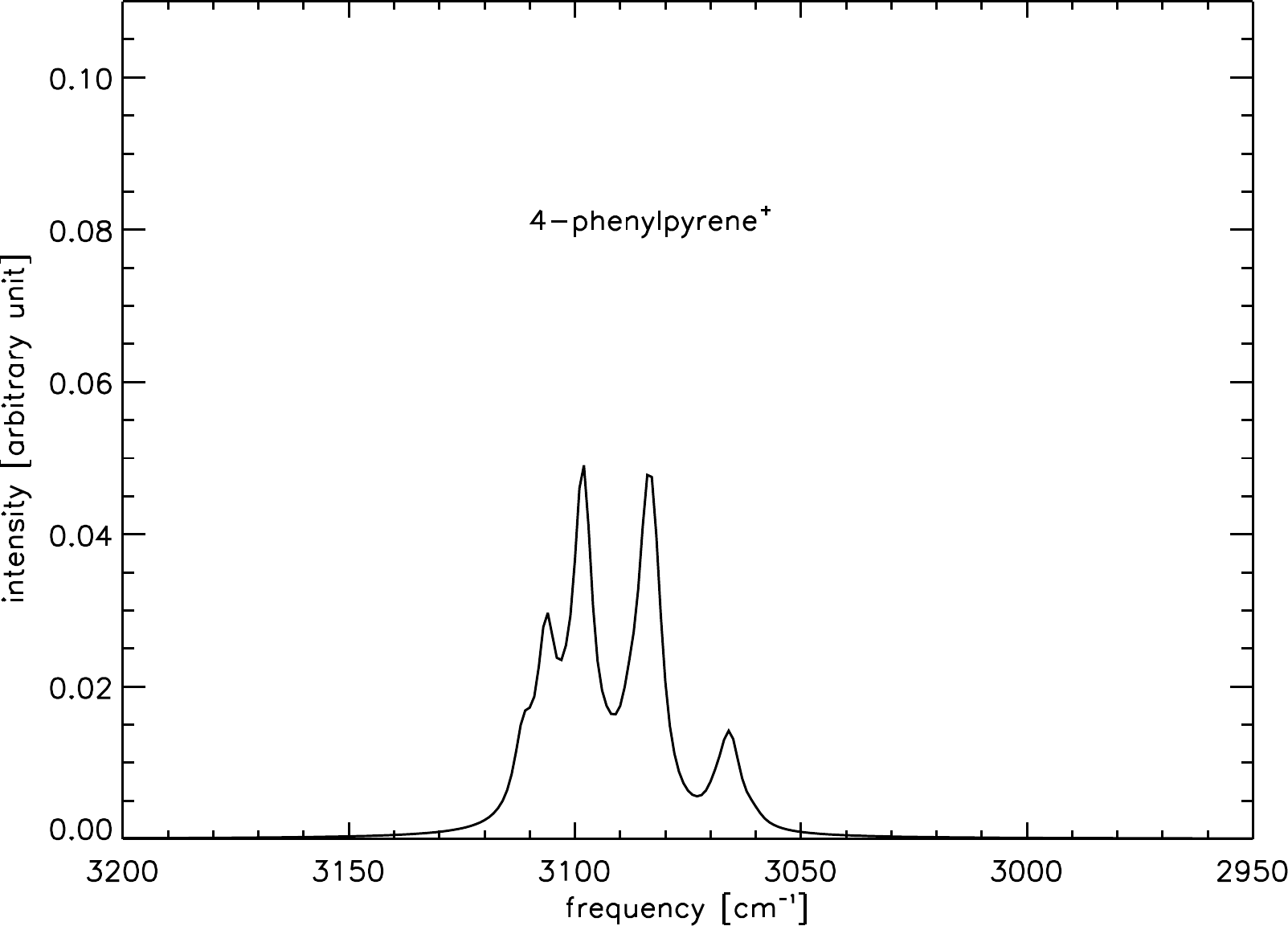}
                     
       \end{tabular}
\caption{The C-H stretch spectral region of infrared spectra in neutrals and cations of phenylpyrenes.}
\end{figure}

\clearpage
\begin{figure}
     \ContinuedFloat
        \centering
        \begin{tabular}{cc}
                      \includegraphics[width=0.4\textwidth]{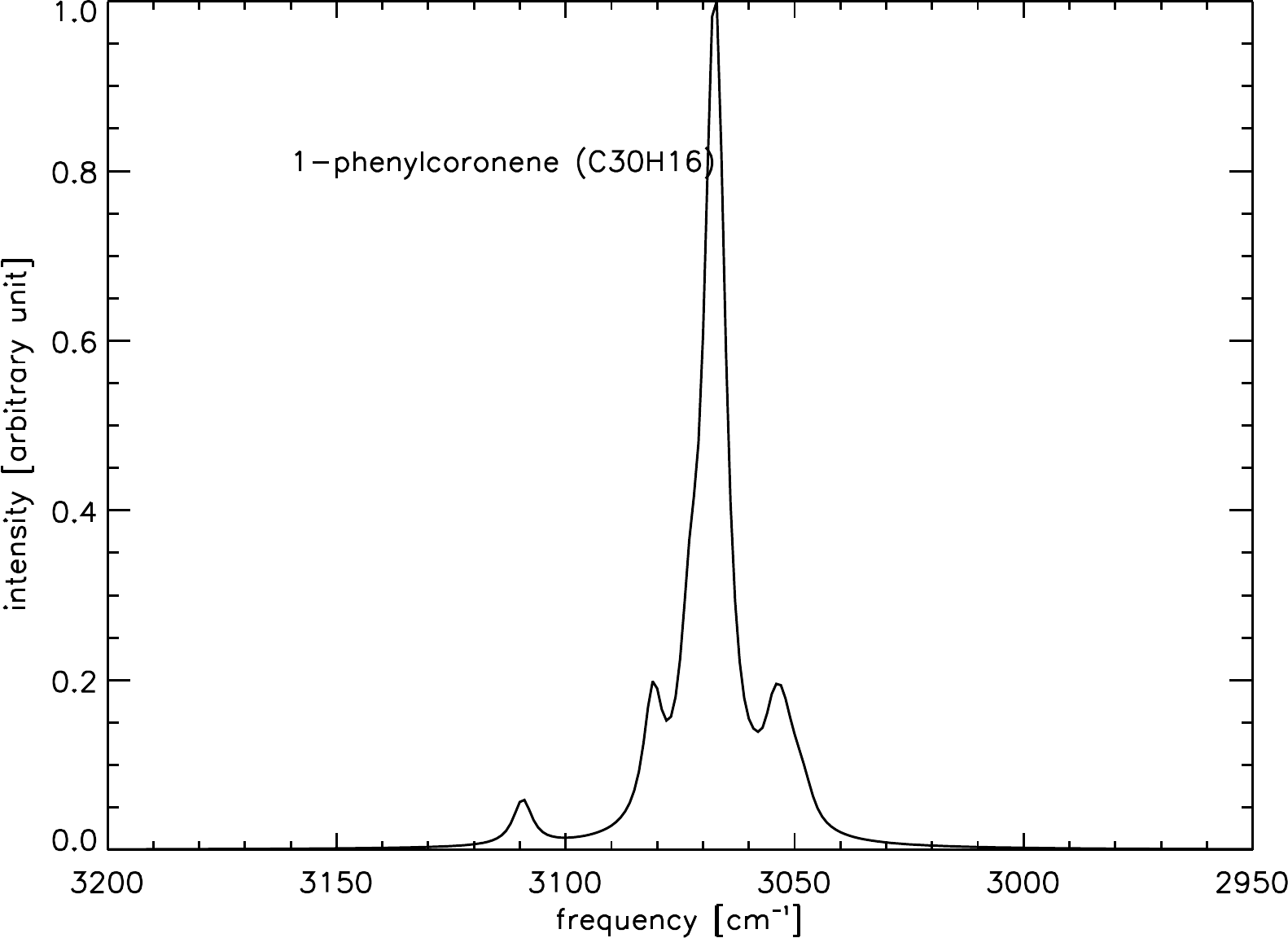}&
                      \includegraphics[width=0.4\textwidth]{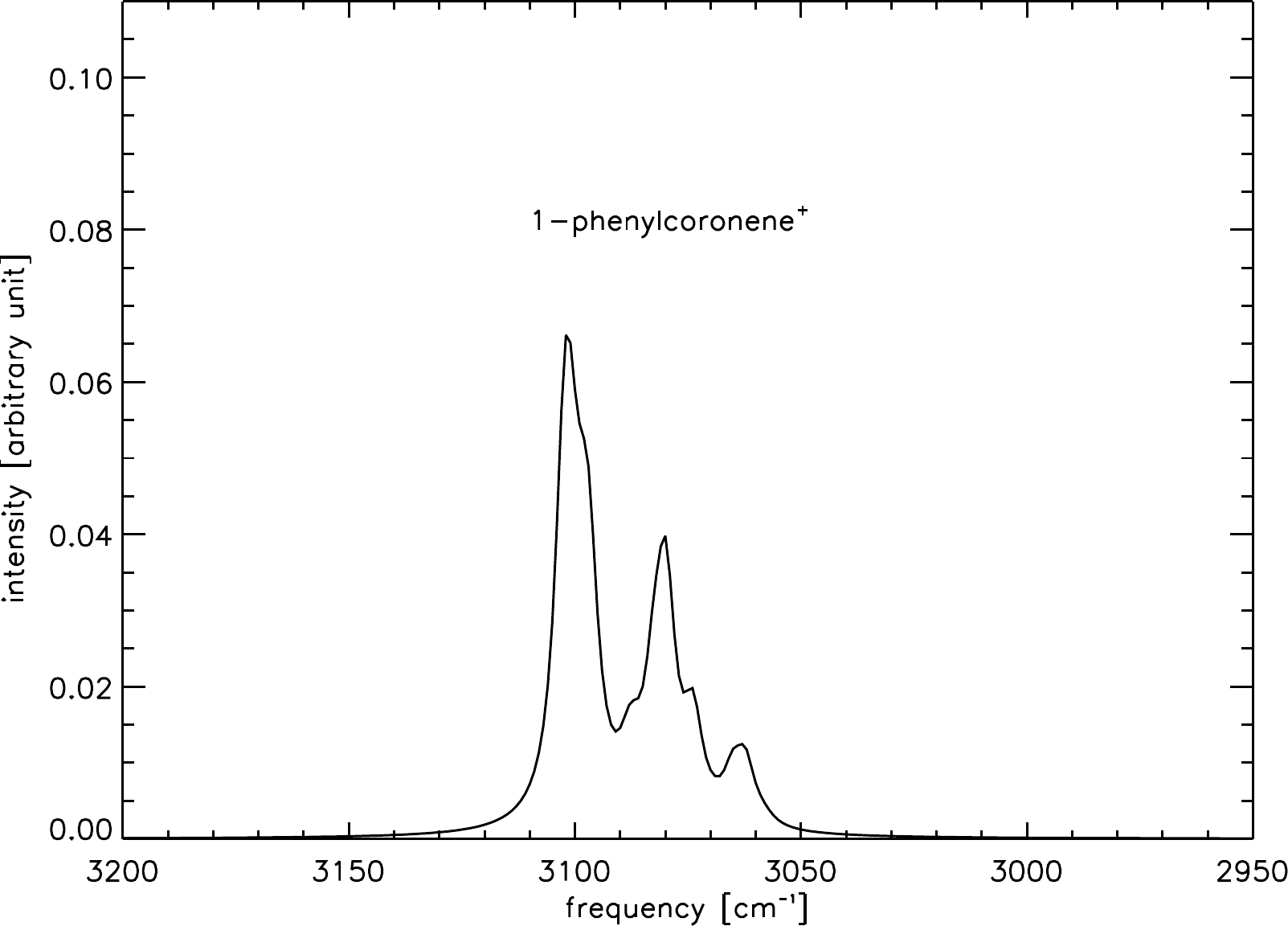}\\
                      \includegraphics[width=0.4\textwidth]{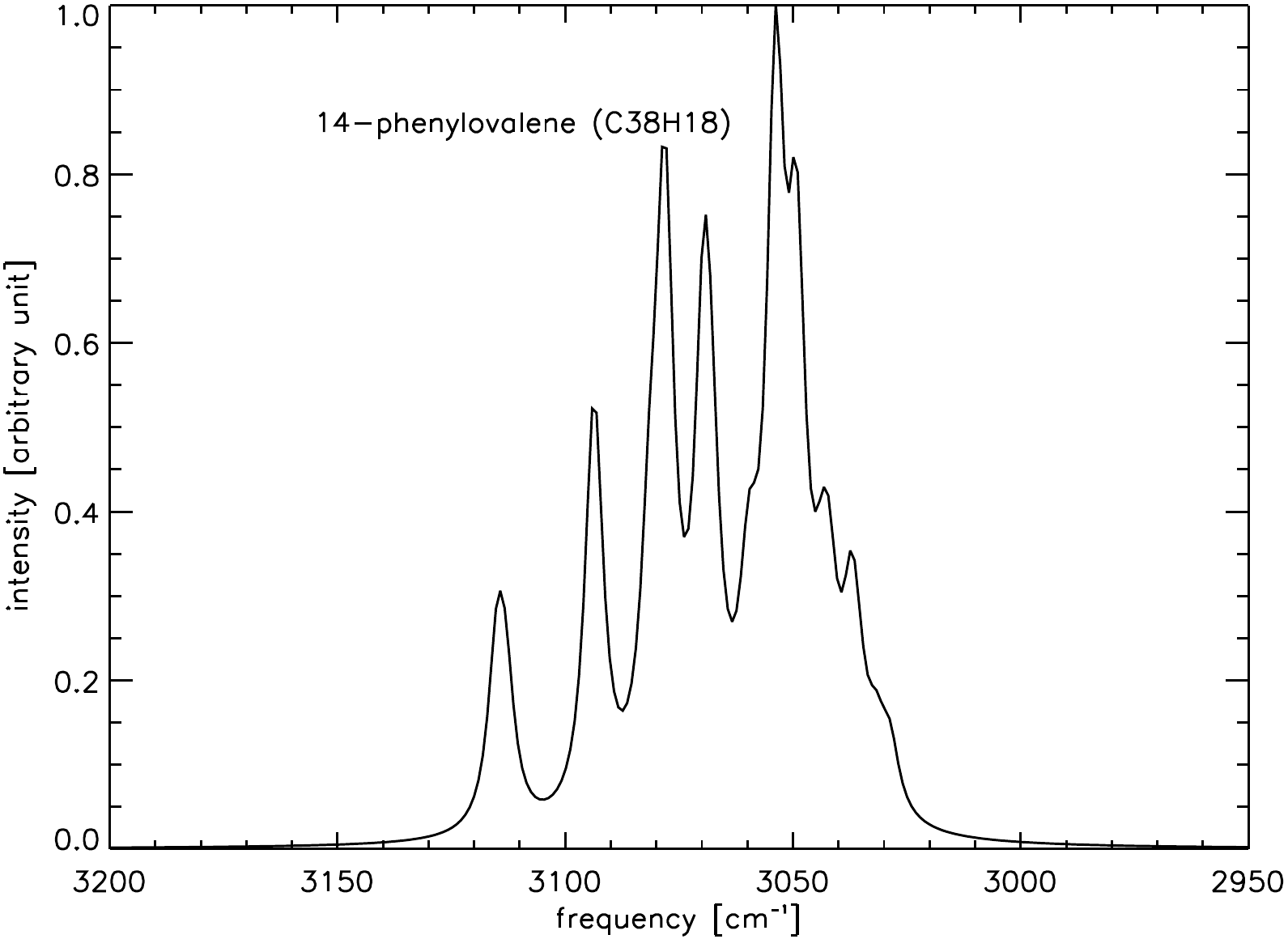}&
                      \includegraphics[width=0.4\textwidth]{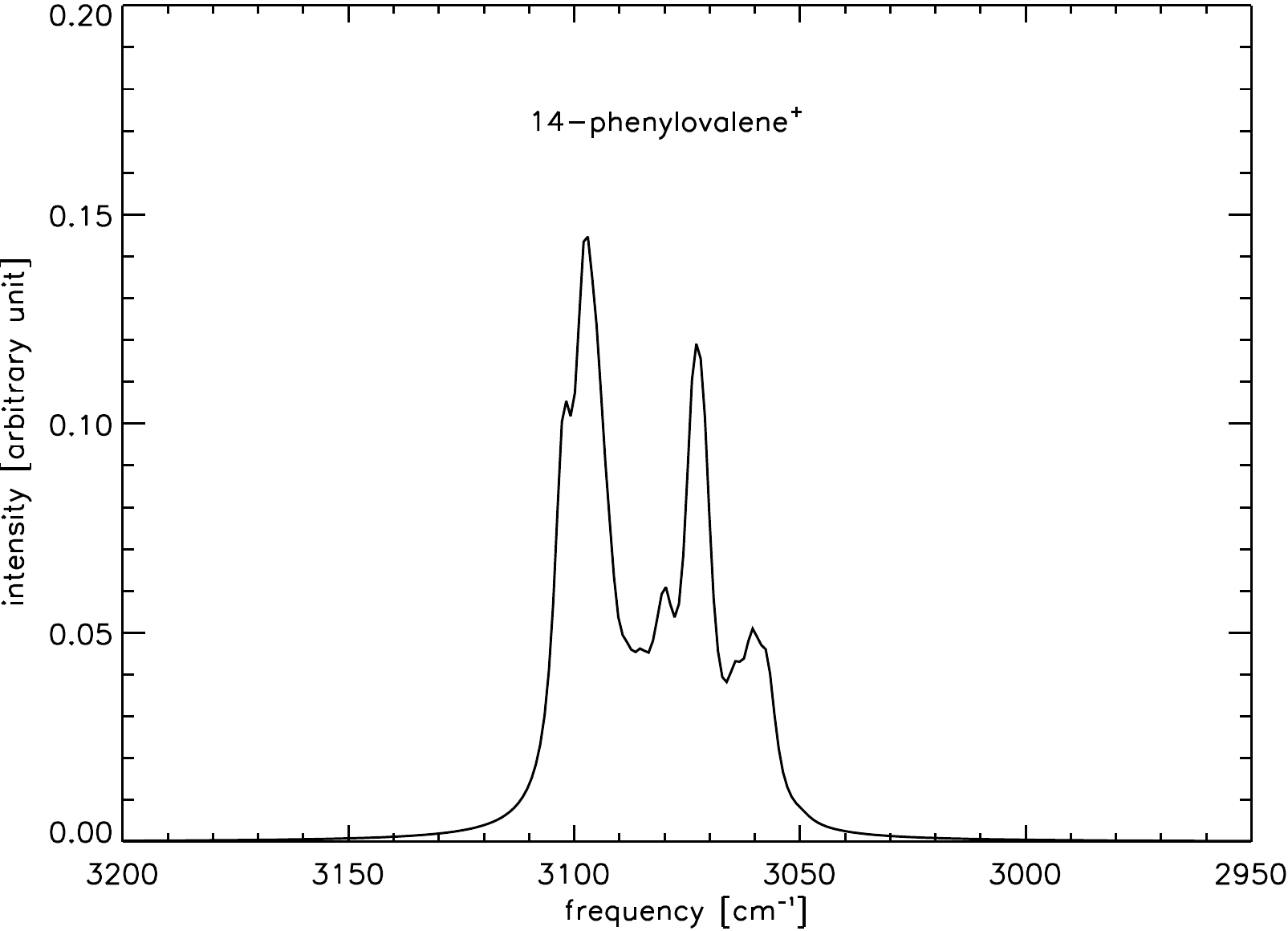}

   \end{tabular}
\caption{The C-H stretch spectral region of infrared spectra in neutrals and cations of phenylcoronene and phenylovalene.}
\end{figure}

\clearpage

\clearpage


\begin{table}
\caption{Optimized geometry-phenyl group torsion angle and energies}
\label{tab1}
\begin{tabular}{cccccc}
\hline
\multicolumn{1}{c}{Phenyl-PAHs}&\multicolumn{2}{c}{torsion angle (degree)}&\multicolumn{2}{c}{optimization energy (Kcal/mol)}&\multicolumn{1}{c}{Energy change}\\
 & & & & &\multicolumn{1}{c}{$\Delta$~E}\\
     &\multicolumn{1}{c}{neutral}&\multicolumn{1}{c}{cation}&\multicolumn{1}{c}{neutral}&\multicolumn{1}{c}{cation}\\
\hline
Biphenyl&39.09&19.69&-463.0040821&-462.7266937&-0.2774 \\
1-phenylnaphthalene&57.48&40.04&-616.5482200&-616.2894685&-0.2588 \\
2-phenylnaphthalene&38.45&26.47&-616.5521073&-616.2921034&-0.2600 \\
2-phenylanthracene&37.3&29.6&-770.0938302&-769.8543519&-0.2395 \\
9-phenylanthracene&80.6&58.7&-770.0865564&-769.8478848&-0.2387 \\
1-phenylpyrene&54.93&46.65&-846.2877170&-846.0481901&-0.2395 \\
2-phenylpyrene&38.12&36.44&-846.2910842&-846.0439777&-0.2471 \\
4-phenylpyrene&53.47&46.94&-846.2876183&-846.0457597&-0.2419 \\
1-phenylcoronene&59.05&48.08&-1152.2268317&-1151.9875017&-0.2381 \\
14-phenylovalene&90.00&90.10&-1458.1499689&-1457.9303203&-0.2196 \\
\hline  

\end{tabular}
\end{table}
\clearpage

\begin{table}
\vspace{-1.0in}
\caption{Comparison of corresponding C-H out-of-plane vibrational frequencies between neutral and ionized Phenyl-PAHs}
\label{tab1a}
\small
\begin{tabular}{ccccccc}
\hline
\multicolumn{1}{c}{Phenyl-PAHs}&\multicolumn{3}{c}{frequency (cm$^{-1}$)}&\multicolumn{2}{c}{absolute intensity}&\multicolumn{1}{c}{intensity}\\
& & & &\multicolumn{2}{c}{(Debye$^2$/AMU $\AA$$^2$)}&\multicolumn{1}{c}{$\Delta_{{fractional-difference}^{*}}$}\\
     &\multicolumn{1}{c}{$\nu_{neutral}$}&\multicolumn{1}{c}{$\nu_{cation}$}&\multicolumn{1}{c}{$\Delta \nu$}&\multicolumn{1}{c}{neutral}&\multicolumn{1}{c}{cation}&\multicolumn{1}{c}{}\\
\hline
Biphenyl&692.0$^{a}$&626.7&65.3&0.76& 1.11 & -0.46\\
 & 732.9$^{a}$&748.1&-15.2&1.07&1.93&-0.51\\
 & 733.2$^{a}$& -- & -- & 0.55 & -- & -- \\
1-phenylnaphthalene& 695.9$^{a}$&676.3&19.6&0.53&0.56&-0.06\\
& 753.6$^{a}$& 752.7& 0.9&0.46&0.19&0.59\\
& 774.0& 760.2&13.8&1.25&1.96&-0.57\\
& 794.9& 800.7& -5.8&0.63&0.32&0.49\\
2-phenylnaphthalene&692.3$^{a}$&664.3&28.0&0.57&0.72&-0.26\\
 & 753.7&770.0&-16.3&0.92&1.53&-0.66\\
 &765.9&752.6&13.3&0.50&0.31&0.38\\
 &813.5&872.2&-58.7&0.55&0.44&0.20\\
2-phenylanthracene&692.7$^{a}$&678.5&14.2&0.51&0.58&-0.14\\
 &735.1&747.9&-12.7&1.05&0.79&0.25\\
 &881.8&909.2&-27.4&0.75&0.90&-0.20\\
9-phenylanthracene& 694.3$^{a}$&692.8&1.5&0.56&0.58&-0.04\\
 & 731.0&743.5&-12.5&1.21&1.16&0.04\\
 & 747.7$^{a}$&751.7&-4.0&0.42&0.56&-0.33\\
 & 876.1&913.1&-37.0&0.50&0.27&0.46\\
1-phenylpyrene&696.4$^{a}$&686.3&10.1&0.65&0.67&-0.03\\
 &747.5&745.5&2.0&0.28&0.13&0.53\\
 &752.0&756.8&-4.8&0.18&0.08&0.55\\
 & 761.0$^{a}$&763.3&-2.3&0.28&0.57&-1.04\\
 &843.9&858.4&-14.5&1.89&1.81&0.04\\
2-phenylpyrene&693.3$^{a}$&691.6&1.7&0.56&0.81&-0.45\\
 & 741.0&734.1&6.9&0.42&0.38&0.10\\
 & 763.5&758.7&4.8&0.29&0.48&-0.66\\
 & 828.2&836.1&-7.9&0.84&1.05&-0.25\\
 & 874.8&889.5&-14.7&1.08&1.00&0.07\\
4-phenylpyrene&696.6$^{a}$&689.4&7.2&0.50&0.28&0.44\\
 & 748.0&749.2&-1.2&0.20&0.16&0.20\\
 & 768.8$^{a}$&766.3&2.5&0.22&0.21&0.05\\
 & 825.5&839.3&-13.8&1.05&0.90&0.14\\
 & 866.6&889.9&-23.3&0.43&0.54&-0.26\\
1-phenylcoronene&696.2$^{a}$&690.8&5.4&0.48&0.69&-0.44\\
 & 757.8$^{a}$&765.0&-7.2&0.28&0.17&0.39\\
 & 846.3&864.6&-18.3&1.47&2.38&-0.62\\
& 846.8& -- & -- & 0.62 & -- & -- \\
 & 884.6&902.0&-17.4&0.70&0.70&0.00\\
14-phenylovalene& 695.2$^{a}$&695.3&-0.1&0.33&0.25&0.24\\
 & 705.7$^{a}$&706.9&-1.2&0.24&0.17&0.29\\
 & 752.0$^{a}$&749.7&2.3&0.50&0.95&-0.90\\
 & 835.8&851.8&-16.0&1.75&2.20&-0.26\\
 & 870.3&897.6&-27.3&1.57&1.19&0.24\\
 
\hline
\multicolumn{7}{p{0.85\textwidth}}{$^{*}$the fractional difference of intensity is defined as $\Delta$=(neutral-cation)/neutral. $^{a}$Phenyl C-H wag mode.}

\end{tabular}
\end{table}

\begin{landscape}
\begin{table}
\hspace{-5.0in}
\caption{Comparison of out-of-plane vibrations of phenyl-PAHs to the C-H out-of-plane vibrations in the corresponding plain PAHs.  
The unit of the frequencies given in the table is in cm$^{-1}$ and the corresponding relative intensities are given in parenthesis. In column A the out-of-plane frequencies and relative intensities of phenyl-PAHs are given and in column B the out-of-plane frequencies and relative intensities of corresponding plain PAHs are given.}
\label{tab1b}
\small
\begin{center}
\begin{tabular}{ccccccccc}
\hline
\multicolumn{1}{c}{Phenyl-PAHs}&\multicolumn{2}{c}{Solo modes frequencies}&\multicolumn{2}{c}{Duo modes frequencies}&\multicolumn{2}{c}{trio modes frequencies}&\multicolumn{2}{c}{Quarto modes frequencies}\\
&\multicolumn{1}{c}{A}&\multicolumn{1}{c}{B}&\multicolumn{1}{c}{A}&\multicolumn{1}{c}{B}&
\multicolumn{1}{c}{A}&\multicolumn{1}{c}{B}&\multicolumn{1}{c}{A}&\multicolumn{1}{c}{B}\\

\hline
1-phenylnaphthalene&--&--&--&--&--&--&774.0(1.00)&--\\
&--&--&--&--&--&--&794.9(0.50)&788.2(1.00)$^{a}$\\
2-phenylnaphthalene&--&--&--&--&--&--&753.7(1.00)&--\\
&--&--&--&--&--&--&765.9(0.54)&-- \\
&--&--&--&--&--&--&813.5(0.60)&788.2(1.00)$^{a}$ \\
2-phenylanthracene&881.8(0.65)&878.3(0.77)$^{a}$&--&--&--&--&735.1(0.91)&729.6(1.00)$^{a}$ \\
9-phenylanthracene&876.1(0.39)&878.3(0.77)$^{a}$&--&--&--&--&731.0(0.95)&729.6(1.00)$^{a}$ \\
1-phenylpyrene&--&--&843.9(1.00)&842.8(1.00)$^{a}$&747.5(0.15)&--&--&-- \\
&--&--&--&--&752.0(0.10)&743.9(0.17)$^{a}$&--&-- \\
2-phenylpyrene&--&--&828.2(0.78)&842.8(1.00)$^{a}$&741.0(0.39)&743.9(0.17)$^{a}$&--&-- \\
&--&--&874.8(1.00)&--&763.5(0.27)&--&--&-- \\
4-phenylpyrene&--&--&825.5(0.90)&842.8(1.00)$^{a}$&748.0(0.17)&743.9(0.17)$^{a}$&--&-- \\
&--&--&866.6(0.37)&--&--&--&--&-- \\
1-phenylcoronene&--&--&846.5(1.00)&857.0(1.00)$^{a}$&--&--&--&-- \\
&--&--&884.6(0.36)&--&--&--&--&-- \\
14-phenylovalene&870.3(0.90)&901.9(0.99)$^{b}$&835.8(1.00)&847.4(0.71)$^{b}$&--&--&--&-- \\
\hline  
\multicolumn{9}{p{1.05\textwidth}}{$^{a}$Data taken from Hudgins1998a \cite{Hudgins1998a} and Hudgins1998b \cite{Hudgins1998b} for the observed Infrared frequencies in the corresponding plain PAHs. $^{b}$ Theoretical data taken from Pathak 2006 \cite{Pathak2006}.}

\end{tabular}

\end{center}
\end{table}
\end{landscape}

\end{document}